\documentclass{aa}

\usepackage{graphicx}
\usepackage{txfonts}
\usepackage{hyperref}
\usepackage{booktabs}
\usepackage{bm}
\usepackage[normalem]{ulem}
\usepackage[dvipsnames]{xcolor}
\usepackage{nicefrac}
\usepackage{mathtools}
\usepackage{hyperref}

\hypersetup{
	  colorlinks,
	  citecolor=blue,
	  linkcolor=blue,
	  urlcolor=blue}

\begin{document}

    \title{A finite-volume scheme for modeling compressible magnetohydrodynamic flows at low Mach numbers in stellar interiors}
    \titlerunning{A new magnetohydrodynamic solver for low-Mach-number flows}

    \author{
        G.~Leidi\inst{1,2}\and
        C.~Birke\inst{3}\and
        R.~Andrassy\inst{1}\and
        J.~Higl\inst{1}\and
        P.~V.~F.~Edelmann\inst{4}\and
        G.~Wiest\inst{1}\and
        C.~Klingenberg\inst{3}\and
        F.~K.~R{\"o}pke\inst{1,5}
        }
    \institute{
        Heidelberger Institut f{\"u}r Theoretische Studien,
        Schloss-Wolfsbrunnenweg 35, D-69118 Heidelberg, Germany\\
        \email{giovanni.leidi@h-its.org}
        \and
        Zentrum f\"ur Astronomie der Universit\"at Heidelberg, Astronomisches
        Rechen-Institut, M\"onchhofstr. 12-14, D-69120 Heidelberg, Germany
        \and
        Department of Mathematics, W\"urzburg University, Emil-Fischer-Str. 40, D-97074 W\"urzburg, Germany
        \and
        Computer, Computational and Statistical Sciences (CCS) Division and Center for Theoretical
        Astrophysics (CTA), Los Alamos National Laboratory, Los Alamos, NM
        87545, USA
        \and
        Zentrum f\"ur Astronomie der Universit\"at Heidelberg, Institut f\"ur
        Theoretische Astrophysik, Philosophenweg 12, D-69120 Heidelberg, Germany
    }

    \date{Received 02 August 2022 / accepted 18 September 2022}
    
    \abstract{Fully compressible magnetohydrodynamic (MHD) simulations are a fundamental tool for investigating the role of dynamo amplification in the generation of magnetic fields in deep convective layers of stars. The flows that arise in such environments are characterized by low (sonic) Mach numbers ($\mathcal{M}_\mathrm{son}\lesssim10^{-2}$). In these regimes, conventional MHD codes typically show excessive dissipation and tend to be inefficient as the Courant-Friedrichs-Lewy (CFL) constraint on the time step becomes too strict. In this work we present a new method for efficiently simulating MHD flows at low Mach numbers in a space-dependent gravitational potential while still retaining all effects of compressibility. The proposed scheme is implemented in the finite-volume $\textsc{Seven-League Hydro}$ (SLH) code, and it makes use of a low-Mach version of the five-wave Harten-Lax-van Leer discontinuities (HLLD) solver to reduce numerical dissipation, an implicit-explicit time discretization technique based on Strang splitting  to overcome the overly strict CFL constraint, and a well-balancing method that dramatically reduces the magnitude of spatial discretization errors in strongly stratified setups. The solenoidal constraint on the magnetic field is enforced by using a constrained transport method on a staggered grid. We carry out five verification tests, including the simulation of a small-scale dynamo in a star-like environment at $\mathcal{M}_\mathrm{son}\sim 10^{-3}$. We demonstrate that the proposed scheme can be used to accurately simulate compressible MHD flows in regimes of low Mach numbers and  strongly stratified setups even with moderately coarse grids.
}

    \keywords{magnetohydrodynamics (MHD) --  methods: numerical}

    \maketitle
%

\section{Introduction}\label{sec:introduction}
The interplay between turbulent convection and shear is fundamental in understanding the role of small- and large-scale dynamo mechanisms in the generation of strong  magnetic fields in stellar interiors.  These processes can only be modeled self-consistently with multidimensional magnetohydrodynamic (MHD) simulations \citep{brun2004,browning2006,browning2008,brown2010,ghizaru2010,kapyla2012,masada2013,karak2015,hotta2015,yadav2016,augustson2016,brun2017,rempel2018,kapyla2021}.  Nowadays, many codes used for astrophysical MHD rely on finite-volume discretization and Godunov-like methods to retain the conservative property of the MHD equations. This method is particularly suited for simulating flows in the transonic and supersonic regimes, which characterize many astrophysical systems. However, stars are objects in nearly magnetohydrostatic equilibrium (MHSE), and the flows arising from such stratifications have a very low sonic Mach number, typically $\mathcal{M}_\mathrm{son}= |\bm{V}|/a \lesssim 10^{-2}$ \citep{kupka2017}, where $\bm{V}$ is the flow speed and $a$ is the adiabatic sound speed. It is well known that conventional finite-volume schemes are not designed to work in such regimes \citep{viallet2011, miczek2015a, dumbser2019, minoshima2020}. First, the approximate Riemann solvers used in many astrophysical MHD codes, such as the Harten-Lax-van Leer \citep[HLLE;][]{einfeldt1991}, Roe \citep{cargo1997}, and Harten-Lax-van Leer discontinuities  \citep[HLLD;][]{miyoshi2005} solvers, show excessive numerical dissipation when the typical Mach number of the flow is below $10^{-2}$. Second, explicit time-steppers have to satisfy the Courant-Friedrichs-Lewy (CFL) stability criterion \citep{courant1928} so that the propagation of fast magnetosonic waves is resolved in time. This poses a severe limitation when simulating low-Mach-number flows. In this regime, the fast magnetosonic waves become parasitic since they transport very little energy and drastically reduce the time step. This makes convectional schemes exceedingly expensive for simulating the evolution of fluid motions and Alfv\'en waves, which are orders of magnitude slower than fast magnetosonic waves in deep layers of stars \citep{brun2005,browning2008,kapyla2010,augustson2016}. Lastly, standard Godunov-type schemes, by construction, cannot preserve stratifications in MHSE. This happens because  hyperbolic fluxes and  gravitational source terms are separately discretized and do not balance exactly in hydrostatic setups, which inevitably leads to the generation of spurious flows even in pure hydrodynamic simulations \citep{edelmann2021a}. This problem becomes even more critical in steep stratifications, where the spatial reconstruction from cell centers results in large jumps in the pressure at the cell interfaces, considerably accelerating the fluid along the gravity vector. Such numerical artifacts can dominate over the physical convective motions, leading to unreliable results.

Difficulties in modeling low-Mach-number MHD flows in stellar interiors are usually overcome by using alternative approaches based on a different formulation of the physical problem. One of them consists in artificially boosting the energy flux (or energy generation) to drive faster convective motions, such that the typical Mach number of the resulting flows falls above the low-Mach regime ($\mathcal{M}_\mathrm{son}  \gtrsim 10^{-2}$), where explicit time-steppers can be used efficiently \citep{kapyla2011,kapyla2012,kapyla2013,viviani2019,kapyla2021}. However, this approach significantly enhances the relative fluctuations of thermodynamic quantities \citep{warnecke2016,kapyla2020}, alters the mixing  at convective boundaries \citep{hotta2017,kapyla2019}, and modifies the spectrum of internal waves in radiative regions of stars \citep{rogers2013,edelmann2019,horst2020a,higl2021}. Another approach consists in solving the set of MHD equations using the anelastic approximation \citep{glatzmaier1984, glatzmaier1985,  brun2004, jones2009, gastine2012a, gastine2012b, smolarkiewicz2013, featherstone2016}, which filters out the fast magnetosonic waves, alleviating the overly strict constraint on the time step. However, such an approximation cannot model the excitation of the compressible  pressure modes. Another way to overcome the constraint on the time step is to drastically reduce the speed of the fast magnetosonic waves \citep{rempel2005,hotta2015}; again, at the cost of modifying the original set of MHD equations.

A numerical scheme that is capable of efficiently solving the fully compressible MHD equations at low sonic Mach numbers in strongly stratified setups is still missing. In this work we present a new method that aims to fill this gap. This can only be accomplished  (i) by reducing the numerical dissipation, (ii) by overcoming the strict CFL condition, and (iii) by preserving the background stratification in MHSE over long timescales. For aspect (i), we use a low-Mach version of the five-wave HLLD solver \citep[][]{minoshima2021}, whose numerical dissipation is independent of the sonic Mach number of the modeled flow in subsonic regimes.

In order to deal with aspect (ii), time-implicit discretization techniques can be used. Most of the fully implicit \citep[e.g.,][]{aydemir1985, charlton1990, chacon2008, lutjens2010} and semi-implicit \citep[e.g.,][]{harned1985,schnack1987,lerbinger1991,glasser1999,jardin2012,fambri2021} MHD schemes presented in the literature are designed to simulate magnetically confined plasmas and low-$\beta$ environments, where $\beta$ is defined as the ratio of the gas pressure to the magnetic pressure: $\beta=p/p_B$. In such strongly magnetized plasmas, both fast magnetosonic and Alfv\'en waves put a strong limit on the time step. On the contrary, plasmas in stellar interiors are characterized by high $\beta$ values \citep{mestel1999}\footnote{Low-$\beta$ environments can be found in the outer layers of active stars, like the solar corona.}, and the only source of stiffness is the generation of fast magnetosonic waves. Thus, Alfv\'en waves do not need to be treated implicitly, which greatly simplifies the numerical problem. Recently, \cite{dumbser2019} developed a semi-implicit conservative method that treats only the fast magnetosonic waves implicitly; however, that scheme cannot be implemented easily within the framework of our hydrodynamic code. In this work we construct an alternative time-marching scheme suitable for modeling high-$\beta$ plasmas at low Mach numbers, based on the approach described by \cite{fuchs2009}, in which the induction equation is solved in a separate step and coupled to the rest of the system  through Strang splitting \citep{strang1968}. As the high speed of the fast magnetosonic waves is mostly determined by the pressure flux in the momentum equation, we solve the subset containing the continuity, momentum, and energy equations implicitly, whereas the induction equation is integrated using an explicit time-stepper. For stability, the time step is now limited by the fastest fluid and Alfv\'en speeds on the grid; it is approximately $1/\mathcal{M}_\mathrm{son}$ longer than that allowed by the CFL condition, which leads to a considerable speed-up when the Mach number of the flow is low. Since the update on the induction equation is performed in a separate step, the flux-Jacobian in the time-implicit part of the algorithm does not need to be evaluated with respect to the magnetic field components. This allows for more flexibility when choosing the method that evolves the magnetic field. In particular, we use a staggered formulation of constrained transport \citep[CT-contact;][]{gardiner2005} to keep $\nabla \cdot \bm{B} = 0$ to machine precision, at least for a specific discretization of the divergence of the magnetic field.

Finally, aspect (iii) is solved by using the deviation well-balancing method \citep{berberich2019,edelmann2021a}, which allows the a priori known background stratification in MHSE to be preserved, dramatically reducing the magnitude of numerical errors and the strength of spurious flows\footnote{A similar approach in which the states are split into a background component and deviations is described in \cite{vogler2005}, \cite{khomenko2006}, \cite{felipe2010}, and \cite{hotta2015}.}. Recently, \cite{cuissa2022} performed fully compressible simulations of stellar magneto-convection at $\mathcal{M}_\mathrm{son} \sim 10^{-3}$ using a well-balancing technique similar to the deviation method. However, their scheme relied on explicit time-steppers, and it was used to simulate only 2.5 convective turnovers. Moreover, they did not cure the excessive dissipation of the HLLD solver at low Mach numbers.

These methods have been implemented in the $\textsc{Seven-League Hydro}$ (SLH) code,
which has already been used in the past to simulate convective boundary mixing, shear instabilities, and wave excitation in stellar interiors, even in regimes of low Mach numbers \citep{miczek2013a, edelmann2014a, miczek2015a, edelmann2016b, edelmann2017a, horst2020a, horst2021a, andrassy2022}. We stress that the current MHD implementation in SLH is not suitable for modeling low-$\beta$ plasmas. For simulating such regimes, a different method should be used instead, which is beyond the scope of this work.

In Sect.~\ref{sec:MHD_equations} we summarize the main properties of the fully compressible MHD equations with gravity. In Sects.~\ref{sec:spatial_discretization} and \ref{sec:time-integration} we provide details on the numerical algorithms implemented in SLH. In Sect.~\ref{sec:tests} several numerical experiments are run with the new MHD scheme in order to check its accuracy and efficiency in simulating flows at low Mach numbers, even in the presence of a steep stratification. Finally, in Sect.~\ref{sec:conclusions}
we draw conclusions and summarize the fundamental aspects of the proposed algorithm.

\section{Equations of compressible ideal MHD with gravity}
\label{sec:MHD_equations}

The MHD scheme implemented in SLH is designed to solve the set of compressible ideal MHD equations with a (time-independent) gravitational source term\footnote{However, other source terms can be added to the system depending on the problem at hand. These include energy generation by nuclear reactions, radiative transport of energy in the diffusion limit, neutrino cooling, and parabolic viscous terms.}:

\begin{align}
        \label{eq:mass_conservation}
   \frac{\partial \rho}{\partial t} + \nabla \cdot (\rho \bm{V}) &= 0,
    \\
    \label{eq:momentum_conservation}
   \frac{\partial  (\rho \bm{V})}{\partial t} + \nabla \cdot  [ \rho \bm{V}
   \otimes \bm{V} + (p+p_B) \bm{I} - \bm{B} \otimes \bm{B}] &= \rho \bm{g},
   \\
   \label{eq:Energy_conservation}
   \frac{\partial (\rho E_\phi)}{\partial t} + \nabla \cdot [( \rho E_\phi + p + p_B )
   \bm{V} - \bm{B}(\bm{B} \cdot \bm{V}) ]  &= 0,
   \\
   \label{eq:induction_equation}
   \frac{\partial  \bm{B}}{\partial t} + \nabla \cdot  ( \bm{V}
   \otimes \bm{B} - \bm{B}
   \otimes \bm{V} )  &= \bm{0},
\end{align}
where $\rho$ denotes the density, $\bm{V}=(V_x,V_y,V_z)$ the velocity field, $\bm{B}=(B_x,B_y,B_z)$ the magnetic field\footnote{Throughout the paper we use the Lorentz-Heaviside notation: $\bm{B}=\bm{b}/\sqrt{4\pi}$.}, $\bm{I}$ the unit tensor, $\bm{g}=(g_x,g_y,g_z)$ the gravitational acceleration, $p$ the gas pressure and $p_B=|\bm{B}|^2/2$ the magnetic pressure. The total energy density $\rho E_\phi$ is defined as
\begin{equation}
\label{total-energy}
\rho E_\phi = \rho e_\mathrm{int} + \frac{1}{2} \rho | \bm{V} |^2 + \frac{1}{2}  | \bm{B} |^2   + \rho e_\phi,
\end{equation}
where $e_\mathrm{int}$ and $e_{\phi}$ are the specific internal and gravitational energies\footnote{If the gravitational potential is time independent, solving the energy equation for $\rho E_{\phi}$ instead of $\rho E$ allows the $\rho \bm{g}\cdot \bm{V}$ source term  to be removed. This leads to more accurate results and better entropy- and energy-conservation properties in simulations of gas dynamics with gravity \citep{muller2020,edelmann2021a}.}.

The system is closed by an equation of state (EoS), which provides the numerical value of the gas pressure. Several different definitions for the EoS can be used in SLH, including a simple ideal gas law, radiation pressure, and a tabulated EoS \citep[Helmholtz EoS;][]{timmes2000} that allows the effects of electron degeneracy and Coulomb corrections to be included, which are often needed to properly describe the thermodynamic conditions found in stellar  material.

\subsection{Eigenstructure of the MHD system and definition of Mach numbers}
\label{sec:eigenstructure}
\begin{figure}
  \centering
  \includegraphics[width=0.5\textwidth]{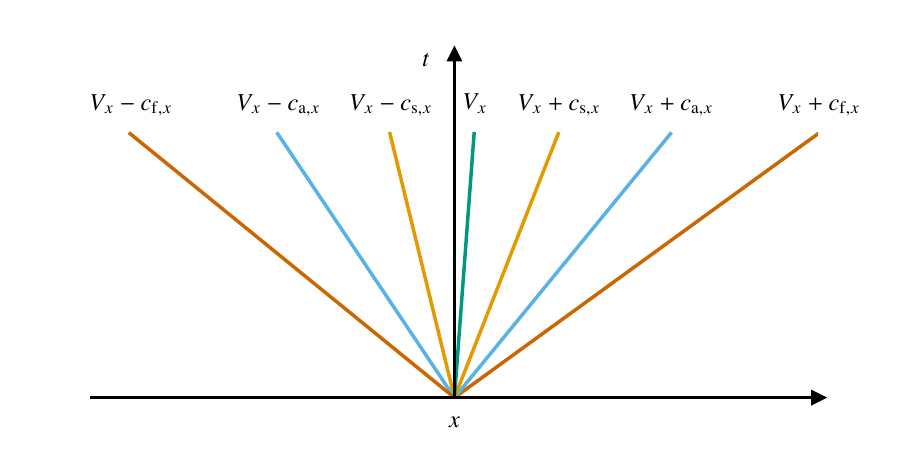}
  \caption{Wave structure of the MHD system.}
\label{fig:riemann-fan}
\end{figure}

The homogeneous MHD system (left-hand side of Eqs. \ref{eq:mass_conservation}-\ref{eq:induction_equation}) reduced to one spatial dimension has seven nonzero eigenvalues\footnote{Here $x$ represents a generic direction.},
\begin{equation}
\lambda_{1,7} = V_x \mp c_{\mathrm{f},x}, \
\lambda_{2,6} = V_x \mp c_{\mathrm{a},x}, \
\lambda_{3,5} = V_x \mp c_{\mathrm{s},x}, \
\lambda_4 = V_x,
\end{equation}
associated with different modes of propagation: left/right fast magnetosonic waves, left/right Alfv\'en waves, left/right slow magnetosonic waves and one entropy wave. $c_{\mathrm{s},x}$, $c_{\mathrm{a},x}$, and $c_{\mathrm{f},x}$ are the slow magnetosonic, Alfv\'en, and fast magnetosonic speeds,
\begin{equation}
c_{\mathrm{f},\mathrm{s},x} = \left[ \frac{1}{2} \left( a^2 + \frac{|\bm{B}|^2}{\rho} \pm \sqrt{\left( a^2+\frac{|\bm{B}|^2}{\rho}\right)^2-4a^2c^2_{\mathrm{a},x}} \right) \right]^ {\frac{1}{2}},
\end{equation}
\begin{equation}
c_{\mathrm{a},x} = |B_x|/\sqrt{\rho},
\end{equation}
with the adiabatic sound speed $a$ defined as
\begin{equation}
a = \left( \frac{\partial  p}{\partial \rho} + \frac{p}{\rho^2}\frac{\partial p}{\partial e_\mathrm{int}}  \right)^{\frac{1}{2}}.
\end{equation}
As illustrated in Fig. \ref{fig:riemann-fan}, the MHD wave pattern has a fixed ordering:
\begin{equation}
\lambda_1 < \lambda_2 < \lambda_3 < \lambda_4 < \lambda_5 < \lambda_6 < \lambda_7.
\end{equation}
Alfv\'en waves correspond to incompressible modes of propagation, as they only carry perturbations in the velocity and magnetic field components orthogonal to the wave vector. Effects of compressibility are due to the propagation of slow and fast magnetosonic waves, while the entropy wave is simple advection of fluid.

In contrast to pure hydrodynamics, the more complex structure of the MHD waves allows several Mach numbers to be defined, depending on the considered reference velocity. In addition to $\mathcal{M}_{\mathrm{son}}$ (see Sect.~\ref{sec:introduction}), the Alfv\'en Mach number is defined as
\begin{equation}
\mathcal{M}_\mathrm{Alf} = \frac{|\bm{V}|}{\sqrt{c^2_{\mathrm{a},x}+c^2_{\mathrm{a},y}+c^2_{\mathrm{a},z}}},
\end{equation}
while the directional slow and fast magnetosonic Mach numbers are given by
\begin{equation}
\mathcal{M}_{\mathrm{slow},\mathrm{fast},x} = \frac{|V_x|}{c_{\mathrm{s},\mathrm{f},x}}.
\end{equation}

\subsection{Low-Mach limit of the MHD system}
\label{sec:low-mach-limit}

Magnetic fields amplified by dynamo mechanisms in deep convective layers of stars are likely to approach equipartition with respect to the kinetic energy content of the flow \citep{brandenburg2005,featherstone2009,augustson2016,hotta2017,kapyla2019}. To model such processes, the MHD system in Eqs. \ref{eq:mass_conservation}-\ref{eq:induction_equation} must be solved in regimes of Mach numbers $\mathcal{M}_{\mathrm{fast},x} \lesssim \mathcal{M}_\mathrm{son} \lesssim 10^{-2}$ and $\mathcal{M}_\mathrm{Alf} \sim 1$. To infer the structure of the solution under such conditions, it is useful to consider the nondimensional form of the fully compressible  MHD equations\footnote{For simplicity, we only consider the homogeneous MHD system in this analysis.},
\begin{align}
    \label{eq:mass_conservation-nd}
   \frac{\partial \hat{\rho}}{\partial \hat{t}} + \hat{\nabla} \cdot (\hat{\rho}\hat{\bm{V}}) &= 0,
    \\
    \label{eq:momentum_conservation-nd}
   \frac{\partial  (\hat{\rho}\hat{\bm{V}})}{\partial \hat{t}} + \hat{\nabla} \cdot  \left[ \hat{\rho} \hat{\bm{V}}
   \otimes \hat{\bm{V}} + \left(\frac{\hat{p}}{\hat{\mathcal{M}}^2_\mathrm{son}}+\frac{\hat{p}_B}{\hat{\mathcal{M}}^2_\mathrm{Alf}}\right) \bm{I} - \frac{ \hat{\bm{B}} \otimes \hat{\bm{B}}}{\hat{\mathcal{M}}^2_\mathrm{Alf}} \right] &= \bm{0},
   \\
   \frac{\partial (\hat{\rho} \hat{E})}{\partial \hat{t}} + \hat{\nabla} \cdot \left[\left( \hat{\rho}\hat{E} + \hat{p} + \hat{p}_B\frac{\hat{\mathcal{M}}^2_\mathrm{son}}{\hat{\mathcal{M}}^2_\mathrm{Alf}} \right)
   \hat{\bm{V}} - \hat{\bm{B}}\left(\hat{\bm{B}} \cdot \hat{\bm{V}}\right)\frac{\hat{\mathcal{M}}^2_\mathrm{son}}{\hat{\mathcal{M}}^2_\mathrm{Alf}} \right]  &= 0,
   \\
   \label{eq:induction-nd}
   \frac{\partial  \hat{\bm{B}}}{\partial \hat{t}} + \hat{\nabla} \cdot  ( \hat{\bm{V}}
   \otimes \hat{\bm{B}} - \hat{\bm{B}}
   \otimes \hat{\bm{V}} )  &= \bm{0}.
\end{align}
Here, the different variables have been rescaled by some reference quantity  representative of the physical system of interest: $x=\hat{x}x_\mathrm{r}$, $\rho=\hat{\rho}\rho_\mathrm{r}$, $\bm{V}=\hat{\bm{V}}V_\mathrm{r}$, $E=\hat{E} a_\mathrm{r}^2$, $p=\hat{p}\rho_\mathrm{r}a_\mathrm{r}^2$, $\bm{B}=\hat{\bm{B}}B_\mathrm{r}$. $\hat{\mathcal{M}}_\mathrm{son}=|V_\mathrm{r}|/a_\mathrm{r}$ and $\hat{\mathcal{M}}_\mathrm{Alf}=|V_\mathrm{r}|/(|B_\mathrm{r}|/\sqrt{\rho_\mathrm{r}})$ are the characteristic sonic and Alfv\'en Mach numbers of the flow.

In the limit of $\hat{\mathcal{M}}_\mathrm{son} \rightarrow 0$, Eqs. \ref{eq:mass_conservation-nd}-\ref{eq:induction-nd} approach the incompressible regime \citep[see][]{matthaeus1988}, in which the gas pressure is constant in space except for fluctuations $\propto\hat{\mathcal{M}}^2_\mathrm{son}$. As this solution does not allow for compressible modes of propagation, only Alfv\'en and entropy waves can transport fluctuations in the state variables across the physical domain. In the regime we are interested in ($\hat{\mathcal{M}}_\mathrm{Alf}\sim1$), these waves travel at similar speeds. However, the incompressible limit is not the only solution to the compressible MHD equations at $\hat{\mathcal{M}}_\mathrm{son} \ll 1$. In fact, both slow and fast magnetosonic waves can be propagated with arbitrary small velocity fluctuations. Fast magnetosonic waves in particular travel at much higher speed than Alfv\'en waves and fluid motions in the low-Mach limit. If the plasma-$\beta$ is high, the large fast magnetosonic speed $c_{\mathrm{f},x}$ is mostly determined by the pressure flux $\hat{p}/\hat{\mathcal{M}}^2_\mathrm{son}$ in Eq. \ref{eq:momentum_conservation-nd}. Since both slow incompressible flows and fast magnetosonic waves are permitted in the limit of $\hat{\mathcal{M}}_\mathrm{son} \rightarrow 0$, the system of fully compressible MHD equations is stiff.

\subsection{Magnetohydrostatic solutions}
\label{sec:hse}

Magnetohydrostatic stratifications are a special class of solutions to the MHD system with gravitational source terms (see Eqs. \ref{eq:mass_conservation}-\ref{eq:induction_equation}), where all the time derivatives are zero and the velocity is zero everywhere.
Under these conditions, the distribution of density, pressure and magnetic field is given by the magnetohydrostatic equation
\begin{equation}
\label{eq:hydrostatic_equilibrium}
    \nabla \cdot [ (p+p_B)\bm{I} - \bm{B}\otimes\bm{B} ] = \rho \bm{g}.
\end{equation}
Any set ($\rho$,$p$,$\bm{B}$) that solves Eq. \ref{eq:hydrostatic_equilibrium} is called a ``magnetohydrostatic solution''. Equation \ref{eq:hydrostatic_equilibrium} is undetermined, so a whole continuum of magnetohydrostatic solutions exists.

The stratification of stars is very well described by MHSE over a large fraction of their lifetime. Large deviations from MHSE are only expected in the late evolutionary stages of massive stars, in atmospheric layers and in stellar winds. Even though their stratification is continuously perturbed by a whole variety of physical processes over fast dynamical timescales, the amplitude of such perturbations remains small, and the overall structure of the star can be considered to be in MHSE. Significant changes to the stratification only happen over the much longer thermal and nuclear timescales \citep{kippenhahn2013}.

\subsection{The solenoidal constraint}
\label{sec:divB}
The magnetic field satisfies the solenoidal constraint
\begin{equation}
\label{eq:divB_constraint}
   \nabla \cdot \bm{B} = 0.
\end{equation}
This constraint has its origin in Maxwell's equation and states that physically no magnetic monopoles can exist. Solutions to Eqs. \ref{eq:mass_conservation}-\ref{eq:induction_equation} automatically satisfy this condition at all times if the initial field obeys the constraint. This can easily be illustrated by rewriting Eq. \ref{eq:induction_equation} into the equivalent form
\begin{equation}
\label{induction2}
    \frac{\partial \bm{B}}{\partial t} + \nabla \times \mathcal{E} = 0,
\end{equation}
where $\mathcal{E}=-\bm{V}\times \bm{B}$ is the electromotive force.
Applying the divergence to Eq. \ref{induction2} results in
\begin{equation}
    \frac{\partial (\nabla \cdot \bm{B} )}{\partial t} = 0.
\end{equation}

\section{Spatial discretization}
\label{sec:spatial_discretization}

The system of partial differential equations (PDEs) described in Sect.~\ref{sec:MHD_equations} takes the general conservative form
\begin{equation}
\label{eq:balance_law}
\frac{\partial \bm{U}}{\partial t} + \frac{\partial \bm{F}(\bm{U})}{\partial x} +
 \frac{\partial \bm{G}(\bm{U})}{\partial y}  +  \frac{\partial \bm{H}(\bm{U})}{\partial z}
= \bm{S}(\bm{U}),
\end{equation}
with the respective vector of conservative variables $\bm{U}$, physical fluxes $\bm{F}$, $\bm{G}$, $\bm{H}$ and source term $\bm{S}$. In SLH, Eq. \ref{eq:balance_law} is solved numerically using the finite-volume method \citep{leveque2002, toro2009a}, which is briefly summarized in the next section.

\subsection{Finite-volume discretization}
\label{sec:finite_volume}

In a first step, the physical system is mapped on a 3D Cartesian grid\footnote{Here we describe the 3D algorithm; however, 1D and 2D Cartesian grids can also be used in SLH.} divided into $N_x\times N_y \times N_z$ cells, whose spatial extent is given by $[x_L,x_R] \times [y_L,y_R] \times [z_L,z_R]$. Each cell in the computational domain is defined by the set of indices $(i,j,k)$, and its volume is given by the product of the spatial resolution elements along each axis: $\Theta_{i,j,k}=\Delta x \Delta y  \Delta z$. Any quantity located at the center of the cell refers to the same indices, while quantities located at the cell boundaries are denoted by sets of indices like $(i+1/2,j,k)$, which in this case defines the interface between cells $(i,j,k)$ and $(i+1,j,k)$.

Integrating Eq. \ref{eq:balance_law} over the cell volume leads to
\begin{equation}
\label{eq:semi-discrete}
\begin{split}
\frac{\partial \hat{\bm{U}}_{i,j,k}}{\partial t} = &- \frac{1}{\Delta x} \left( \hat{\bm{F}}_{i+1/2,j,k} - \hat{\bm{F}}_{i-1/2,j,k} \right) \\
&- \frac{1}{\Delta y} \left( \hat{\bm{G}}_{i,j+1/2,k} - \hat{\bm{G}}_{i,j-1/2,k} \right) \\
&- \frac{1}{\Delta z} \left( \hat{\bm{H}}_{i,j,k+1/2} - \hat{\bm{H}}_{i,j,k-1/2} \right) \\
& + \hat{\bm{S}}_{i,j,k},
\end{split}
\end{equation}
where $\hat{\bm{U}}_{i,j,k}$ is the volume-averaged vector of conserved quantities
\begin{equation}
\hat{\bm{U}}_{i,j,k} = \frac{1}{\Theta_{i,j,k}}\int_{\Theta_{i,j,k}} \bm{U} d\Theta.
\end{equation}
The same procedure applies to $\hat{\bm{S}}_{i,j,k}$, while the surface-averaged fluxes are defined as
\begin{equation}
\hat{\bm{F}}_{i+1/2,j,k} = \frac{1}{A_{i+1/2,j,k}}\int_{A_{i+1/2,j,k}} \bm{F}\cdot\hat{\bm{n}} dA,
\end{equation}
where $A_{i+1/2,j,k}$ is the area of the interface $(i+1/2,j,k)$ and $\hat{\bm{n}}$ is the normal to the surface pointing outward from the cell.

The right-hand side of Eq. \ref{eq:semi-discrete} can be discretized in space if suitable numerical values for $\hat{\bm{F}}_{i+1/2,j,k}$ and $\hat{\bm{S}}_{i,j,k}$ are provided. For the latter, a typical choice consists in substituting the volume-averaged quantity with its value in the center of the cell, which is accurate to second order:
\begin{equation}
\hat{\bm{S}}_{i,j,k} \simeq  \bm{S}_{i,j,k}.
\end{equation}
The computation of numerical fluxes, in contrast, needs more care, and upwind techniques must be used to achieve stability. The resulting system of ordinary differential equations (ODEs) is then discretized in time using the methods described in Sect.~\ref{sec:time-integration}.

\subsection{Numerical flux function}
\label{sec:flux_function}
\begin{figure}
  \centering
  \includegraphics[width=0.5\textwidth]{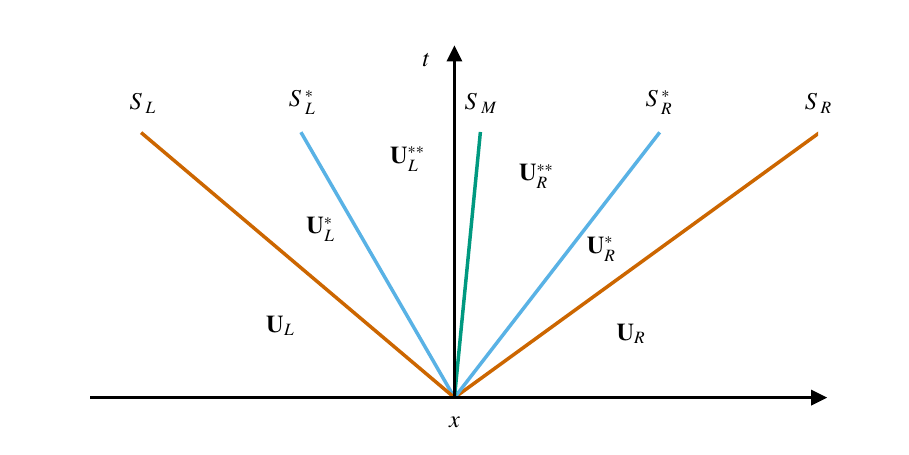}
  \caption{Wave structure of HLLD-type solvers: only the fast magnetosonic ($S_L$, $S_R$), Alfv\'en ($S^*_L$, $S^*_R$), and entropy ($S_M$) waves are considered in the computation of the states across the Riemann fan. The slow magnetosonic waves are discarded.}
\label{fig:riemann-fan-lhlld}
\end{figure}

In order to get a proper estimate of the fluxes $\hat{\bm{F}}_{i+1/2,j,k}$, we use the Godunov method \citep{godunov1959}. First, a pair of left and right states $\bm{U}^L_{i+1/2,j,k}$, $\bm{U}^R_{i+1/2,j,k}$ is reconstructed\footnote{Several spatial reconstruction routines are implemented in SLH, from simple constant extrapolation to the Piecewise Parabolic Method of \cite{colella1984}. These reconstruction schemes can be applied to both conservative and primitive variables.} (through 1D sweeping)  to the center of each cell boundary, starting from the cell-centered states $\hat{\bm{U}}_{i,j,k}$. These states define a 1D Riemann problem, which can then be solved (either exactly or approximately) to provide the value of a flux function $\bm{\mathcal{F}}(\bm{U}^L_{i+1/2,j,k},\bm{U}^R_{i+1/2,j,k})$. The surface-averaged flux is then approximated (to second-order accuracy) as
\begin{equation}
\label{eq:riemann-solver}
\hat{\bm{F}}_{i+1/2,j,k} \simeq \bm{\mathcal{F}}(\bm{U}^L_{i+1/2,j,k},\bm{U}^R_{i+1/2,j,k}).
\end{equation}
Many MHD Riemann solvers used nowadays are designed to work in supersonic regimes. In order to achieve numerical stability, such solvers need to add upwind numerical diffusion terms to the physical fluxes\footnote{The physical fluxes are usually computed in Riemann solvers as some variation of the central flux $(\bm{F}_L+\bm{F}_R)/2$.}, which smear out any discontinuity present in the flow on a timescale comparable to the cell crossing time of the shock. The choice of these terms depends on the specific approximate Riemann solver used. In particular, the diffusion term associated with the pressure flux (see Eq. \ref{eq:momentum_conservation}) usually takes the form \citep[see, e.g.,][]{einfeldt1991,cargo1997,miyoshi2005}
\begin{equation}\label{eq:numerical-diffusion}
D^x_p \propto -\bar{\rho}\bar{c}_\mathrm{f}(V^R_{x}-V^L_{x}),
\end{equation}
where $\bar{\rho}$ and $\bar{c}_\mathrm{f}$ are suitable averages of the density and the fast magnetosonic speed at the cell interface. However, in low-Mach regimes, discontinuities in the flow are only transported by the linearly degenerate entropy and Alfv\'en waves. These modes propagate with small speeds (see Sect.~\ref{sec:low-mach-limit}), and by the time they cross one cell in the computational grid they are strongly dissipated by the action of the numerical term in Eq. \ref{eq:numerical-diffusion}. This effect can also be explained by noticing that the pressure-diffusion coefficient scales as $\mathcal{O}(1/\mathcal{M}_{\mathrm{fast},x})$, so that it overwhelms the physical flux proportional to $\mathcal{O}(1)$ at low sonic Mach numbers. To remove this excessive dissipation, we use the low-dissipation HLLD solver (LHLLD) of \cite{minoshima2021}. This is a variation of the original five-wave HLLD solver (see Fig. \ref{fig:riemann-fan-lhlld}) of \cite{miyoshi2005}. LHLLD introduces a Mach-dependent parameter $\phi\propto \mathcal{M}_{\mathrm{fast},x}$ in the intermediate state of the total pressure $p_T = p + p_B$ :
\begin{equation}
\label{eq:intermediate_pressure}
\begin{split}
    p_\mathrm{T}^* =& \frac{ (S_R-V^R_{x}) \rho_R p^L_{\mathrm{T}} - (S_L-V^L_{x}) \rho_L p^R_{\mathrm{T}} }{ (S_R-V^R_{x}) \rho_R - (S_L-V^L_{x}) \rho_L} \\
    &+ \phi \frac{  \rho_L \rho_R (S_R-V^R_x) (S_L-V^L_x) (V^R_x-V^L_x) }{ (S_R-V^R_x) \rho_R - (S_L-V^L_x) \rho_L }.
    \end{split}
\end{equation}
In this context, $S_L$ and $S_R$ are conservative estimates of the speeds $\lambda_{1,7}$. In SLH they are evaluated as
\begin{equation}
\begin{split}
    S_L = &\min (V^L_x,V^R_x) - \max (c^L_{\mathrm{f},x},c^R_{\mathrm{f},x}), \\
    S_R = &\max (V^L_x,V^R_x) + \max (c^L_{\mathrm{f},x},c^R_{\mathrm{f},x}).
\end{split}
\end{equation}
The low-Mach fix $\phi$ is computed according to the following formulas:
\begin{equation}
\label{low Mach-fix}
\begin{split}
    c^L_\mathrm{u} = &  \left[ \frac{1}{2} \left( \frac{|\bm{B}_L|^2}{\rho_L} + |\bm{V}_L|^2 + \sqrt{\left(\frac{|\bm{B}_L|^2}{\rho_L}+|\bm{V}_L|^2\right)^2-4\frac{|\bm{V}_L|^2B^2_x}{\rho_L}} \right) \right]^ {\frac{1}{2}}, \\
    c^R_\mathrm{u} = & \left[ \frac{1}{2} \left( \frac{|\bm{B}_R|^2}{\rho_R} + |\bm{V}_R|^2 + \sqrt{\left(\frac{|\bm{B}_R|^2}{\rho_R}+|\bm{V}_R|^2\right)^2-4\frac{|\bm{V}_R|^2B^2_x}{\rho_R}} \right) \right]^ {\frac{1}{2}}, \\
    \chi = &\max \Bigg\{ \frac{c^L_\mathrm{u}}{c^L_{\mathrm{f},x}},\frac{c^R_\mathrm{u}}{c^R_{\mathrm{f},x}} \Bigg\}, \\
    \phi = & \chi(2-\chi).
\end{split}
\end{equation}
Since the fast magnetosonic wave speeds $c^{L,R}_{\mathrm{f},x}$ and consequently also $S_{L,R}$ scale as $\mathcal{O}(1/\mathcal{M}_{\mathrm{fast},x})$, the second term in Eq. \ref{eq:intermediate_pressure} would scale as $\mathcal{O}(1/\mathcal{M}_{\mathrm{fast},x})$ if  $\phi=1$, as in the original formulation of the HLLD solver. As previously described, this would lead to excessive numerical dissipation for small values of $\mathcal{M}_{\mathrm{fast},x}$. Instead, by computing $\phi$ according to Eq. \ref{low Mach-fix}, the dissipation term becomes independent of the fast magnetosonic Mach number, since $\phi \propto \mathcal{M}_{\mathrm{fast},x}$.
This modification does not affect the other properties of the HLLD solver, such as preserving positivity of density and internal energy \citep{miyoshi2005, minoshima2021}. We note that the combined diffusion coefficient in Eq. \ref{eq:intermediate_pressure} has a residual scaling $\mathcal{O}(1/\mathcal{M}_\mathrm{Alf})$, which would still introduce too much dissipation in very sub-Alfv\'en regimes. However, these are far from our main astrophysical applications.

\subsection{Well-balancing method}
\label{sec:well_balancing}

As already noted in Sects.~\ref{sec:introduction} and \ref{sec:finite_volume}, hyperbolic fluxes and gravitational source terms are discretized with different methods. As a consequence, Godunov-type schemes do not automatically preserve magnetohydrostatic solutions on a discrete grid exactly. Therefore, whenever a stratification needs to be enforced to be in MHSE on the computational grid, we use the deviation well-balancing method \citep{berberich2019,edelmann2021a}. The main ingredient of this method is an a priori known target state $\bm{\tilde{U}}$ that is a magnetohydrostatic solution to Eq. \ref{eq:balance_law},
\begin{equation}
\label{eq:target_solution_equation}
 \frac{\partial \bm{F}(\bm{\tilde{U}})}{\partial x} +
 \frac{\partial \bm{G}(\bm{\tilde{U}})}{\partial y}  +  \frac{\partial \bm{H}(\bm{\tilde{U}})}{\partial z}
= \bm{S}(\bm{\tilde{U}}),
\end{equation}
with $\bm{\tilde{V}}=0$.
Subtracting Eq. \ref{eq:target_solution_equation} from the original balance law in Eq. \ref{eq:balance_law} yields a system of PDEs for the deviations from the target solution  $\Delta \bm{U} = \bm{U} - \bm{\tilde{U}}$:
\begin{equation}
\label{eq:deviation_balance_law}
\begin{split}
\frac{\partial( \Delta \bm{U})}{\partial t} + &\left( \frac{\partial \bm{F}(\tilde{\bm{U}}+\Delta \bm{U})}{\partial x} - \frac{\partial \bm{F}(\bm{\tilde{U}})}{\partial x} \right)  \\
 + &\left( \frac{\partial \bm{G}(\tilde{\bm{U}}+\Delta \bm{U})}{\partial y} - \frac{\partial \bm{G}(\bm{\tilde{U}})}{\partial y} \right) \\
+ &\left( \frac{\partial \bm{H}(\tilde{\bm{U}}+\Delta \bm{U})}{\partial z} - \frac{\partial \bm{H}(\bm{\tilde{U}})}{\partial z} \right)
= \bm{S}(\tilde{\bm{U}}+\Delta \bm{U})-\bm{S}(\bm{\tilde{U}}).
\end{split}
\end{equation}
Now, to obtain a well-balanced method, Eq. \ref{eq:deviation_balance_law} is discretized according to the finite-volume method described in Sect.~\ref{sec:finite_volume}, which leads to the semi-discrete form
\begin{equation}
    \label{eq:deviation_discretization}
    \begin{split}
    \frac{\partial (\Delta \bm{U})_{i,j,k}}{\partial t} =
    &- \frac{1}{\Delta x} \left( \bm{F}_{i+1/2,j,k}^{dev} -     \bm{F}_{i-1/2,j,k}^{dev} \right) \\
    &- \frac{1}{\Delta y} \left( \bm{G}_{i,j+1/2,k}^{dev} - \bm{G}_{i,j-1/2,k}^{dev} \right) \\
    &- \frac{1}{\Delta z} \left( \bm{H}_{i,j,k+1/2}^{dev} - \bm{H}_{i,j,k-1/2}^{dev} \right) \\
    & + \bm{S}_{i,j,k}^{dev}.
\end{split}
\end{equation}
In this formulation, the deviation fluxes and source terms are defined by
\begin{align}
    \bm{F}_{i+1/2,j,k}^{dev} &= \hat{\bm{F}}_{i+1/2,j,k} - \bm{F}\left( \bm{\tilde{U}}_{i+1/2,j,k}\right), \\
    \bm{G}_{i,j+1/2,k}^{dev} &= \hat{\bm{G}}_{i,j+1/2,k} - \bm{G}\left( \bm{\tilde{U}}_{i,j+1/2,k}\right), \\
    \bm{H}_{i,j,k+1/2}^{dev} &= \hat{\bm{H}}_{i,j,k+1/2} - \bm{H}\left( \bm{\tilde{U}}_{i,j,k+1/2}\right), \\
    \bm{S}_{i,j,k}^{dev} &= \hat{\bm{S}}_{i,j,k} - {\bm{S}}(\tilde{\bm{U}}_{i,j,k}),
\end{align}
where $\hat{\bm{F}}_{i+1/2,j,k}$ is computed according to Eq. \ref{eq:riemann-solver} in the states
\begin{equation}
    \bm{U}^{L,R}_{i+1/2,j,k}=\bm{\tilde{U}}_{i+1/2,j,k}+\Delta \bm{U}^{L,R}_{i+1/2,j,k},
\end{equation}
while $\bm{F}\left( \bm{\tilde{U}}_{i+1/2,j,k}\right)$ corresponds to the physical fluxes in Eq. \ref{eq:balance_law} evaluated in the target solution at the cell boundary. The deviations $\Delta \bm{U}_{i,j,k}$, rather than the states $\bm{U}_{i,j,k}$, are reconstructed to the boundary of the cell\footnote{Deviations in the primitive variables can also be reconstructed if the corresponding equilibrium values are provided at the cell centers and at the cell boundaries.}. This guarantees that magnetohydrostatic solutions are preserved on the discrete grid, since in that case $\Delta \bm{U}_{i,j,k}=\bm{0}$, which leads to
\begin{equation}
    \hat{\bm{F}}_{i+1/2,j,k} = \mathcal{\bm{F}}\left( \bm{\tilde{U}}_{i+1/2,j,k},\bm{\tilde{U}}_{i+1/2,j,k}\right)=\bm{F}\left( \bm{\tilde{U}}_{i+1/2,j,k}\right).
\end{equation}
Thus, the resulting method is well-balanced.
Moreover, by removing the numerical errors arising from the magnetohydrostatic stratification, this method allows low-Mach flows to be simulated in stratified setups, which only cause small deviations from the MHSE state and would be completely dominated by spurious flows otherwise.

\subsection{Constrained transport method}
\label{sec:constrained_transport}

The divergence-free constraint described in Sect.~\ref{sec:divB} is not automatically satisfied if the induction equation is solved with Godunov-type schemes. As a result, magnetic monopoles are created locally at each time step and  tend to accumulate, as they cannot be transported away by any of the MHD waves. If not properly treated, these artifacts can accelerate the flow along the magnetic field lines, generate wrong field topologies, and ultimately lead to severe stability problems \citep{brackbill1980}.

Different strategies have been presented in the literature to cure this problem \citep[for a review of these methods, see][]{toth2000}. Among these, the eight-wave formulation \citep{powell1994, powell1999} modifies the MHD equations by including additional source terms that are proportional to $\nabla \cdot \bm{B}$. The modified system has an additional nonzero eigenvalue $\lambda_8=V_x$, which transports jumps in the normal component (to the cell interface) of the magnetic field, so numerical monopoles are advected with the flow and do not accumulate over time.

Other solutions rely on divergence cleaning schemes \citep{dedner2002}, where the divergence constraint is coupled to the MHD system using a generalized Lagrangian multiplier, $\psi$. This allows numerical monopoles to be transported with the maximum available speed on the grid and divergence errors to be damped at the same time.

One downside of both the eight-wave formulation and divergence cleaning is that they are not conservative and they cannot enforce any discretization of $\nabla \cdot \bm{B}$ to zero. Furthermore, these methods are most effective when open boundaries are used, so that the magnetic monopoles can leave the domain. However, this is rarely the case for simulations of stellar interiors, where impermeable boundaries are often used to avoid a significant mass loss from the system.

Constrained transport methods based on a staggered formulation, instead, conserve the magnetic flux through the boundaries of each cell and force one particular discretization of $\nabla \cdot \bm{B}$ to remain zero within round-off errors \citep{evans1988, dai1998, balsara1999, toth2000, londrillo2004, gardiner2008, mignone2021}. Although a conservative scheme cannot guarantee that the discretized Lorentz force is orthogonal to the magnetic field lines in each cell of the computational grid \citep{toth2000}, the magnitude of the parallel component of the force acting on the fluid is much smaller than in other methods.

The key point of staggered constrained transport methods is to compute the surface integral of Eq. \ref{induction2} over cell boundaries using Stokes's theorem, which leads to the finite-area equation\footnote{Here the calculation is made over the cell boundary $(i+1/2,j,k)$.}
\begin{equation}
\label{induction:finite-area}
\begin{split}
\frac{\partial \hat{B}_{x,i+1/2,j,k}}{\partial t} = - &  \frac{1}{\Delta y} \left( \hat{\mathcal{E}}_{z,i+1/2,j+1/2,k} - \hat{\mathcal{E}}_{z,i+1/2,j-1/2,k} \right) \\
+ & \frac{1}{\Delta z} \left( \hat{\mathcal{E}}_{y,i+1/2,j,k+1/2} - \hat{\mathcal{E}}_{y,i+1/2,j,k-1/2} \right).
\end{split}
\end{equation}
Here, $\hat{B}_{x,i+1/2,j,k}$ is the surface-averaged magnetic field component normal to the cell boundary
\begin{equation}
\hat{B}_{x,i+1/2,j,k} = \frac{1}{A_{i+1/2,j,k}}\int_{A_{i+1/2,j,k}} \bm{B}\cdot\hat{\bm{n}} dA,
\end{equation}
while the line-averaged electromotive force is defined as
\begin{equation}
\label{emf}
    \hat{\mathcal{E}}_{z,i+1/2,j+1/2,k} = \frac{1}{\Delta z_{i+1/2,j+1/2,k}}\int_{\Delta z_{i+1/2,j+1/2,k}} \mathcal{E}_z dz.
\end{equation}
Analogous formulas can be derived for the other components of the magnetic field and the electromotive force.

In order to solve Eq. \ref{induction:finite-area} numerically, a proper estimate for $\hat{B}_{x,i+1/2,j,k}$ and $\hat{\mathcal{E}}_{z,i+1/2,j+1/2,k}$ must be provided. For the former, we approximate the surface-averaged quantity with its value at the center of the cell boundary,
\begin{equation}
\hat{B}_{x,i+1/2,j,k} \simeq B_{x,i+1/2,j,k},
\end{equation}
which is accurate to second order. In contrast to the standard finite-volume approach, here the magnetic field component normal to the interface is stored at cell boundaries, while the line-averaged electromotive force is evaluated at cell edges. Thus, the operation is performed on a staggered grid. Since the parallel magnetic field still needs to be reconstructed to compute the flux function, its value at cell-center locations is estimated as a simple arithmetic average between the neighboring cell interfaces:
\begin{equation}
\begin{split}
    \hat{B}_{x,i,j,k} = &\frac{1}{2} \left( \hat{B}_{x,i-1/2,j,k} + \hat{B}_{x,i+1/2,j,k} \right), \\
    \hat{B}_{y,i,j,k} = &\frac{1}{2} \left( \hat{B}_{y,i,j-1/2,k} + \hat{B}_{y,i,j+1/2,k} \right), \\
    \hat{B}_{z,i,j,k} = &\frac{1}{2} \left( \hat{B}_{z,i,j,k-1/2} + \hat{B}_{z,i,j,k+1/2} \right).
\end{split}
\end{equation}

To compute the line-averaged electromotive force in Eq. \ref{emf}, in SLH we use the CT-contact algorithm of \cite{gardiner2005}. In this method, the electric field at cell edges is computed as a simple arithmetic average of the four neighboring face-centered electromotive force components, with the addition of a diffusion term that helps removing spurious oscillations when the magnetic field is advected. For instance, $\hat{\mathcal{E}}_{z,i+1/2,j+1/2,k}$ is approximated to second-order accuracy by
\begin{equation}
\label{contact-ct-emf}
\begin{split}
    \hat{\mathcal{E}}_{z,i+1/2,j+1/2,k} \simeq &\ \frac{1}{4}  \Big (
    \bar{\mathcal{E}}_{z,i+1/2,j,k} + \bar{\mathcal{E}}_{z,i+1/2,j+1,k} \\
    &+\bar{\mathcal{E}}_{z,i,j+1/2,k} + \bar{\mathcal{E}}_{z,i+1,j+1/2,k}
    \Big ) \\
    &+ \frac{\Delta y}{8}
    \Bigg\{
    \left(\frac{\partial \mathcal{E}_z}{\partial y}\right)_{i+1/2,j+1/4,k} - \left(\frac{\partial \mathcal{E}_z}{\partial y}\right)_{i+1/2,j+3/4,k}
    \Bigg\} \\
     &+ \frac{\Delta x}{8}
    \Bigg\{
    \left(\frac{\partial \mathcal{E}_z}{\partial x}\right)_{i+1/4,j+1/2,k} - \left(\frac{\partial \mathcal{E}_z}{\partial x}\right)_{i+3/4,j+1/2,k}
    \Bigg\},
\end{split}
\end{equation}
where $\bar{\mathcal{E}}_z$ can be computed from the solution to the Riemann problem in Eq. \ref{eq:riemann-solver}. The calculation for the $x-$ and $y$-component is again analogous. The upwind diffusion term enters in the derivatives of the electromotive force in Eq. \ref{contact-ct-emf}, which are obtained according to the sign $s_{i+1/2,j,k}$ of the entropy (contact) waves at the cell interfaces:
\begin{equation}
\begin{split}
    \left(\frac{\partial \mathcal{E}_z}{\partial y}\right)_{i+1/2,j+1/4,k} = & \\
    & \frac{1+s_{i+1/2,j,k}}{2}\left(\frac{\bar{\mathcal{E}}_{z,i,j+1/2,k}-\mathcal{E}^{cc}_{z,i,j,k}}{\Delta y/2} \right) + \\
    & \frac{1-s_{i+1/2,j,k}}{2}\left(\frac{\bar{\mathcal{E}}_{z,i+1,j+1/2,k}-\mathcal{E}^{cc}_{z,i+1,j,k}}{\Delta y/2} \right).
\end{split}
\end{equation}
Here $\mathcal{E}^{cc}_{z,i,j,k}=(-\bm{V}_{i,j,k} \times \bm{B}_{i,j,k})_z$ represents the $z$-component of the cell-centered electromotive force. The discretization of the line-averaged electromotive force leads to a semi-discrete form of Eq. \ref{induction:finite-area} that can be integrated numerically in time. Any time-stepper that solves the resulting system of ODEs can keep the cell-volume average of $\nabla \cdot \bm{B}$,
\begin{equation}
    \label{divb}
    \begin{split}
    (\nabla \cdot \bm{B})_{i,j,k} = &
    \frac{\hat{B}_{x,i+1/2,j,k}-\hat{B}_{x,i-1/2,j,k}}{\Delta x} + \\
    &\frac{\hat{B}_{y,i,j+1/2,k}-\hat{B}_{y,i,j-1/2,k}}{\Delta y} + \\
    &\frac{\hat{B}_{z,i,j,k+1/2}-\hat{B}_{z,i,j,k-1/2}}{\Delta z},
\end{split}
\end{equation}
within rounding errors.

\section{Time integration algorithm}
\label{sec:time-integration}

The CFL constraint in time-explicit marching schemes restricts the time step to the crossing time of the fastest wave resulting from the underlying PDEs over a grid cell. In low-Mach-number flows, the fast magnetosonic wave speeds become very large, so that the time step needs to be reduced accordingly. Thus, simulating the evolution of slow fluid motions and Alfv\'en waves becomes expensive. In these regimes, implicit methods, in which the time step is not limited by stability conditions but only by the desired accuracy, represent an attractive alternative. When using such methods, the time step should be restricted to the shortest advection and Alfv\'en crossing time over one grid cell. If the sonic Mach number is small enough, the possibility of larger chosen time steps then outweighs the disadvantage of higher computational costs for a single time step by using the implicit solver.

As outlined in Sect.~\ref{sec:introduction}, we split the induction equation (see Eq. \ref{eq:induction_equation}) from the continuity, momentum and energy equations (see Eqs. \ref{eq:mass_conservation}-\ref{eq:Energy_conservation}), based on the approach described by \cite{fuchs2009}. This allows different spatial and temporal discretizations to be used depending on the problem at hand. In regimes of low Mach numbers and high $\beta$ values, the stiffness is mostly generated by the pressure flux $\propto 1/\hat{\mathcal{M}}^2_\mathrm{son}$ in the momentum equation, while the nondimensional form of the induction equation does not depend on the Mach number of the flow (see Sect.~\ref{sec:low-mach-limit}). This suggests that implicit time discretization only needs to be applied to the subset of continuity, momentum and energy equations, whereas the induction equation can be solved with explicit time-steppers. These two updates can be combined to second-order accuracy with Strang splitting \citep{strang1968}:
\begin{equation}
\label{strang}
    \bm{U}^{n+1} = \mathcal{I}^{(\frac{1}{2}\Delta t)} \mathcal{H}^{(\Delta t)} \mathcal{I}^{(\frac{1}{2} \Delta t)} \bm{U}^n.
\end{equation}
Here, $\mathcal{I}$ represents a linear operator that updates only the magnetic field with an explicit marching scheme, while the nonlinear operator $\mathcal{H}$ updates density, momentum and total energy (including source terms) using an implicit stepper. In each sub-step of Strang splitting, the discretization of the fluxes, source terms, and electromotive force is performed according to the methods described in Sect.~\ref{sec:spatial_discretization}.

In SLH, several implicit time-steppers can be used to solve the semi-discrete form of Eq. \ref{eq:semi-discrete}, such as first-order backward-Euler, higher-order ESDIRK schemes, and Crank-Nicolson. The resulting nonlinear system of equations is solved iteratively with a root-finding Raphson-Newton algorithm, which relies on the analytic formulation of the flux-Jacobian. Iterative linear solvers (such as BiCGSTAB(l), GMRES, and Multigrid) are used in combination with preconditioning techniques to solve each sub-step of the nonlinear solver\footnote{For more details on the implementation of implicit time stepping in SLH, see \cite{miczek2013a} and \cite{miczek2015a}.}. In contrast, the semi-discrete form of the induction equation (see Eq. \ref{induction:finite-area}) is solved with the time-explicit SSP-RK2 method of \cite{shu1988}.

Numerical experiments performed with the proposed implicit-explicit Strang splitting (IESS) approach suggest that the maximum time step allowed for stability is approximately determined by
\begin{equation}
\label{cflua-timestep}
    \Delta t = \min_{\Omega=(i,j,k)} \Bigg\{ \frac{\Delta x}{|V_{x,\Omega}|+c_{\mathrm{a},x,\Omega}}, \frac{\Delta y}{|V_{y,\Omega}|+c_{\mathrm{a},y,\Omega}}, \frac{ \Delta z}{|V_{z,\Omega}|+c_{\mathrm{a},z,\Omega}} \Bigg\},
\end{equation}
so that the propagation of fluid motions  and Alfv\'en waves is well resolved in time. This time step is approximately $1/\mathcal{M}_\mathrm{son}$ larger than that allowed by the conventional CFL condition if the plasma-$\beta$ is high, which considerably reduces the computational effort when simulating low-Mach-number flows. The price one has to pay is that the propagation of fast magnetosonic waves is not well resolved in time. Another advantage of IESS is that it can easily be implemented within the framework of the SLH code, which already had fully implicit time integration capabilities to solve the compressible Euler equations.

A single step of the described time-marching scheme can be summarized in the following way.\ First, $\Delta t$ is obtained from Eq. \ref{cflua-timestep} given $\bm{B}^n$, $\rho^n$, $\rho \bm{V}^n$, and $\rho E_{\phi}^n$. If gravity is not present, $e_{\phi}$ does not appear in Eq. \ref{total-energy}.

Second, SSP-RK2 and CT-contact are used to solve the induction equation over the first half of the time step, $\Delta t /2$. This results in an intermediate solution for the magnetic field, $\bm{B}^{n+1/2}$.

Third, this intermediate solution, $\bm{B}^{n+1/2}$, is used to solve the continuity, momentum, and energy equations over the full time step, $\Delta t$, with an implicit time-stepper. If gravity is present, then the well-balancing method described in Sect.~\ref{sec:well_balancing} can be used. Any other source term is also considered in this step. This allows the solution for density, momentum, and energy to be obtained at the next step, $\rho^{n+1}$, $\rho \bm{V}^{n+1}$, and $\rho E_{\phi}^{n+1}$.

Fourth, $\bm{B}^{n+1/2}$, $\rho^{n+1}$, $\rho \bm{V}^{n+1}$, and $\rho E_{\phi}^{n+1}$ are used to solve the induction equation over $\Delta t/2$. This yields the magnetic field at the final step $\bm{B}^{n+1}$.

The proposed MHD scheme is extremely modular, so different time-steppers, spatial reconstruction schemes and approximate Riemann solvers can be used in each sub-step of the algorithm, and well-balancing can be switched off if required. For instance, in addition to LHLLD, the original five-wave HLLD solver of \cite{miyoshi2005} is also implemented in SLH. The performance and accuracy of both Riemann solvers are checked in some of the numerical experiments described in Sect.~\ref{sec:tests}. Finally, in case slightly subsonic or transonic regimes need to be modeled with SLH, a fully un-split SSP-RK2 explicit time-stepper can be used.

\section{Numerical tests}
\label{sec:tests}
In order to assess the accuracy and performance of the newly implemented MHD algorithm, we have to rely on numerical experiments. Since the main purpose of the scheme is to be able to simulate MHD flows at low sonic Mach numbers in strong stratifications, we decide not to show the typical tests commonly run by other MHD codes. These usually include shock-tubes, supersonic vortices and magnetic blasts, which, however, are designed to test the shock-capturing capabilities of a numerical scheme. Instead, we ran a series of verification benchmarks that are more suited for testing the low-Mach properties of an MHD code.

As a first test, we solved the homogeneous MHD equations in three different cases (i, ii, and iii). In order to check the convergence and scaling of the methods for the whole MHD wave family, we performed a 1D linear analysis (i). The scaling was also checked against the advection of a stable MHD vortex in a wide range of Mach numbers (ii). Such a setup is particularly important as it resembles the typical vortex structures present in magneto-convection. The simulations were also run in fully-explicit mode using SSP-RK2, which allows the speed-up of IESS to be quantified as a function of the Mach number.

The ability of accurately evolving shear instabilities is fundamental in the context of simulations of turbulence as they generate additional vorticity, which leads to the cascade of energy. For this reason, we ran simulations of a magnetized Kelvin--Helmholtz instability (iii). We followed the growth and evolution of the instability in a resolution study from low-Mach to slightly subsonic regimes. A comparison between the HLLD and LHLLD solvers was performed to show the advantage of using low-dissipation fluxes over conventional methods in regimes of low Mach numbers.

Then we considered two setups in which gravity is present (iv, v).
To check the entropy-conservation properties of the scheme based on the deviation well-balancing method, we modeled the rise of a parcel of fluid with higher entropy content than the (isentropic) background stratification, that is,\ a ``hot bubble'' (iv). By changing the magnitude of the entropy perturbation, we simulated different rise velocities of the bubble, down to Mach numbers of $\mathcal{M}_{\mathrm{son}}\sim 10^{-4}$. To quantify the magnitude of the numerical errors generated by an unbalanced stratification, we also simulated the rise of the bubble at $\mathcal{M}_{\mathrm{son}}\sim 10^{-2}$ without well-balancing.

Finally, we simulated a fully 3D small-scale dynamo (SSD) amplification in a star-like environment at moderate grid resolutions (v). By changing the rate at which energy is injected in the system, we simulated progressively slower flows down to $\mathcal{M}_\mathrm{son} \sim 10^{-3}$.

For all of the following tests, an ideal gas EoS was used with $\gamma=5/3$ except when specified otherwise. Within the framework of the IESS time-marching scheme described in Sect.~\ref{sec:time-integration}, ESDIRK23 was chosen to treat the implicit part of the algorithm. This guarantees second-order accuracy in time. The time step in Eq. \ref{cflua-timestep} was reduced  by $20\%$ to get a more conservative stability criterion. Finally, unlimited linear reconstruction, which is second-order accurate in space, was applied to primitive variables. Overall, the proposed scheme is (globally) second-order accurate.


\subsection{Linear analysis}
\label{linear-analysis}

In this test we followed the propagation of linear modes for all the MHD waves (see Sect.~\ref{sec:MHD_equations}) as a way to get quantitative estimates of diffusion errors and to check the scaling of the numerical scheme. The setup is based on \cite{stone2008}, which we modified by considering much larger values of the gas pressure to increase the fast magnetosonic speed relative to the Alfv\'en and entropy wave speeds. Such a stiff system is characteristic of low-Mach flows in high-$\beta$ environments (see Sect.~\ref{sec:low-mach-limit}).

The homogeneous MHD equations were solved on a periodic 1D Cartesian grid divided into $N$ cells, with the spatial domain ranging from 0 to $L=1$. For the  $i$-th wave, the solution at $t=0$ was obtained by perturbing a uniform medium $\bm{w}_0=(1,0,0,0,10^3/\gamma,1,\sqrt{2},1/2)$ with  $\delta \bm{w}=A \bm{R}_i(\bm{w}_0) \sin(2 \pi x / L)$, where $\bm{w}$ is the vector of primitive variables $(\rho, V_x,V_y,V_z,p,B_x,B_y,B_z)$, $A=10^{-4}$ is the amplitude of the perturbation, $x$ is the spatial coordinate and $\bm{R}_i$ is the $i$-th column of the right-eigenvector matrix \citep[see Appendix A.3 in][]{stone2008}. The chosen values for $\bm{w}_0$ are such that $c_{\mathrm{f},x}\simeq40.85$, $c_{\mathrm{a},x}=1$ and $c_{\mathrm{s},x}\simeq0.99$. For the entropy mode, we set $V_{x}=1$, so the sonic Mach number of the wave is $\mathcal{M}_\mathrm{son}\simeq0.032$.  The simulations were run for one crossing time defined as $t_\mathrm{c}=L/\lambda_i$, where $\lambda_i$ is the wave speed. The $L_{1}$ error was then computed for each primitive variable $w_k$ as
\begin{equation}\label{eq:L1}
    L_{1,i}(w_k) = \frac{1}{N}\sum_j | w_{k,j}(t=t_\mathrm{c})-w_{k,j}(t=0)|,
\end{equation}
and the global error associated with the $i$-th wave was then computed as
\begin{equation}\label{error}
    \delta_i = \frac{1}{A}\sqrt{\sum_k \left(\frac{ L_{1,i}(w_k)}{\xi(w_{0,k})}\right)^2},
\end{equation}
with
\begin{equation}
    \xi(w_{0,k}) =
    \begin{cases*}
      w_{0,k}, &  for $w_{0,k}>0$, \\
      1,  & otherwise.
    \end{cases*}
\end{equation}
In Eq. \ref{eq:L1}, $j$ is the spatial index. The tolerance of the Raphson-Newton algorithm was set to $10^{-10}$, so that the errors computed using Eq. \ref{error} were not dominated by the finite convergence of the nonlinear solver.


\begin{figure}
  \centering
  \includegraphics[width=0.5\textwidth]{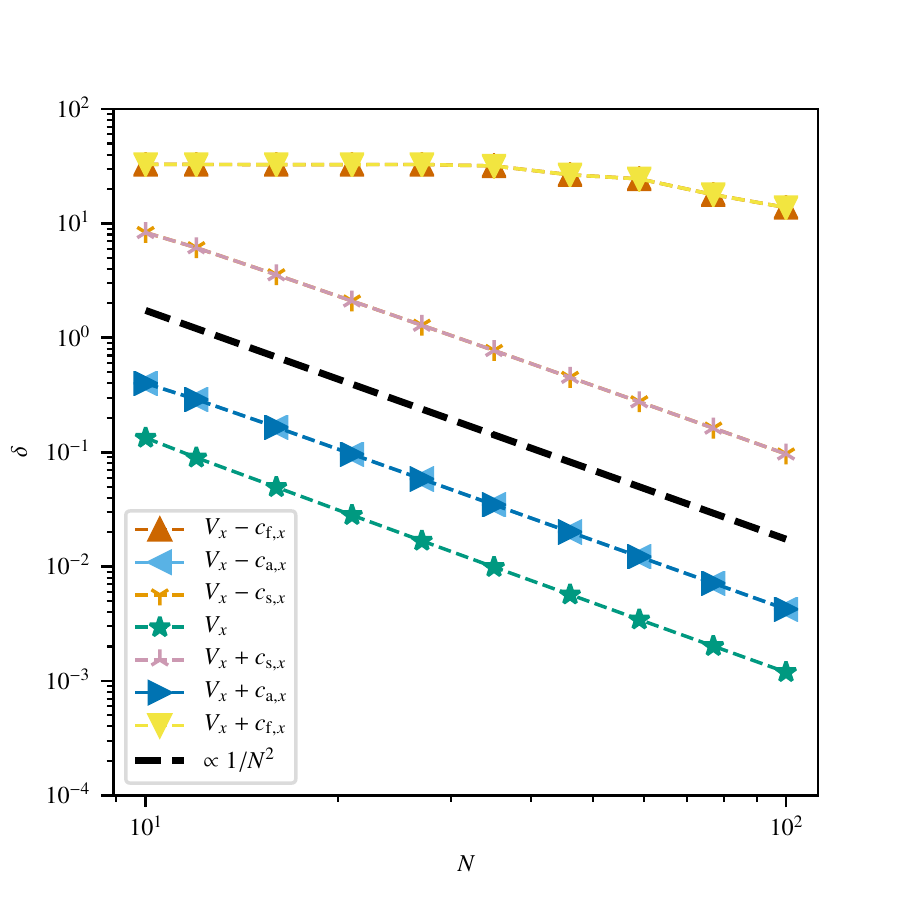}
  \caption{Global error as a function of the grid resolution for the seven MHD waves after one crossing time, $t_\mathrm{c}$. The dashed black line represents the second-order scaling.}
\label{fig:A-scaling}
\end{figure}

\begin{figure}
  \centering
  \includegraphics[width=0.5\textwidth]{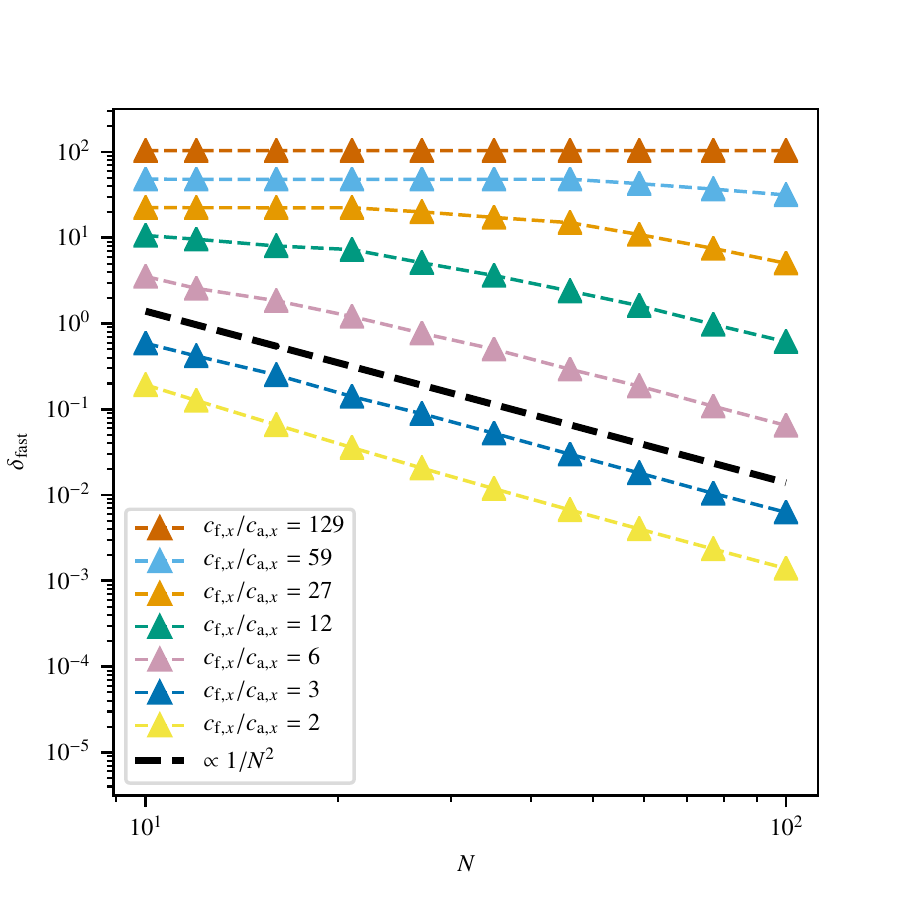}
  \caption{Global error as a function of the grid resolution for the left-going fast magnetosonic wave. Different colors are associated with different values of  $c_{\mathrm{f},x}/c_{\mathrm{a},x}$. The dashed black line is the second-order scaling.}
\label{fig:p-vary}
\end{figure}

Figure \ref{fig:A-scaling} shows the change of the global error as a function of $N$ (ranging from $10$ to $100$) for all the seven MHD waves.  Leftward and rightward propagating waves have identical errors.  Since the MHD scheme relies on second-order methods to treat both the spatial and the temporal parts, the scheme converges with second-order accuracy for all waves except the fast magnetosonic waves, which are characterized by much larger errors. This is expected since IESS allows the MHD equations  to be integrated over much longer time steps than the CFL constraint (see Sect.~\ref{sec:time-integration}). As a consequence, the propagation of fast magnetosonic waves is not properly resolved in time and discretization errors strongly deteriorate the numerical solution. This effect is further quantified in Fig. \ref{fig:p-vary}, where we show the global error associated with the left-going fast magnetosonic wave as a function of the grid resolution for different values of the gas pressure $p$, such that  $c_{\mathrm{f},x}/c_{\mathrm{a},x}=2,3,6,12,27,59,129$. Overall, the errors tend to decrease for smaller values of $c_{\mathrm{f},x}/c_{\mathrm{a},x}$, since the time step in Eq. \ref{cflua-timestep} gets closer to the CFL time step and fast magnetosonic waves are progressively better resolved. Moreover, the simulations run with $c_{\mathrm{f},x}/c_{\mathrm{a},x} \leq 27$ converge with second-order accuracy on the grids considered in this study.


\subsection{Balsara vortex}
\label{test:balsara-vortex}
In the previous section we demonstrated that the current MHD scheme is capable of simulating linear waves with the expected (second-order) scaling with respect to resolution on 1D grids. In order to check the scaling in 2D and to test the low-Mach capabilities of the scheme, we considered the MHD vortex first described by \cite{balsara2004}. This is an exact stationary solution of the ideal 2D homogeneous MHD equations, in which the distribution of the centrifugal acceleration, magnetic tension, gas and magnetic pressure gradients is such that the vortex is stable. The spatial domain is $(x,y) \in [-5,5]\times [-5,5]$, and we used $64\times64$ grid cells with periodic boundaries in both directions. The initial conditions are given by
\begin{equation}
\begin{split}
    &(V_x,V_y)  =   \tilde{V}e^{\frac{1-r^2}{2}}(-y,x), \\
    &(B_x,B_y)  =     \tilde{B}e^{\frac{1-r^2}{2}}(-y,x), \\
    &p          =  1+\left[ \frac{\tilde{B}^2}{2}(1-r^2)-\frac{\tilde{V}^2}{2}\right]e^{1-r^2}, \\
    &\rho  = 1,
\end{split}
\end{equation}
with $r^2=x^2+y^2$. $\tilde{V}$ is the maximum rotational velocity of the vortex and $\tilde{B}$ sets the value of the maximum Alfv\'en speed on the grid. To make this problem numerically more challenging, the vortex is advected along the diagonal of the computational grid, with $|\bm{V}_\mathrm{adv}|=\tilde{V}$. The vortex is evolved for one advective crossing time $t_\mathrm{adv}=10\sqrt{2}/\tilde{V}$, after which it returns to the initial position. In this time interval, the vortex rotates 2.25 times.

\begin{figure*}[h!]
  \centering
  \includegraphics[width=\textwidth]{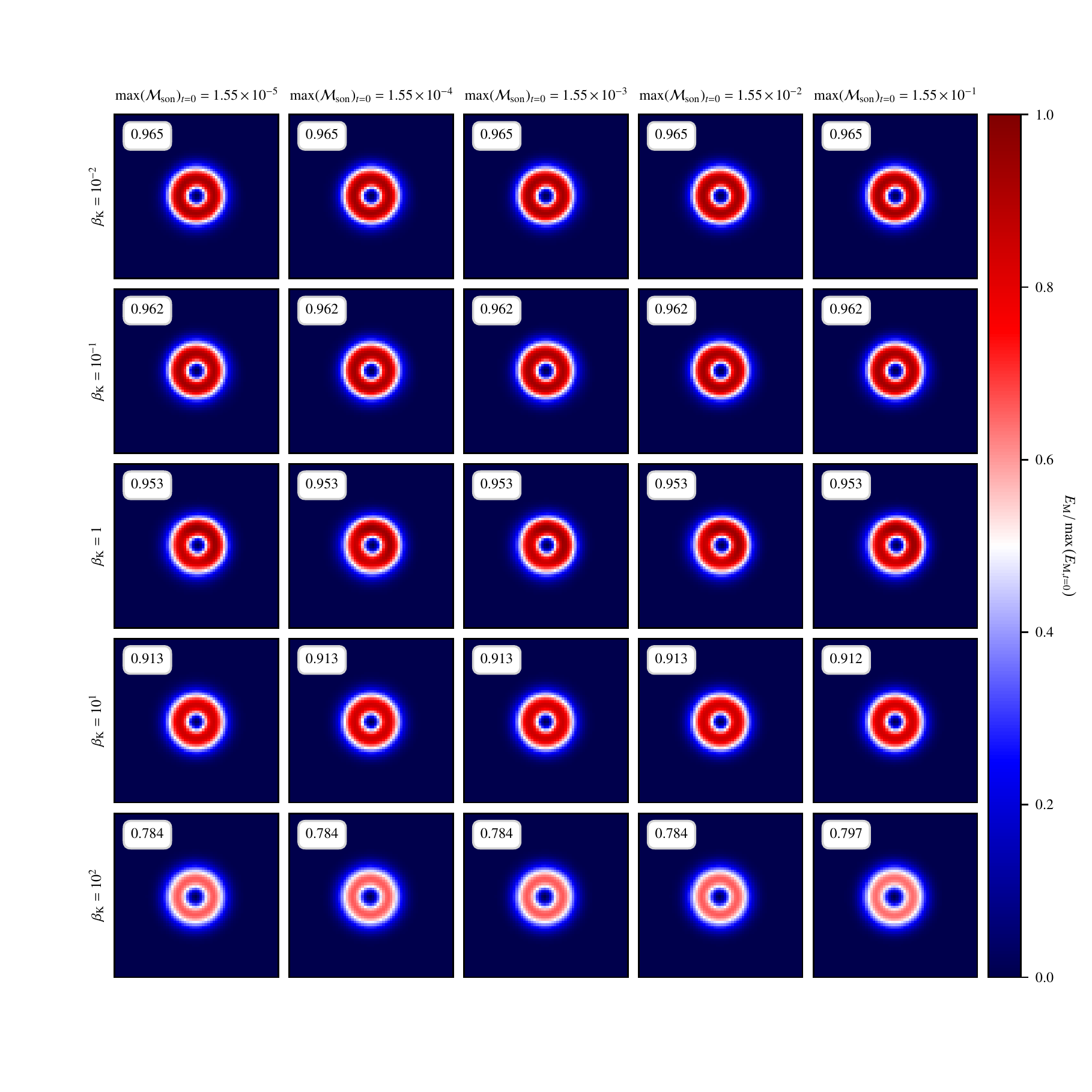}
  \caption{Magnetic energy distribution of the Balsara vortex after one advective crossing time, $t_\mathrm{adv}$, normalized by the maximum magnetic energy at $t=0$. The ratio of the magnetic to the (rotational) kinetic energy of the vortex is varied along the $y$ axis (in descending order), while the initial maximum  rotational velocity, $\tilde{V}$, varies along the $x$ axis. The inset in each subplot shows the ratio of the  final to the initial magnetic energy. The vortex run with $\tilde{V}=10^{-1}$ and $\beta_\mathrm{K}=10^2$ (bottom-right corner) has a maximum Mach number  $\max(\mathcal{M}_\mathrm{son})_{t=0}=1.65\times10^{-1}$. In that system, the gas pressure drops in the regions around the center of the vortex to balance the large magnetic and centrifugal forces, which ultimately decreases the sound speed where the velocity is maximum.}
\label{fig:parameter-study}
\end{figure*}

We ran the grid of models
\begin{equation}
\begin{split}
    & \left( \tilde{V} \right) \times \left( \beta_\mathrm{K} \right) =  \\
    & \left( 10^{-5},10^{-4},10^{-3},10^{-2},10^{-1} \right) \quad \times \\
    & \left( 10^{-2},10^{-1},1,10^{1},10^{2} \right),
\end{split}
\end{equation}
with $\beta_\mathrm{K}=\tilde{B}^2/\tilde{V}^2$ being the ratio of the  magnetic to the rotational kinetic energy, which is constant across the domain. Given this choice of parameters, the initial maximum Mach number $\mathcal{M}_\mathrm{son}$ ranges from $1.55\times10^{-5}$ to $1.55\times10^{-1}$, so this parameter study covers both low Mach numbers and slightly subsonic regimes, in both weakly and strongly magnetized fluids.

Figure \ref{fig:parameter-study} shows the magnetic energy distribution after one advective crossing time $t_\mathrm{adv}$. Numerical dissipation converts a fraction of kinetic and magnetic energy into internal energy, but the shape of the vortex is well preserved in all runs. The dissipation rate is virtually independent of $\mathcal{M}_\mathrm{son}$. In contrast, dissipation of magnetic energy depends on the value of $\beta_\mathrm{K}$. As already pointed out in Sect.~\ref{sec:flux_function}, the pressure-diffusion coefficient in LHLLD has a residual scaling $\mathcal{O}(1/\mathcal{M}_\mathrm{Alf})$. A larger value of $\beta_\mathrm{K}$ corresponds to lower $\mathcal{M}_\mathrm{Alf}$, which then increases the magnitude of the numerical dissipation. The velocity field is progressively more diffused out and becomes less efficient in sustaining the magnetic field through induction against numerical resistivity.

\begin{figure}
  \centering
  \includegraphics[width=0.5\textwidth]{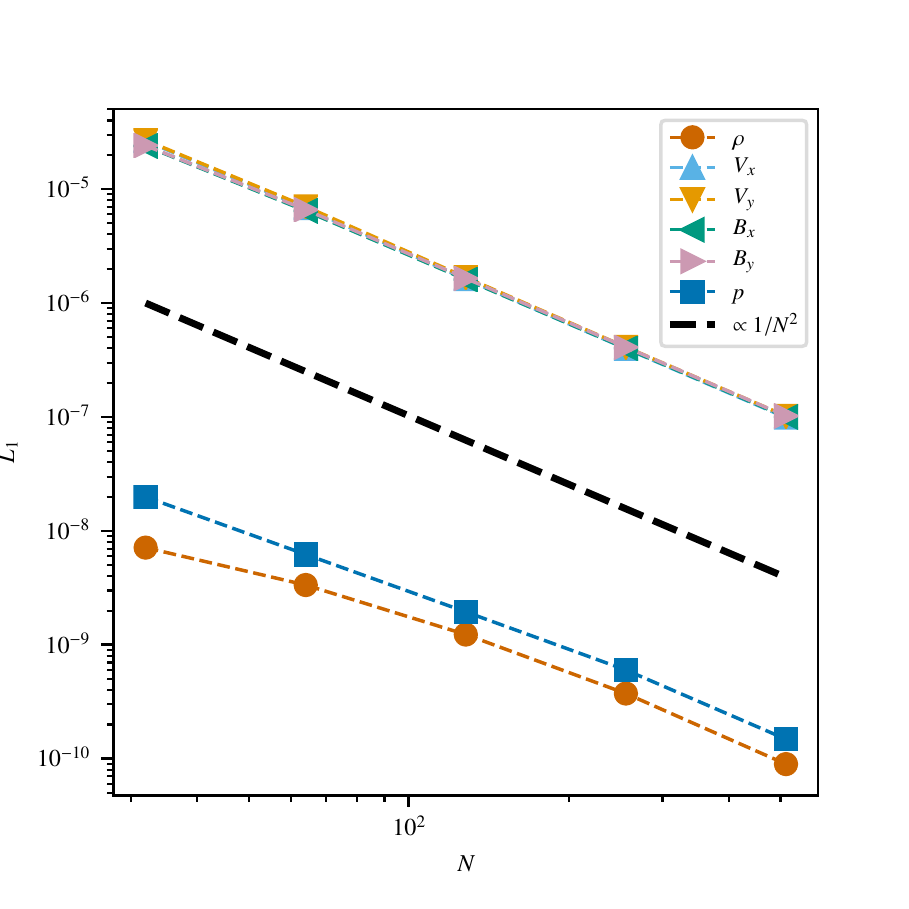}
  \caption{Convergence of the $L_1$ error in the Balsara vortex for each primitive variable as a function of resolution. For these simulations, $\tilde{V}=10^{-3}$ and $\beta_\mathrm{K}=1$. The dashed black line is the second-order scaling.}
\label{fig:F-scaling}
\end{figure}

To check the convergence of the scheme in 2D, we ran a vortex with $\tilde{V}=10^{-3}$ (corresponding to $\max(\mathcal{M}_\mathrm{son})_{t=0}=1.55\times10^{-3}$) and $\beta_\mathrm{K}=1$ at different resolutions\footnote{We took these values as representative of the typical conditions found in stellar convection zones close to equipartition regimes \citep{augustson2016}.}. At the end of the simulation, the $L_1$ error was computed for each primitive variable $w_k$ as
\begin{equation}
    L_{1}(w_k) = \frac{1}{N^2}\sum_{i,j} | w_{k,i,j}(t=t_\mathrm{adv})-w_{k,i,j}(t=0)|,
\end{equation}
where $i,j$ are the spatial indices. Figure \ref{fig:F-scaling} shows the convergence of the $L_1$ error for different grids from $N=32$ up to $N=512$ cells per dimension. Convergence is second order for all primitive variables.

\begin{figure*}[h!]
  \centering
  \includegraphics[width=\textwidth]{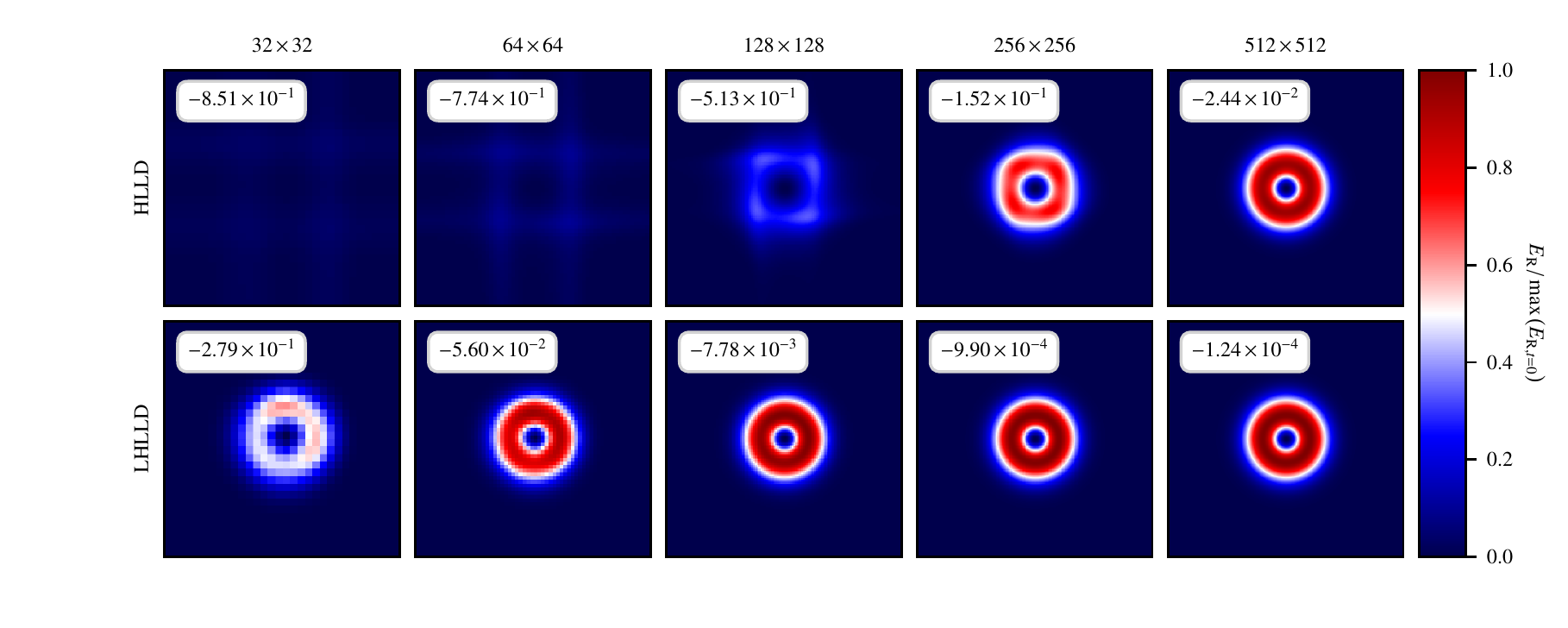}
  \caption{Distribution of the rotational kinetic energy (normalized by the maximum initial value) of the Balsara vortex after one advective time, $t_\mathrm{adv}$, at $\tilde{V}=10^{-3}$ and $\beta_\mathrm{K}=1$. The top panels show the vortices obtained with the HLLD flux function as a function of resolution, while the plots in the bottom panels are obtained with LHLLD. The insets show the fraction of rotational kinetic energy that has been dissipated by the end of the simulation: $(E_{\mathrm{R},t=t_\mathrm{adv}})_\mathrm{tot} / (E_{\mathrm{R},t=0})_\mathrm{tot}-1$.}
\label{fig:kinetic-energy}
\end{figure*}

To compare the amount of numerical dissipation introduced by a standard and a low-Mach MHD flux function,  we reran this last set of simulations with HLLD. In Fig. \ref{fig:kinetic-energy} we show the final rotational kinetic energy distribution obtained with the two methods:
\begin{equation}
E_\mathrm{R}=\frac{1}{2} \rho\left[\left(V_x-\tilde{V}/\sqrt{2}\right)^2+\left(V_y-\tilde{V}/\sqrt{2}\right)^2\right].
\end{equation}
At low resolution, HLLD considerably stretches the vortex and a large fraction of kinetic energy is dissipated into internal energy. In contrast, simulations run with LHLLD show mild dissipation and dispersion errors are only visible at the lowest resolutions. All simulations converge with increasing resolution, but the kinetic energy conservation in the vortex simulated with HLLD is still two orders of magnitude worse than that obtained with LHLLD at the highest resolution considered in this study.

\begin{figure}
  \centering
  \includegraphics[width=0.5\textwidth]{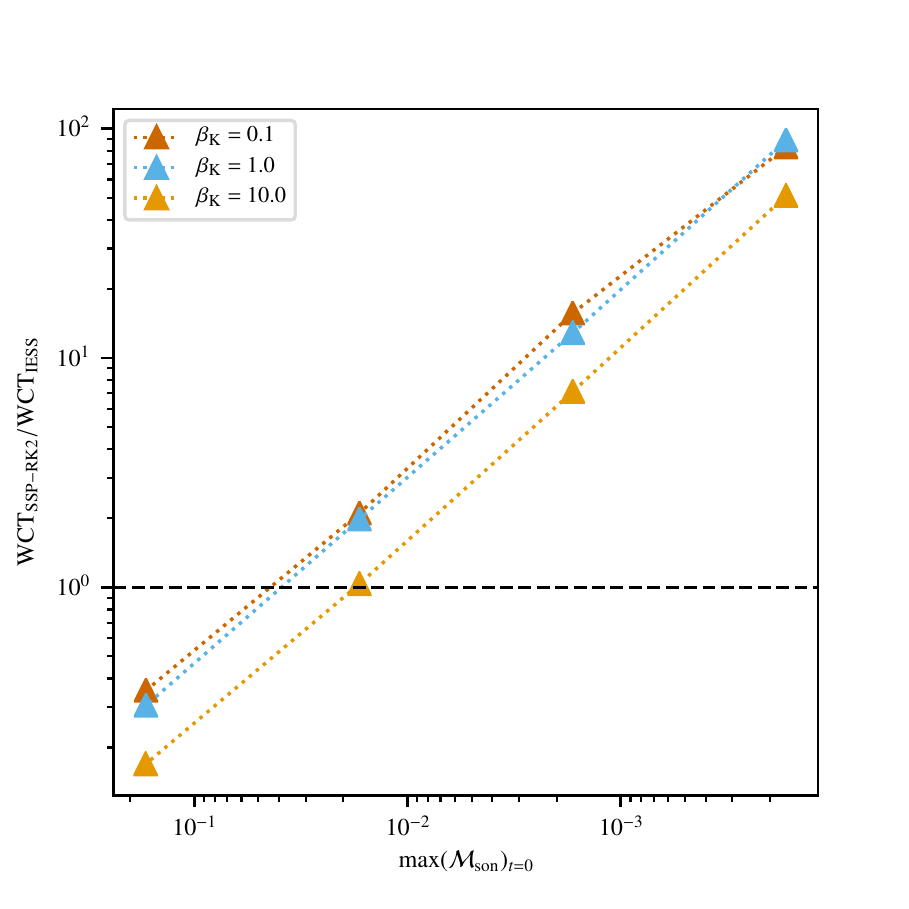}
  \caption{Ratio of the wall clock times obtained with SSP-RK2 and IESS as a function of initial maximum sonic Mach number of the Balsara vortex for different magnetic to (rotational) kinetic energy ratios. The dashed black line is drawn to represent the same relative efficiency.}
\label{fig:efficiency}
\end{figure}

As explained in Sect.~\ref{sec:time-integration}, one advantage of IESS is that the MHD equations can be integrated on time steps longer than that allowed by the CFL condition without sacrificing stability.  However, a single step of the proposed scheme is much more expensive than a single step of a more standard time-explicit marching scheme, as a large nonlinear system has to be solved iteratively with a Raphson-Newton method.  Because of these competing effects, we expect the IESS scheme to be more efficient than an explicit time-stepper below a certain Mach number. To determine this threshold, we ran sets of simulations with the parameters
\begin{equation}
\begin{split}
    &(\tilde{V})\times(\beta_\mathrm{K}) =  \\
    &(10^{-4},10^{-3},10^{-2},10^{-1}) \quad \times\\
    &(10^{-1},1,10^{1}),
\end{split}
\end{equation}
using both IESS and the explicit SSP-RK2\footnote{For the time-explicit simulations, the CFL time step is reduced by $20\%$.} on $40\times40$ grid cells. Every other sub-step of the Godunov method (like the spatial reconstruction, the LHLLD flux function and constrained transport) remained unchanged, so the only difference was in the time discretization. At the end of each simulation, the ratio of the wall-clock times $\mathrm{WCT}_\mathrm{SSP-RK2}/ \mathrm{WCT}_\mathrm{IESS}$ was taken as a measure of the relative efficiency between the marching schemes\footnote{No snapshots were saved throughout the simulations to minimize the cost of I/O operations.}. The results are shown in Fig. \ref{fig:efficiency}. As expected, the speed-up of IESS increases as the Mach number of the vortex is decreased. The simulations with $\beta_\mathrm{K}=10$ are slower than the other cases, as the larger Alfv\'en speed considerably reduces the time step estimate in Eq. \ref{cflua-timestep}, while no significant difference is seen between $\beta_\mathrm{K}=0.1$ and $\beta_\mathrm{K}=1.0$. IESS overtakes SSP-RK2 at $\max(\mathcal{M}_\mathrm{son})_{t=0}\simeq 4\times10^{-2}$ for $\beta_\mathrm{K}=(0.1,1)$ and $\max(\mathcal{M}_\mathrm{son})_{t=0}\simeq 2\times10^{-2}$ for $\beta_\mathrm{K}=10$. At $\max(\mathcal{M}_\mathrm{son})_{t=0}=10^{-3}$, IESS is ten to twenty times faster than SSP-RK2. This justifies the implementation efforts of a partially implicit time discretization algorithm for modeling slow flows.

\subsection{Magnetized Kelvin--Helmholtz instability}
\label{sec:khi}
For the following test, we ran MHD simulations of a Kelvin--Helmholtz instability. This is the primary instability that arises when there is a velocity shear within a continuous fluid, and it is the main source of vorticity that leads to the energy cascade in 3D turbulent flows. An accurate representation of this process is therefore a fundamental requirement for any numerical scheme to be used for simulating magneto-convection. We considered a 2D domain with $(x,y) \in [0,2]\times[-0.5,0.5]$, mapped on a  $2N\times N$ grid. The horizontal velocity profile is given by
\begin{equation}
    V_x=\mathcal{M}_x[1-2\eta(x)],
\end{equation}
with
\begin{equation}
    \eta(x) =
    \begin{cases*}
      \frac{1}{2} \big\{ 1+\sin \left[ 16\pi(y+0.25)\right] \big\}, & for $y > -\frac{9}{32}$ and $y < -\frac{7}{32}, $ \\
      1,       & for $y \ge -\frac{7}{32}$ and $y \le \frac{7}{32}, $ \\
      \frac{1}{2}\big\{1-\sin\left[16\pi(y-0.25) \right] \big\}, &  for $y > \frac{7}{32}$ and $y < \frac{9}{32}$, \\
      0,  & otherwise.
    \end{cases*}
\end{equation}
The parameter $\mathcal{M}_{x}$ is the maximum sonic Mach number of the horizontal flow, and $p=1$ and $\rho=\gamma$, so initially the adiabatic sound speed $a$ is 1 everywhere. In this test, $\gamma=1.4$. The magnetic field at $t=0$ is uniform and horizontal  ($B_x=0.1\mathcal{M}_x$), and the minimum Alfv\'en Mach number $\mathcal{M}_\mathrm{Alf}$ is 11.82 for all values of $\mathcal{M}_{x}$.

It is well known that magnetic fields aligned with the shear flow have a stabilizing effect because they exert a restoring force on the perturbed interface \citep{chandrasekhar1961}. With a too strong field, the instability may reach saturation when the flow is still essentially laminar or it may be suppressed completely. Instead, weak magnetic stresses do not considerably affect the initial growth of the instability, so the flow can develop the typical vortex structures present in the pure hydrodynamic case. This leads to a much more complex evolution in the nonlinear phase \citep{frank1996}. For this setup, nearly laminar flows are expected only when $\min(\mathcal{M}_\mathrm{Alf})_{t=0}\lesssim 1.1$, as shown in Fig. \ref{fig:C-stability}.

\begin{figure}
  \centering
  \includegraphics[width=0.5\textwidth]{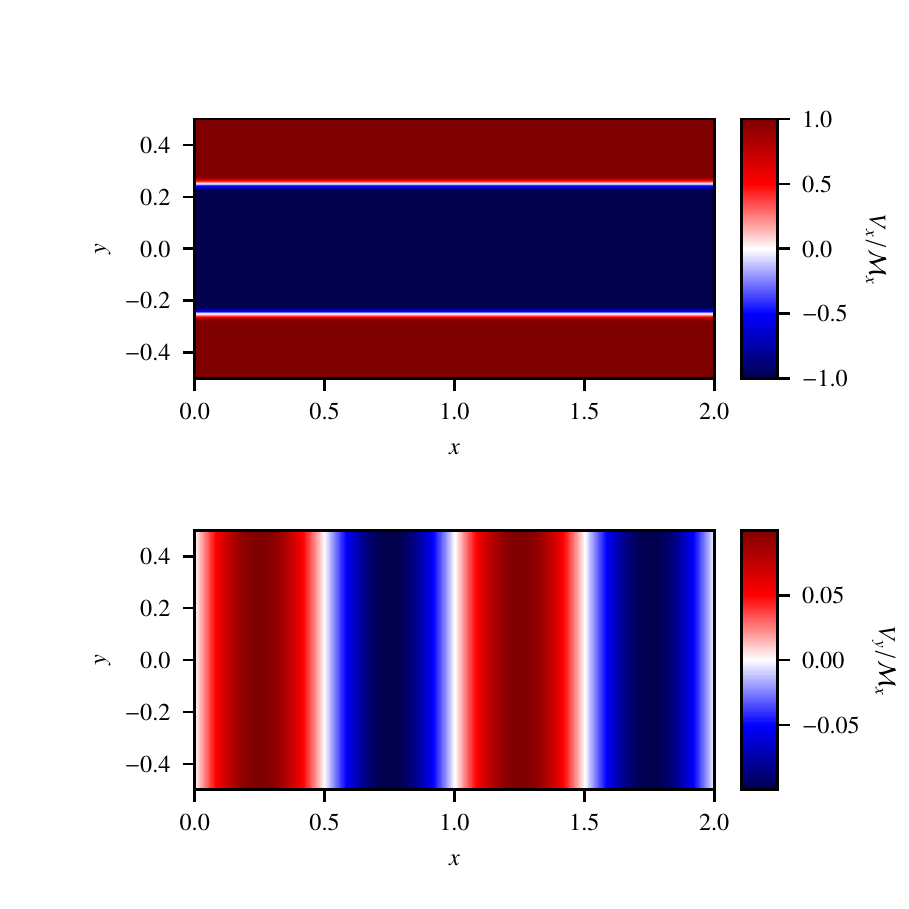}
\caption{Initial setups of $V_x$ and $V_y$ (here rescaled by $\mathcal{M}_x$) used for simulating the growth of the Kelvin--Helmholtz instability.}
\label{fig:C-ICs}
\end{figure}

The instability is started by adding a perturbation to the $y$-velocity component in the initial state, $V_y=0.1\mathcal{M}_x\sin(2\pi x)$ (see Fig. \ref{fig:C-ICs}). The initial conditions are periodic in both directions. The evolution of the Kelvin--Helmholtz instability was studied for a wide range of Mach numbers and grid resolutions:
\begin{equation}
\begin{split}
    & \left( \mathcal{M}_{x} \right) \times \left( N \right) =  \\
    & \left( 10^{-4},10^{-3},10^{-2},10^{-1} \right) \quad \times \\
    & \left( 32, 64, 128, 256, 512, 1024 \right) .
\end{split}
\end{equation}
The final time reached by each simulation was set according to the initial amplitude of the shear flow ($t_\mathrm{max}=4.8/\mathcal{M}_x$). The chosen initial conditions are such that the interface across the shear flow is smooth and resolved, which leads to convergent results at least in the early stages of the evolution of the flow. As in the previous test, we compared the results obtained with both the HLLD and LHLLD solvers.

\begin{figure*}[h!]
  \includegraphics[width=\textwidth]{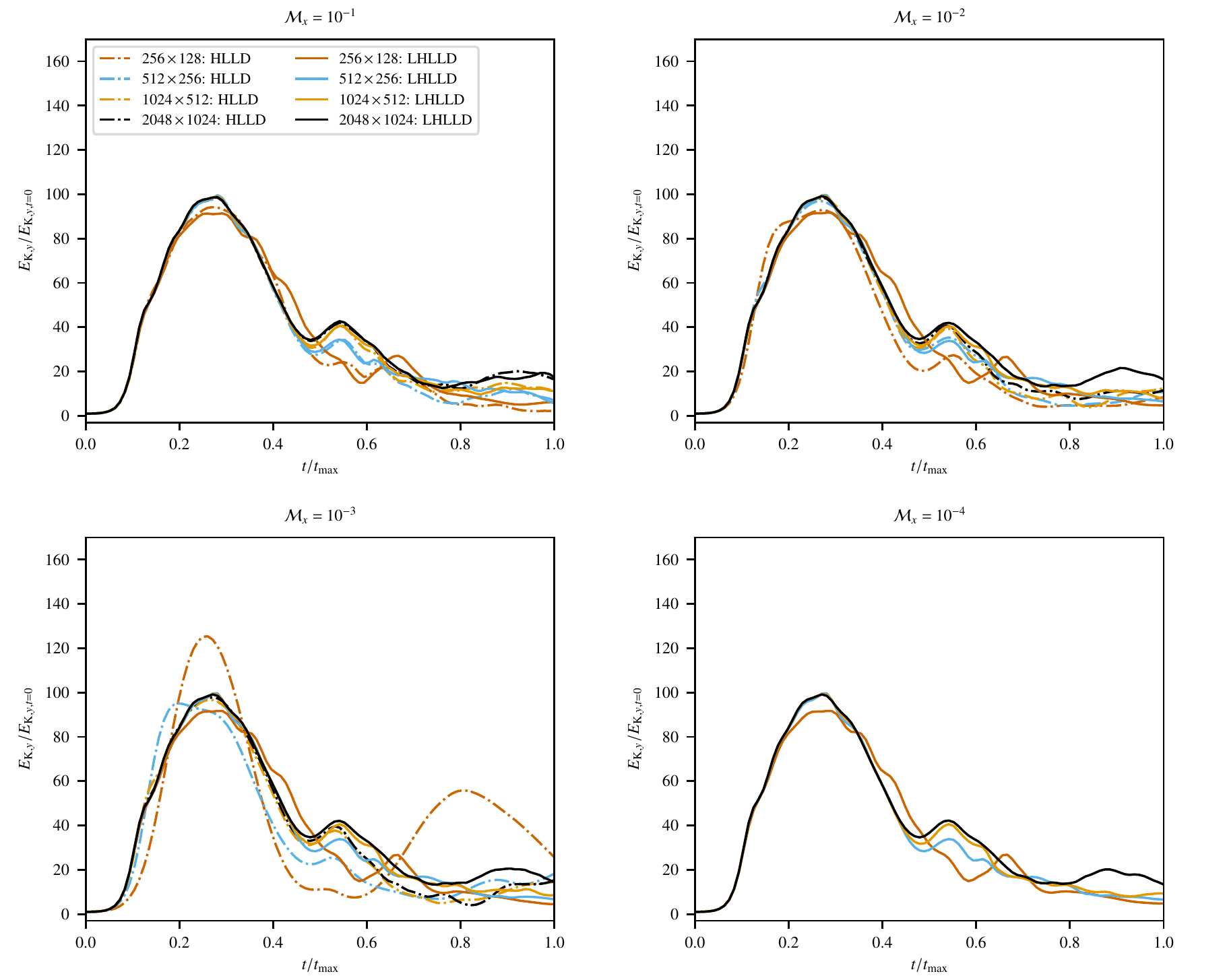}
\caption{Time evolution of the $y$-direction kinetic energy rescaled by its initial value in the magnetized Kelvin--Helmholtz instability test problem. Each panel corresponds to a  different initial Mach number, $\mathcal{M}_x$. Different colors are used for different grid resolutions (the $64\times32$ and $128\times64$ grids cells have been left out for clarity). Results obtained with the HLLD solver are represented by dot-dashed lines, while solid lines are used for LHLLD. The solid black line in each panel is the reference solution.  As explained in the text, the nonlinear solver does not converge when using HLLD at $\mathcal{M}_x=10^{-4}$ for $N>64$.}
 \label{fig:C-Ekiny}
\end{figure*}

\begin{figure*}[h!]
  \includegraphics[width=\textwidth]{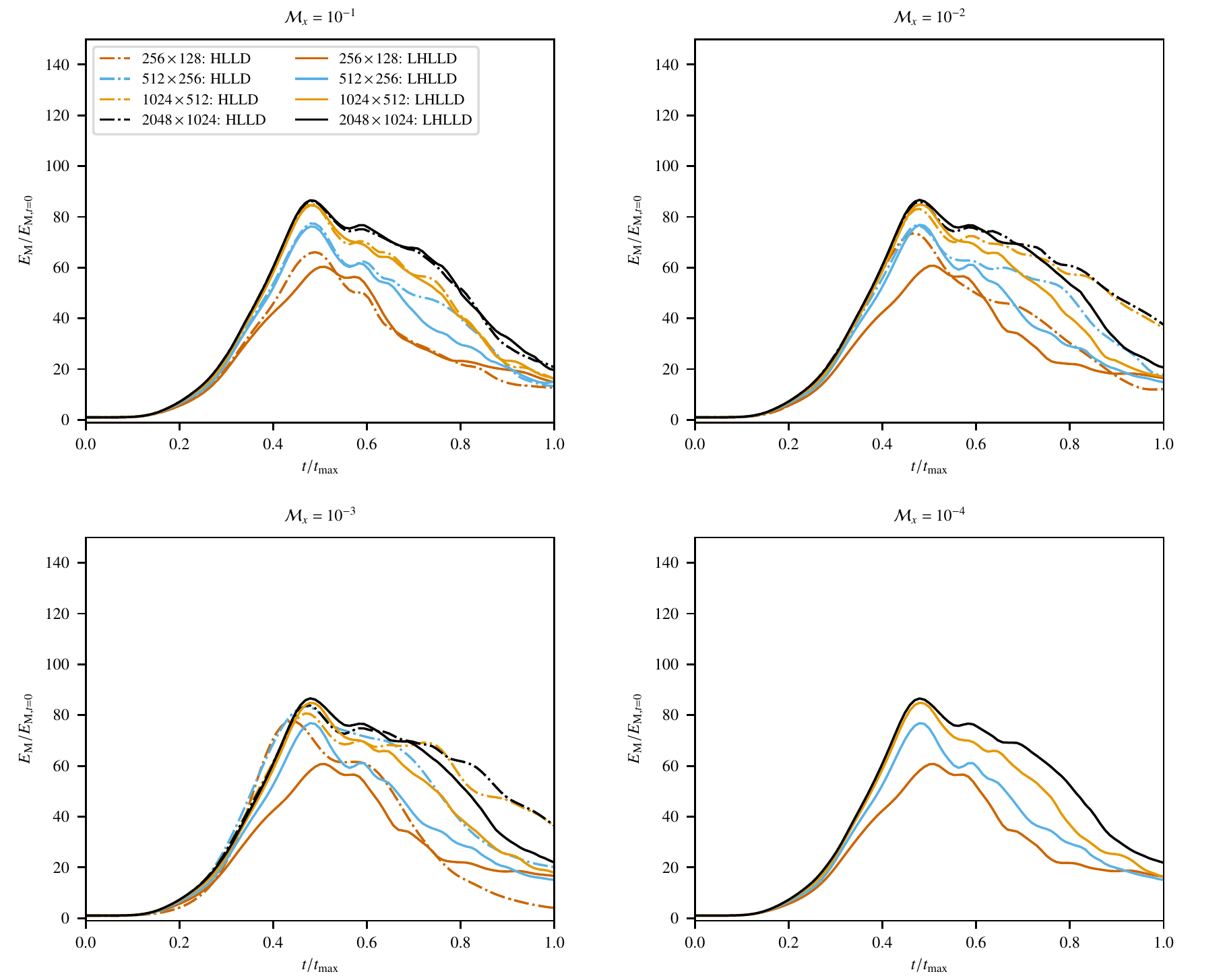}
\caption{Same as Fig. \ref{fig:C-Ekiny} but showing the total magnetic energy divided by its initial value.}
\label{fig:C-Eb}
\end{figure*}

Figures \ref{fig:C-Ekiny} and \ref{fig:C-Eb} show the time evolution of the $y$-direction kinetic energy $E_{\mathrm{K},y}=\sum_{ij}(\rho V^2_y)_{ij}/2$ and the total magnetic energy $E_\mathrm{M}=\sum_{ij}|\bm{B}_{ij}|^2/2$ for all the simulations considered in this study. As in the previous problem, $i$,$j$ are the spatial indices. Because of stretching and wrapping of the field lines within the vortices, the magnetic energy slowly increases with time at the expense of the kinetic energy content of the flow. After the primary rolls reach the top and bottom boundaries ($t/t_\mathrm{max}\simeq0.25$), $E_{\mathrm{K},y}$ saturates due to the periodicity of the grid and starts to decrease. The secondary vortices keep winding up the magnetic field lines until Lorentz forces start to feedback on the velocity field, breaking down these inner structures. The two original shear interfaces get closer to each other (see Fig. \ref{fig:C-reconnection}) until a strong numerical reconnection event happens at $t/t_\mathrm{max}\simeq0.45$, which violently decouples the primary rolls and causes a secondary peak in $E_{\mathrm{K},y}$ at $t/t_\mathrm{max}\simeq0.5$. After this time, other reconnection events break down the flow into smaller structures, and both the magnetic and the kinetic energy are slowly dissipated away by the action of numerical resistivity and viscosity.

Since in this case we solved the ideal MHD equations, there is no characteristic scale on which magnetic and kinetic energy are dissipated into heat, so numerical effects play a significant role on progressively smaller scales at higher resolution. Thus, the amplification and dissipation of magnetic energy hardly converge for the resolutions considered in this study. The initial growth of $E_{K,y}$, in contrast, is not much influenced by the initial weak field, and it is mostly determined by the strength of the shear flows and the width of the shear interface, which is resolved. As a consequence, $E_{K,y}$ converges until the major numerical reconnection event affects the velocity field.
As shown in Fig. \ref{fig:C-Ekiny}, the HLLD solver requires more resolution to reach convergence as the setup is run at progressively lower sonic Mach numbers. Eventually, the Mach-dependent pressure-diffusion coefficient in Eq. \ref{eq:numerical-diffusion} completely dominates the evolution of the flow and deteriorates the numerical solution. For this reason, at $\mathcal{M}_x=10^{-4}$ we were able to successfully run with HLLD only the $64\times32$ and $128\times64$ grids, while for higher resolutions the nonlinear solver failed to converge.

\begin{figure*}[h!]
  \includegraphics[width=\textwidth]{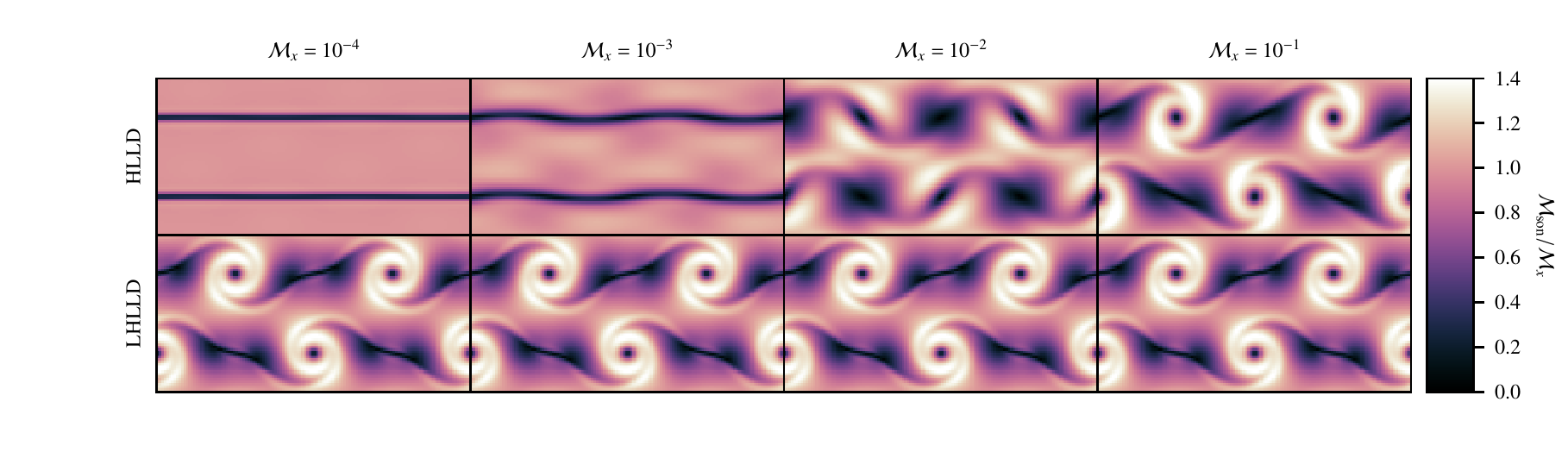}
\caption{Distribution of the sonic Mach number in the Kelvin--Helmholtz instability test at $t/t_\mathrm{max}=1/6$ obtained with the HLLD (top panels) and the LHLLD (bottom panels) solvers on a $128\times64$ grid for different values of $\mathcal{M}_x$. All panels are rescaled by the corresponding value of $\mathcal{M}_x$. }
\label{fig:C-lowres}
\end{figure*}

The effects of numerical dissipation are also shown in Fig. \ref{fig:C-lowres}, where the distributions of the sonic Mach number obtained with HLLD and LHLLD are compared at fixed resolution ($128\times64$ cells) for different values of $\mathcal{M}_x$ at $t/t_\mathrm{max}=1/6$. While in moderately subsonic regimes the large-scale structures in the flow are qualitatively similar, for lower Mach numbers HLLD introduces progressively more dissipation and the instability is eventually halted. When LHLLD is used instead, the morphology of the flow seems to be independent of the Mach number.

\begin{figure}
  \centering
  \includegraphics[width=0.5\textwidth]{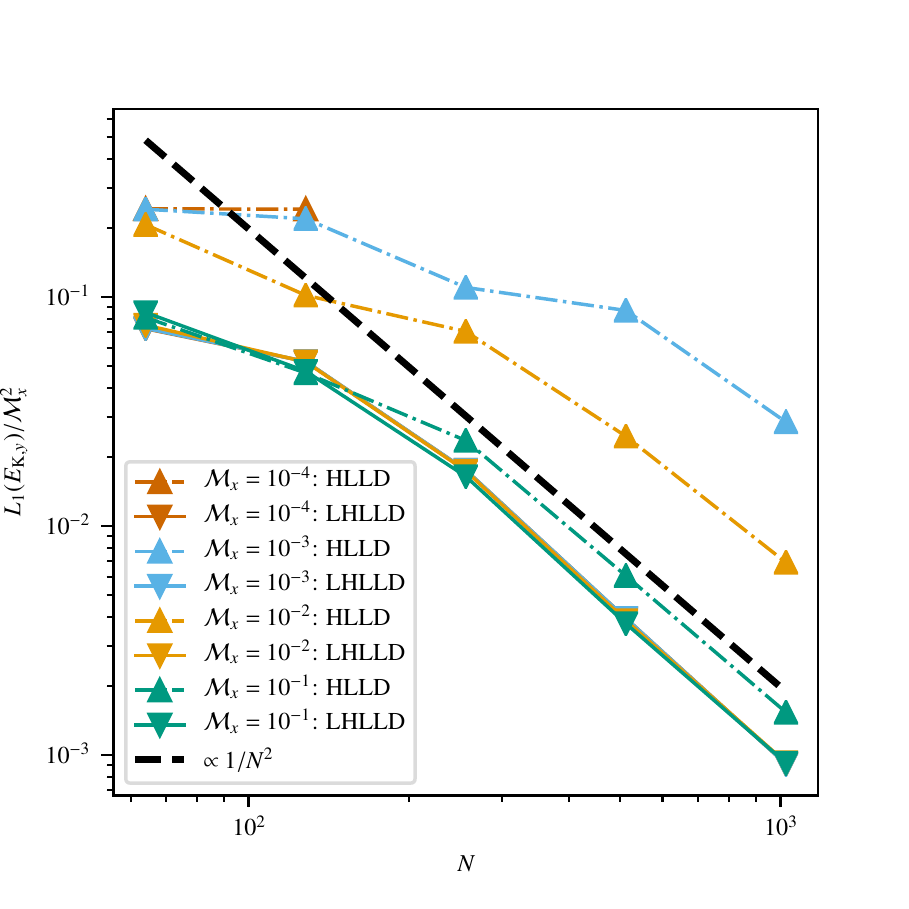}
\caption{Convergence with resolution $N$ of the $L_1$ error associated with $E_{K,y}$ rescaled by $\mathcal{M}^2_x$ in the simulations of the Kelvin--Helmholtz instability. Different colors are used for different initial sonic Mach numbers $\mathcal{M}_x$ using the LHLLD (solid lines) and HLLD (dot-dashed lines) solvers. The dashed black line is the second-order scaling.}
\label{fig:C-convergence}
\end{figure}

Finally, we performed a quantitative convergence study by computing the $L_1$ error associated with $E_{K,y}$ at $t/t_\mathrm{max}=1/6$. At this time, the first rolls have developed to considerable vertical wavelengths (see Fig. \ref{fig:C-lowres}) so that the instability has already entered the nonlinear regime, and the flow is expected to converge as shown in Fig. \ref{fig:C-Ekiny}. The $L_1$ error was computed against a reference solution, which was taken from the highest grid resolution runs considered in this test ($N=1024$) using the LHLLD solver. All simulations (including the reference solutions) were down-sampled to a $64\times32$ grid, so that the errors could directly be computed for different resolutions. This analysis was repeated for different values of $\mathcal{M}_x$ using both HLLD and LHLLD. The results are shown in Fig. \ref{fig:C-convergence}. The errors are rescaled by $\mathcal{M}^2_x$ so that curves corresponding to different sonic Mach numbers lie on the same scale. Overall, the convergence is second-order with $N$ for all simulations. LHLLD provides almost identical (rescaled) errors at given resolution in different regimes of Mach numbers. This is expected because the numerical dissipation introduced by this solver does not depend on $\mathcal{M}_\mathrm{son}$, thanks to the low-Mach fix in Eq. \ref{low Mach-fix}. Instead, the errors computed for the HLLD runs show a clear dependence on the sonic Mach number, and the errors get larger for slower flows. In particular, at $\mathcal{M}_x=0.1$, HLLD needs approximately $1.2$ times the resolution of LHLLD to achieve the same accuracy, which justifies the use of HLLD in this regime of Mach numbers. Instead, when $\mathcal{M}_x=10^{-2}$ and $\mathcal{M}_x=10^{-3}$, HLLD needs respectively  twice or four times the resolution to be as accurate as the low-dissipation flux, which increases the amount of computing time  by $8$  or $64$. Thus, the use of a low-Mach approximate Riemann solver becomes indispensable for providing accurate results in regimes of low sonic Mach numbers with moderate grid resolutions, which would be unfeasible with more standard solvers.

\subsection{Hot bubble}
\label{sec:hot-bubble}

Flows in deep stellar convection zones are usually characterized by the presence of slow parcels of fluid that move in a stratification that is unstable against convection. In the absence of volume heating and cooling processes, these packets of fluid preserve their entropy content until they mix with the surroundings. Therefore, a numerical scheme designed to simulate such flows should have good entropy-conservation properties. However, entropy conservation is hard to achieve if the density, temperature and pressure stratifications span several orders of magnitude and if the flows are very slow, since their entropy content would only be slightly higher or lower than the adiabatic surroundings\footnote{Better entropy-conservation properties can be achieved by directly evolving the specific entropy instead of $\rho E_\phi$. However, this approach does not conserve the total energy.}. Under these conditions, discretization errors caused by  an imperfect balance of the background MHSE stratification can dominate the dynamics and deteriorate the numerical solution. As shown in Sect.~\ref{sec:well_balancing}, the magnitude of such errors can be drastically reduced by using well-balancing techniques.

In this section we check the entropy-conservation properties of the MHD scheme implemented in SLH by running simulations of the hot bubble setup described by \cite{edelmann2021a}, where a bubble of higher entropy content with respect to the surroundings buoyantly rises in an adiabatic stratification. The physical domain is mapped on a 2D Cartesian grid ($N_x=2/3\times N_y$), and the background stratification is in MHSE. Boundary conditions are periodic everywhere and the gravitational acceleration takes the form
\begin{equation}
    g_y(y) = g_0 \sin(k_y y),
\end{equation}
where $g_0=-1.09904373\times10^5$ cm s$^{-2}$, $k_y=2\pi / L_y$, $y$ is the vertical spatial coordinate, and $L_y$ is the vertical extent of the grid. The value of $g_0$ is set such that the ratio of the maximum to the minimum gas pressure\footnote{More details on how to compute the pressure profile can be found in \cite{edelmann2021a}.} $p(y)$ is 100, which corresponds to $4.6$ pressure scale heights. The entropy profile inside the bubble is given by
\begin{equation}
    A = A_0 \Bigg\{ 1+\left(\frac{\Delta A}{A}\right)_{t=0} \cos\left(\frac{\pi}{2}\frac{r}{r_0}\right)^2 \Bigg\},
\end{equation}
where $A_0$ is background entropy, $r_0$ is the radius of the bubble, $r$ is the distance from the center of the bubble and $\left(\Delta A / A\right)_{t=0}$ is the initial entropy perturbation. The density is
\begin{equation}
    \rho(y)=\left(\frac{p(y)}{A}\right)^{1/\gamma},
\end{equation}
so that the (initial) buoyant acceleration of the bubble is proportional to the entropy perturbation,
\begin{equation}\label{eq:buoyancy}
    a_\mathrm{b} =\frac{\Delta \rho}{\rho}g_y \propto \left(\frac{\Delta A}{A}\right)_{t=0}.
\end{equation}
We ran the models for the set of parameters
\begin{equation}
\begin{split}
    &\left(\frac{\Delta A}{A}\right)_{t=0} \times  \left(N_y\right)= \\ &\left(10^{-7},10^{-5},10^{-3},10^{-1}\right) \quad  \times \\
    &\left( 96,192,384,768 \right),
\end{split}
\end{equation}
and we set the maximum time such that in each run the bubble raised approximately the same distance $l$. This allowed different regimes of sonic Mach numbers  to be simulated, as the velocity, $V$, reached by the bubble over a length, $l$, scales as
\begin{equation}
    V \propto (a_\mathrm{b}l)^{1/2}.
\end{equation}
This ultimately leads to the relation
\begin{equation}\label{mach-entropy-scaling}
    \mathcal{M}_\mathrm{son} \propto \left( \frac{\Delta A}{A} \right)^{1/2}_{t=0}.
\end{equation}
A uniform horizontal magnetic field was added to the system, and its strength was rescaled depending on the entropy perturbation,
\begin{equation}
    B_x = B_0\left( \frac{\Delta A}{A} \right)^{1/2}_{t=0}.
\end{equation}
This ensures that the relative magnitude of magnetic stresses compared to the ram pressure of the bubble remains the same for all simulations, and that the morphology of the flow is unaltered.  $B_0=47.3$ was chosen such that the final Alfv\'en Mach number at the position of largest entropy is in the range $\mathcal{M}_\mathrm{Alf}\simeq2-3$ depending on the grid resolution. Thus, magnetic fields are dynamically important but not strong enough to suppress buoyancy.

In Fig. \ref{fig:entropy} we show the final entropy excess for all the simulations run in the parameter study. The center of the bubble accelerates faster than other regions as it is the point with maximum entropy, and the acceleration profile across the bubble leads to the development of shear at its outer edges. As the bubble rises in the stratification, the magnetic field lines are stretched into thin tubes, which locally amplifies the magnetic energy (see Fig. \ref{fig:magnetic-pressure}). The amount of amplification depends on the numerical resistivity and so on resolution. In contrast to the pure hydrodynamic case studied by \cite{edelmann2021a}, here the presence of a magnetic field suppresses the formation of vortices at the sides of the bubble. Overall, the entropy content of the bubble is well preserved even on the coarsest grid, but some negative entropy fluctuations are present at the very top of the bubble. These negative fluctuations are numerical artifacts. In fact, the entropy fluctuations may locally increase as a fraction of magnetic and kinetic energy is dissipated into internal energy, but they cannot become negative  physically. These artifacts do not depend on the entropy perturbation,  and they are limited to a very narrow region in the spatial domain that tends to shrink as the resolution is increased. All models converge upon grid refinement.

\begin{figure}
  \centering
  \includegraphics[width=0.5\textwidth]{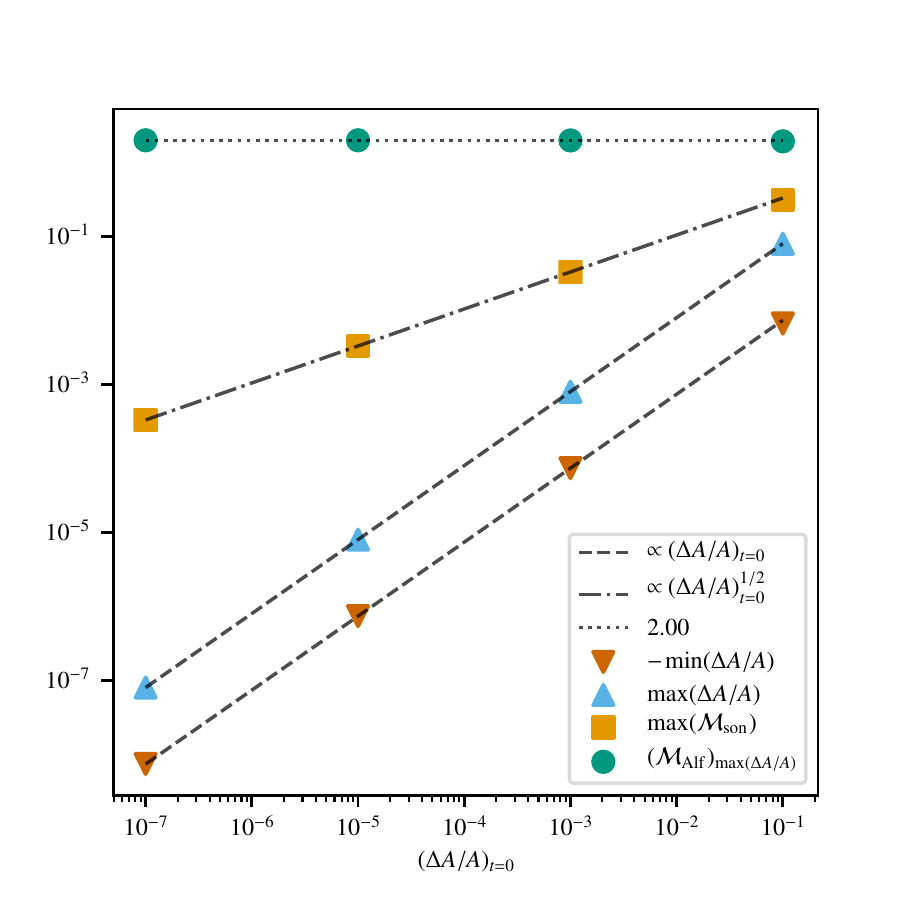}
  \caption{Maximum sonic Mach number, minimum and maximum entropy fluctuations and Alfv\'en speed of the hot bubble as a function of the initial entropy perturbation obtained on a $64\times96$ grid. The black lines represent the physical scalings.}
\label{fig:D-scaling}
\end{figure}

According to Eq. \ref{mach-entropy-scaling}, the sonic Mach number of the bubble is expected to scale as the square root of the initial entropy perturbation. Any deviation from this relation, which has been obtained on the basis of physical arguments, can be due to difficulties in modeling slow flows in a stratified setup and the build-up of significant numerical errors. In Fig. \ref{fig:D-scaling} we show this scaling for the coarsest grid resolution. All data points overlap with the theoretical curve, and the minimum Mach number $\mathcal{M}_\mathrm{son}$ achieved in this parameter study is $3.32\times10^{-4}$ (see also Fig. \ref{fig:mach}). The ratio of the rising velocity of the bubble to the Alfv\'en speed (in the point of maximum entropy) does not depend on the amplitude of the entropy perturbation. Since the initial magnetic field is proportional to $\left(\Delta A / A \right)^{1/2}_{t=0}$, the amount of amplification due to induction only depends on the velocity of the bubble $V$ and the timescale over which magnetic induction operates ($\propto 1/V$).

\begin{figure*}[h!]
  \centering
  \includegraphics[width=\textwidth]{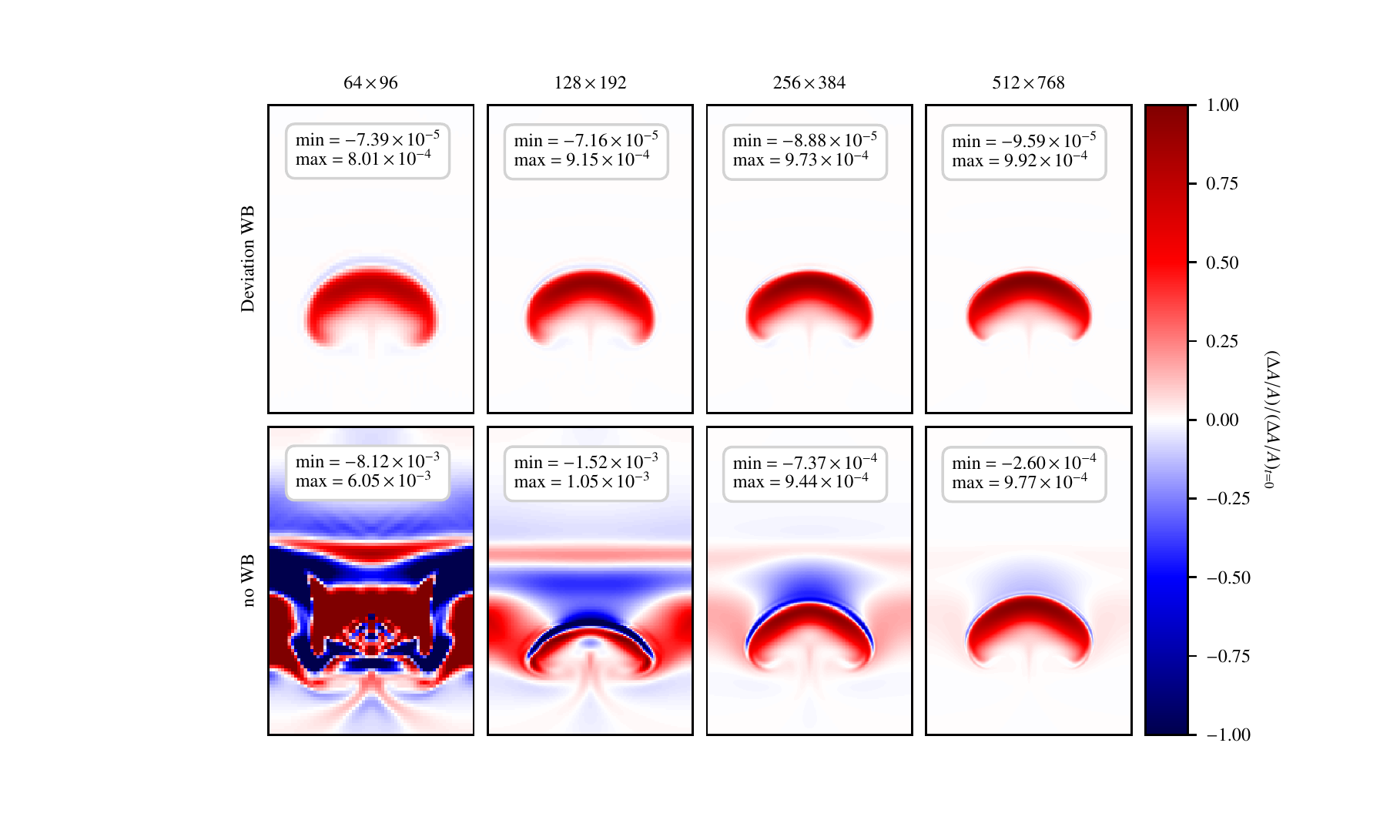}
  \caption{Final distribution of the entropy fluctuations of the hot bubble for $(\Delta A/A)_{t=0}=10^{-3}$ at different grid resolutions. The entropy fluctuations are rescaled by $(\Delta A/A)_{t=0}$. The top row shows the results obtained with deviation well-balancing, whereas no well-balancing method was used in the simulations shown in the bottom row. The insets show the minimum and maximum values of the entropy fluctuation in each panel.}
\label{fig:wb}
\end{figure*}

Finally, to quantify the strength of the spurious flows that are expected to arise if the stratification is left unbalanced, in Fig. \ref{fig:wb} we show a comparison between simulations obtained with and without deviation well-balancing, where the vertical resolution $N_y$ ranges from $96$ to $768$. For this comparison, we fixed $\left(\Delta A/A \right)^{1/2}_{t=0}=10^{-3}$ such that the maximum sonic Mach number of the bubble is approximately $3\times10^{-2}$. The unbalanced simulations develop large entropy fluctuations, both negative and positive, which strongly deteriorate the numerical solution. As the grid is refined, the simulations tend to converge, but wide regions of negative entropy fluctuations are still present even on the finest grid. Thus, this test demonstrates that well-balancing techniques are fundamental to correctly simulate the evolution of small entropy perturbations in steep isentropic stratifications and to reduce the effects of numerical errors when using moderately coarse grids.

\subsection{Small-scale dynamo in a stratified setup}
\label{sec:ssd}

The previous tests demonstrated that the proposed MHD implementation can accurately simulate slow flows even in strongly stratified setups. As this numerical method will mostly be applied to simulate stellar interiors, it seems natural to test the scheme for dynamo amplification, which is the main cause for the generation of strong magnetic fields in a wide mass range of stars \citep[see][and references therein]{brun2017}. In this section we focus on simulations of SSDs, where the magnetic energy is amplified on scales comparable to or smaller than the scales at which turbulence is forced \citep{meneguzzi1981,schekochihin2004, brandenburg2005,schekochihin2007,iskakov2007}, in contrast to large-scale dynamos where most of the magnetic energy is at scales larger than the forcing scale \citep{brun2004,kapyla2008,charbonneau2013,augustson2016}. Even though the efficiency of the dynamo amplification depends on many physical parameters, including the magnetic Prandtl number $P_\mathrm{m}=\nu/\eta$ \citep{schekochihin2004,schekochihin2007,graham2010,brandenburg2011,brandenburg2014}, here we do not perform a parameter study for $P_\mathrm{m}$, since in the current MHD scheme the viscosity ($\nu$) and resistivity ($\eta$) coefficients are not fixed, but are intrinsic to the underlying numerical methods, so they are not easy to constrain. Instead, we aim to check if it is possible to excite an SSD using the SLH code at low sonic Mach numbers.

We built the initial conditions based on the work of \cite{andrassy2022}, who performed a pure hydrodynamic study of a 3D convection zone with a stable layer on top of it, where the convective flows had a typical maximum Mach number $\mathcal{M}_\mathrm{son}\sim 0.1$. The stratification of that model resembled oxygen shell burning in a massive star, even though some simplifications were adopted. Among these, an ideal gas EoS was used and effects of neutrino cooling were ignored. Here, we modified that setup even further by retaining only the convection zone and by decreasing the rate of energy injection to test our method in the low-Mach-number regime. Removing the stable layer greatly simplifies the problem at low Mach numbers, as the propagation of internal gravity waves does not need to be resolved. Since the wavelength of these modes becomes shorter for progressively slower convective flows \citep{sutherland2010,edelmann2021a}, high grid resolutions would be necessary to capture this process accurately at low Mach numbers, which would make the simulations very expensive.

For our experiment, we used $N_x\times N_y\times N_z=2N\times N\times2N$ grid cells and the spatial domain (normalized by a characteristic length $L=4\times10^8 \, \mathrm{cm}$) is $-1\leq x\leq1$, $1\leq y\leq2$, $-1\leq z\leq1$. Periodic boundaries were used in the horizontal directions, while reflecting boundaries were used in the vertical direction. The initial stratification is adiabatic and in MHSE, and it is given by the polytropic relation
\begin{equation}
    \frac{\mathrm{d}\ \mathrm{ln}\ p}{\mathrm{d}\ \mathrm{ln}\ \rho} = \gamma.
\end{equation}
The stratification spans 2.2 pressure scale heights. The gravitational acceleration takes the form
\begin{equation}
    g(y) = g_0 f_g(y) y^{-5/4},
\end{equation}
where $g_0=1.414870$, and
\begin{equation}
    f_g(y) =
    \begin{cases*}
      \frac{1}{2}\Big\{ 1+ \sin \left[ 16\pi\left( y-\frac{1}{32}\right) \right]\Big\}, &  for $1\leq y < 1+\frac{1}{16}$, \\
      1, &  for $1+\frac{1}{16} \leq y < 2-\frac{1}{16}$, \\
      \frac{1}{2}\Big\{ 1+ \sin \left[ 16\pi\left( y-\frac{1}{32}\right) \right]\Big\}, &  for $2-\frac{1}{16}\leq y < 2$, \\
      0, & otherwise.
    \end{cases*}
\end{equation}
Moreover, in contrast to \cite{andrassy2022}, we are not interested in studying convective boundary mixing, so we only used a single species with mean molecular weight $\mu=1$.

Convection is driven by a constant in time heat source $\dot{q}(y)$ placed close to the bottom boundary,
\begin{equation}
    \dot{q}(y)  =
    \begin{cases}
       b\dot{q}_0 \sin\left( 8\pi y \right),& \mathrm{for}\ 1 \le y \le 1 +
       \frac{1}{8}, \\[0.5em]
        0,& \mathrm{for}\ 1 + \frac{1}{8} < y \le 2,
    \end{cases}
\end{equation}
where $\dot{q_0}=3.795720\times10^{-4} \, \mathrm{erg} \, \mathrm{cm}^{-3}\, \mathrm{s}^{-1}$ and $b$ is a nondimensional factor that allows the strength of the convective flows to be controlled through the scaling \citep[see, e.g.,][]{kippenhahn2013,andrassy2020,horst2021a,edelmann2021a,kapyla2021}
\begin{equation}
\label{scaling-mach-boosting}
   \mathcal{M}_\mathrm{son}\propto \dot{q}^{1/3}  \propto b^{1/3}.
\end{equation}
We considered the following grid of models:
\begin{equation}
\label{grid-models}
\begin{split}
    & \left( b \right) \times \left( N \right) =  \\
    & \left( 10^{-6},10^{-3},1 \right) \quad \times \\
    & \left( 32, 64, 96 \right).
\end{split}
\end{equation}
We ran the simulations at nominal luminosity ($b=1$) with SSP-RK2, where the convective flows are characterized by a  maximum $\mathcal{M}_\mathrm{son}$ of $0.1$. In this regime of sonic Mach numbers, IESS is less efficient than explicit time-steppers (see Fig. \ref{fig:efficiency}). In contrast, the cases with $b=10^{-3}$ and $b=10^{-6}$ were run with IESS, since according to Eq. \ref{scaling-mach-boosting}, the maximum $\mathcal{M}_\mathrm{son}$ is $10^{-2}$ and $10^{-3}$, respectively.

In order to initiate an SSD, a weak seed magnetic field was added to the system\footnote{As observed by \cite{seta2020}, the evolution of the dynamo in the nonlinear regime does not depend on the form of the seed field, as long as its magnitude is weak enough to not affect the development of convection.},
\begin{equation}
    B_x(t=0)=10^5 b^{1/3},
\end{equation}
where the dependence on $b$ is such that the timescale on which the magnetic energy reaches saturation (expressed in units of convective turnovers) does not depend on the value of $b$. The convective turnover timescale was estimated as
\begin{equation}
    \tau_\mathrm{conv} = 2\frac{L}{\langle\sigma_V\rangle_t},
\end{equation}
where
\begin{equation}
    \sigma_V= \sqrt{\sigma_{V,x}^2 + \sigma_{V,y}^2 + \sigma_{V,z}^2}
\end{equation}
is averaged over time, and $\sigma_{V,x}^2$, $\sigma_{V,y}^2$, $\sigma_{V,z}^2$ represent the standard deviation of each velocity component computed over the whole domain.  All simulations were run until $t/\tau_\mathrm{conv}=70$, with $\tau_\mathrm{conv} \simeq 5000,500,50$ s for $b=10^{-6},10^{-3},1$, respectively. The initial MHSE density stratification was perturbed to initiate convection\footnote{Details on how to compute the density perturbation can be found in \cite{andrassy2022}.}. This perturbation leads to the development of buoyant structures that rise in the stratification until they hit the top boundary, after which they quickly become turbulent (see Fig. \ref{fig:vcuts}). Convection fully develops after one convective turnover.
\begin{figure*}[h!]
  \centering
  \includegraphics[width=\textwidth]{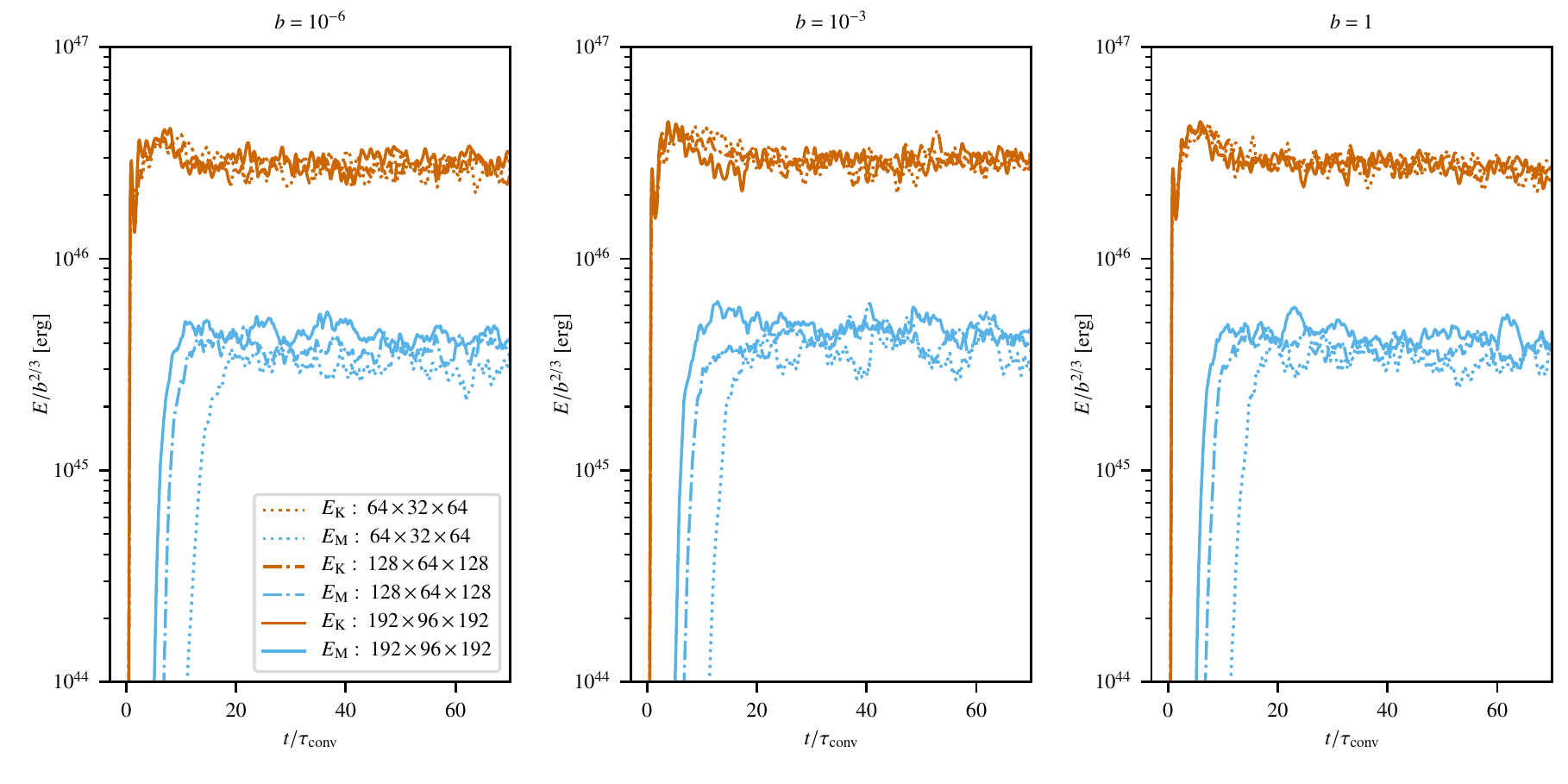}
  \caption{Time evolution of the magnetic energy (sky blue) and kinetic energy (vermillion) in the simulations of the SSD for different grid resolutions (dotted: $64\times32\times64$, dot-dashed:$128\times64\times128$, solid: $192\times96\times192$). Each panel shows the results obtained with a specific value of $b$ (from left to right: $b=10^{-6},10^{-3},1$). The time is expressed in units of the convective turnover, while the magnetic and kinetic energy curves are rescaled by $b^{2/3}$ to take into account the different energy contents of the flows.}
\label{fig:kin-sat}
\end{figure*}

\begin{figure}
  \centering
  \includegraphics[width=0.5\textwidth]{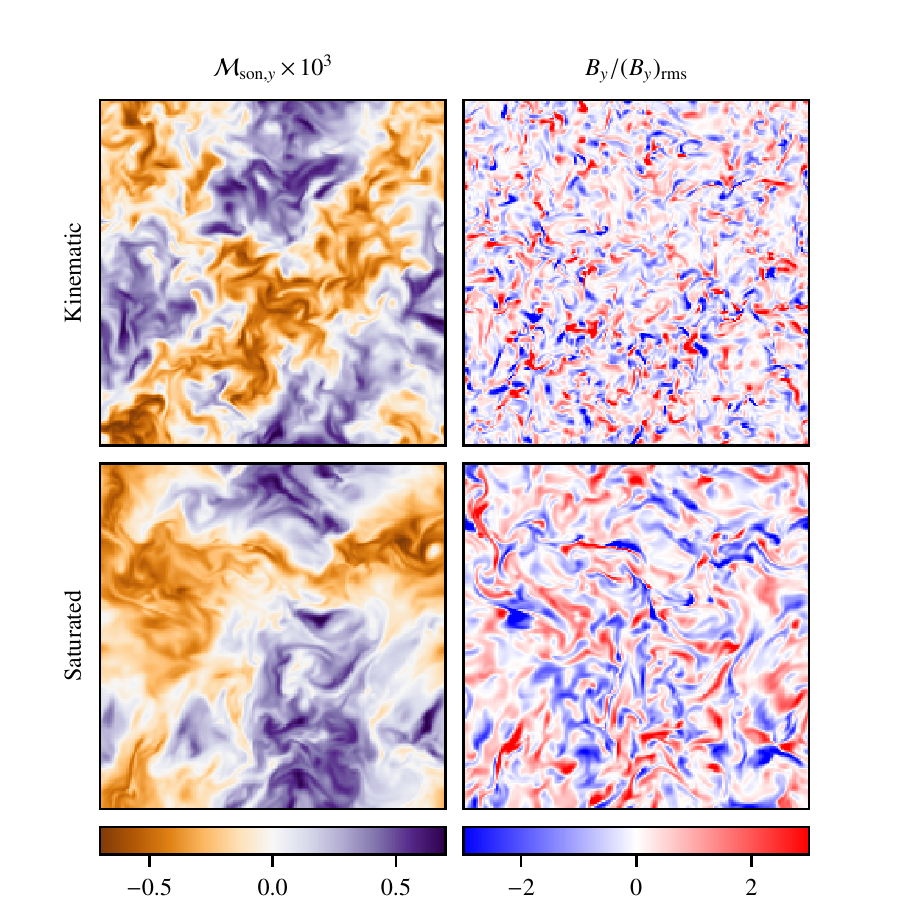}
  \caption{Horizontal slices in the $y=1.5$ plane taken in the kinematic (upper plots) and saturated (lower plots) regimes of the SSD with $b=10^{-6}$ on the $192\times96\times192$ grid. The panels on the left show the vertical sonic Mach number $\mathcal{M}_{\mathrm{son},y}=V_y/a$ multiplied by $10^3$, while the plots on the right show the vertical magnetic field rescaled by the root mean square value across the plane.}
\label{fig:cuts}
\end{figure}

Figure \ref{fig:kin-sat} shows the time evolution of the kinetic and the magnetic energy for all the simulations run in the grid of models considered in this test. In the kinematic phase, the magnetic field is irrelevant to the dynamics, and it is amplified exponentially by the action of a dynamo process, with most of the magnetic energy distributed close to the resistive scale. As visible in the horizontal cuts shown in Fig. \ref{fig:cuts}, the magnetic field distribution is characterized by small-scale structures with mixed polarity, while the velocity field is distributed on slightly larger scales, which suggests that $P_\mathrm{m}\gtrsim 1$. The growth rate $\gamma_\mathrm{M}= \mathrm{d}(\mathrm{ln} E_\mathrm{M}) / \mathrm{d} \tau_\mathrm{conv}$ increases with resolution  (see also Fig. \ref{fig:time-evol-lin}), which is compatible with SSD amplification. In particular, we find that $\gamma_\mathrm{M}\propto \Delta x^{-0.8}$. The dependence of the growth rate on $\Delta x$ is weaker than $\gamma_\mathrm{M}\propto \Delta x^{-4/3}$, which is typically observed in simulations of SSDs in solar and stellar convection zones \citep{graham2010,rempel2014,hotta2015,riva2022,cuissa2022}, and steeper than  $\gamma_\mathrm{M}\propto \Delta x^{-2/3}$, which is predicted by the Kazantsev dynamo theory \citep{kazantsev1968,brandenburg2005}.

When the magnetic field becomes strong enough, the Lorentz force starts to influence the evolution of the turbulent flows, damping the velocity on the small scales (see the bottom panels in Fig. \ref{fig:cuts}). A statistically steady state configuration is then reached where the magnetic energy achieves sub-equipartition values $E_\mathrm{M}/E_\mathrm{K}\simeq0.1-0.2$. In all simulations, an SSD is successfully excited, and the mean value of the amplified magnetic energy increases with resolution (see Table \ref{table}). In fact, the size of the resistive scale is smaller on finer grids, which in turn increases the maximal stretching rate of the field lines and the SSD becomes more efficient. In contrast, no systematic difference is observed in the magnetic to the kinetic energy ratio in simulations run with the same resolution but different values of $b$. This is due to the fact that, thanks to the use of the LHLLD solver, the size of the viscous and resistive scales does not depend on the sonic Mach number of the flow, which is mostly determined by the chosen value of $b$. At given resolution, the SSD amplifies the magnetic field on the same spatial (resistive) scales, so the evolution of $E_\mathrm{M}/E_\mathrm{K}$ becomes virtually independent of $\mathcal{M}_\mathrm{son}$ (and so of $b$) if the time is rescaled by the convective turnover $\tau_\mathrm{conv}$, except for statistical fluctuations caused by the chaotic nature of the turbulent flows.

\begin{table}
\caption{Time averages of $E_\mathrm{M}/b^{2/3}\ [10^{45}\times\mathrm{erg}]$ over the time interval $20<t/\tau_\mathrm{conv}<70$ for the different resolutions, $N$, and boost factors, $b$, considered in this study. The errors represent one standard deviation over the time series.}              
\label{table}      
\centering                                      
\begin{tabular}{c | c | c | c}
\toprule
 & $b=10^{-6}$ & $b=10^{-3}$ & $b=1$ \\
\midrule
$N=32$ & $3.13\pm0.34$ & $3.65\pm0.60$ & $3.33\pm0.37$\\      
$N=64$ & $3.73\pm0.32$ & $4.31\pm0.55$ & $3.78\pm0.31$\\      
$N=96$ & $4.40\pm0.44$ & $4.67\pm0.39$ & $4.40\pm0.47$\\      
\bottomrule
\end{tabular}
\end{table}

\begin{figure}
  \centering
  \includegraphics[width=0.5\textwidth]{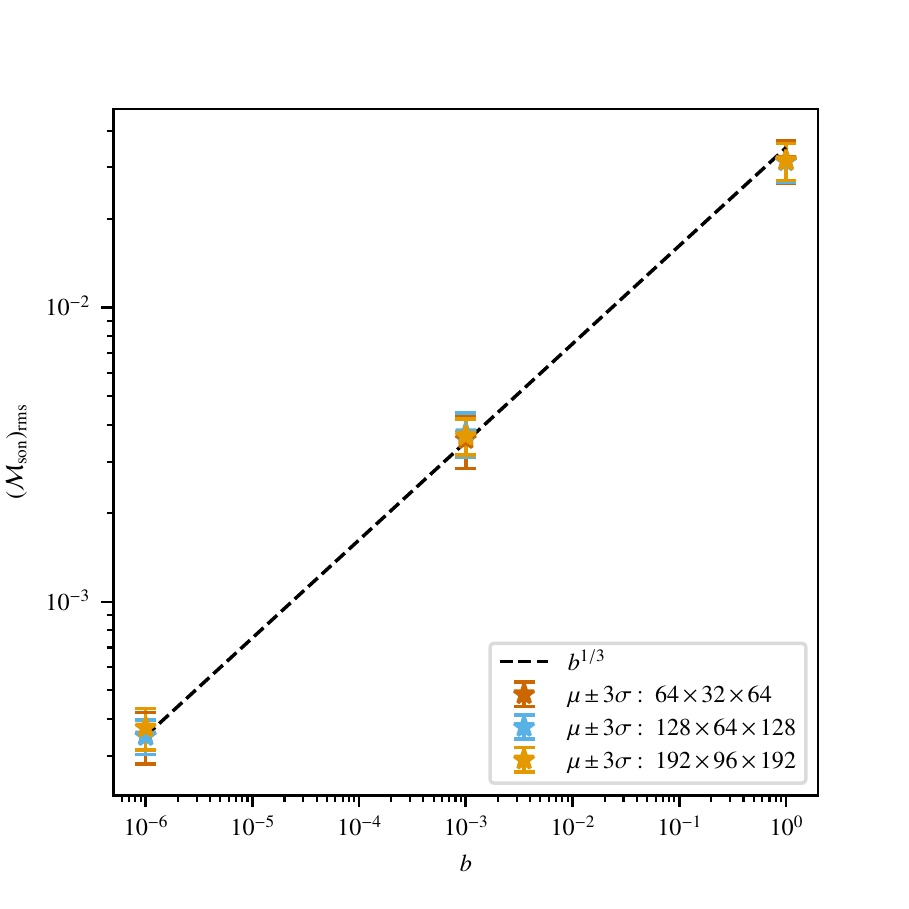}
  \caption{Root mean square of the sonic Mach number as a function of the driving luminosity $b$. The data points are averages computed in the time interval $20< t/\tau_\mathrm{conv} <40$. The error bars represent three standard deviations over the time series, while the dotted black line is the $b^{1/3}$ scaling.}
\label{fig:bthirds}
\end{figure}

Figure \ref{fig:bthirds} shows the root mean square sonic Mach number averaged over 20 convective turnovers in the saturated regime as a function of $b$. As noted in \cite{edelmann2021a}, numerical errors introduced by an unbalanced stratification can cause deviations from the scaling in Eq. \ref{scaling-mach-boosting}. In this case, the use of deviation well-balancing and LHLLD allows a good agreement between the computed Mach numbers and the scaling law to be reached, within three standard deviations. This proves that the convective flows are correctly simulated and are not dominated by numerical errors. The smallest $(\mathcal{M}_\mathrm{son})_\mathrm{rms}$ achieved in these runs is approximately $3.7\times10^{-4}$, which is close to what typically found in simulations of core-convective stars \newline \citep{augustson2016,edelmann2019,horst2020a,higl2021}.

\begin{figure}
  \centering
  \includegraphics[width=0.5\textwidth]{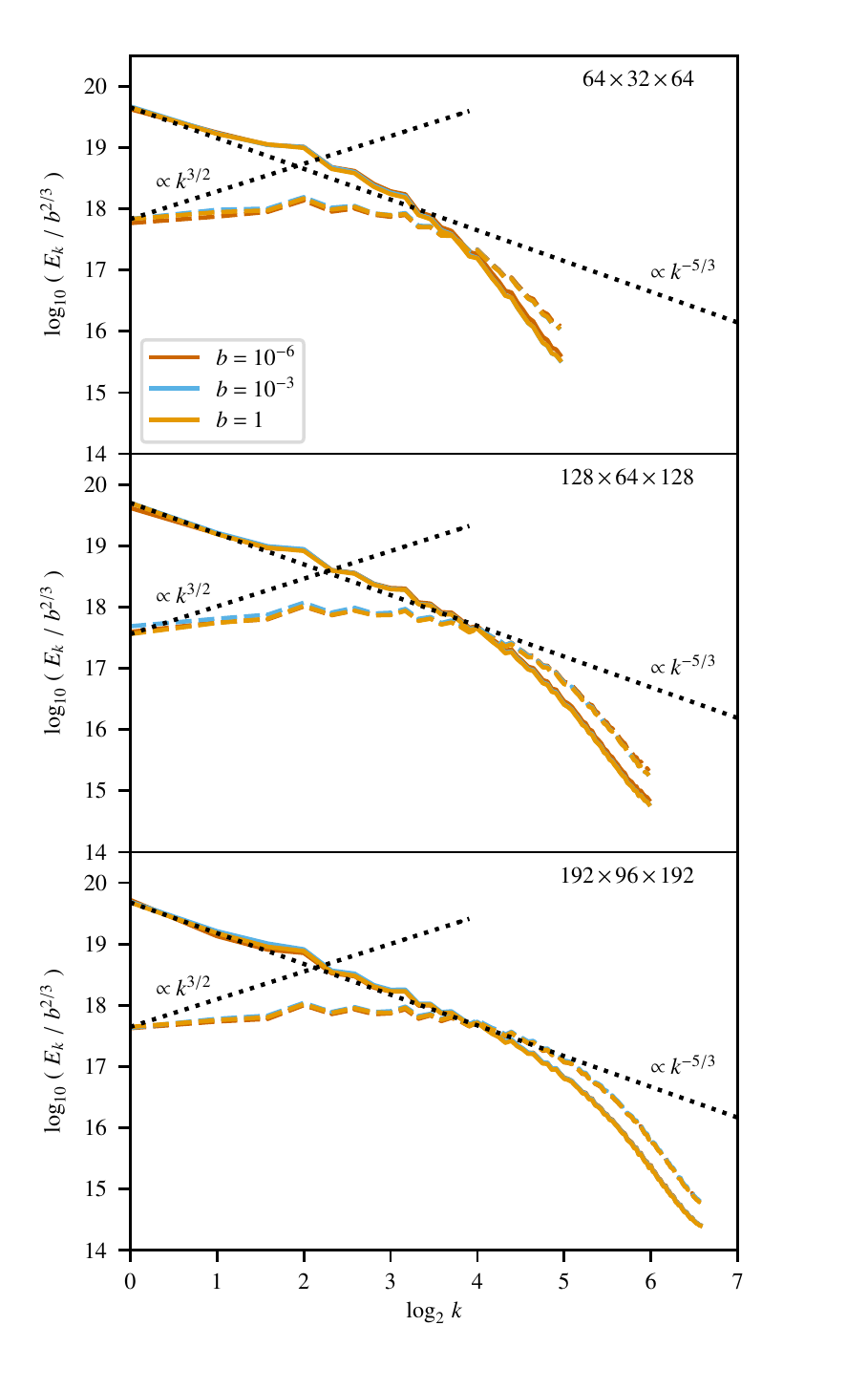}
  \caption{Kinetic (solid lines) and magnetic (dashed lines) energy spectra as a function of the wave number $k$ obtained in an horizontal slice of the convective box at $y=1.5$. All spectra are averaged over the time interval $20< t/\tau_\mathrm{conv} <40$ and are rescaled by $b^{2/3}$. The three panels show the results for different numbers of grid cells (from top to bottom: $64\times32\times64$, $128\times64\times128$, $192\times96\times192$). The dotted black lines are the Kolmogorov ($k^{-5/3}$) and Kazantsev ($k^{3/2}$) scalings.}
\label{fig:spectra}
\end{figure}

Finally, Fig. \ref{fig:spectra} shows the kinetic and magnetic energy spectra (taken in the midplane of the box) in the saturated stage for different values of $b$ and grid resolutions. Both spectra have been rescaled by $b^{2/3}$ to take into account the different energy contents of the flows achieved with different values of $b$. The kinetic energy spectra converge to the $k^{-5/3}$ Kolmogorov law \citep{kolmogorov1941} upon grid refinement, and the dissipation range shifts toward progressively larger wave numbers $k$. The magnetic energy distributions peak in the inertial range, as expected in SSD simulations, and on the large scales they show a shallower dependence on $k$ than the Kazantsev isotropic dynamo theory, $k^{3/2}$ \citep{kazantsev1968}. This can be explained by the fact that, in this setup, turbulence is not isotropic, and large-scale anisotropic convective flows are present because of the steep stratification and the use of closed vertical boundaries (see Fig. \ref{fig:vcuts}). Magnetic and kinetic energy achieve equipartition at the bottom of the inertial range, except for the simulations run on $64\times32\times64$ grid cells, in which equipartition is reached only in the dissipation range. The maximum magnetic to kinetic energy ratio is around $3$ in the dissipation range for the highest resolution considered in this study. Again, since the numerical diffusion of the MHD scheme is Mach-independent, the shape and amplitude of the rescaled spectra do not depend on $b$ on any of the three grids. Thus, this test indicates that SLH is capable of correctly simulating fully compressible magneto-convection and SSDs in regimes of low sonic Mach numbers even with moderate grid resolutions.

\section{Summary and conclusions}
\label{sec:conclusions}

In this work we have presented a new finite-volume scheme for solving the fully compressible MHD equations with gravity in regimes of low sonic Mach numbers and high-$\beta$ environments, which is suitable for simulating magneto-convection and dynamo processes in deep layers of stars.
This method relies on a low-dissipation MHD Riemann solver \citep[LHLLD;][]{minoshima2021} to avoid the excessive numerical dissipation typical of high-resolution, shock-capturing solvers as $\mathcal{M}_\mathrm{son} \rightarrow 0$.

The strict CFL condition on the time step is overcome by using an implicit-explicit time discretization algorithm, for which the induction equation is integrated using an explicit time-stepper, while the rest of the MHD system is integrated implicitly. The solutions to the two subsets of equations are coupled through Strang splitting following the prescription of \cite{fuchs2009}. The combined marching scheme has a less restrictive condition on the time step, which is limited only by the fastest fluid and Alfv\'en speeds instead of the fast magnetosonic speed, leading to a considerable speed-up in regimes of low sonic Mach numbers.

Whenever required, a magnetohydrostatic solution can be enforced on the discrete grid with the deviation well-balancing method \citep{berberich2019, edelmann2021a}. This technique leads to better entropy-conservation properties of the numerical scheme, even in cases where the pressure and density stratifications span several orders of magnitude across the computational domain. Finally, the $\nabla \cdot \bm{B} = 0$ condition is enforced using the CT-contact method \citep{gardiner2005}. This new scheme is implemented in the SLH code, and it has been tested in five numerical experiments.

First, we checked the global convergence of the methods by following the propagation of linear modes for all the MHD waves on a 1D grid. This test proves that the scheme is globally second-order accurate.

For the second test, we ran simulations of an MHD vortex advected along the diagonal of a square 2D grid. The characteristic velocities involved in the problem were varied such that the maximum Mach number, $\mathcal{M}_\mathrm{son}$, ranged from  $1.55\times10^{-5}$ to  $1.55\times10^{-1}$. This experiment showed that the MHD scheme also scales with second-order accuracy on 2D grids and that the numerical dissipation is independent of $\mathcal{M}_\mathrm{son}$, even though it becomes larger for lower $\mathcal{M}_\mathrm{Alf}$. However, we observed a considerable dissipation only when the magnetic energy of the vortex was $100$ times its rotational kinetic energy. This regime is far from our main astrophysical applications. The dissipation of kinetic energy for both the LHLLD and the standard HLLD solvers has been quantified for different resolutions at a maximum Mach number $\mathcal{M}_\mathrm{son} = 1.55\times10^{-3}$. Even though all the results converged upon grid refinement, conservation of rotational kinetic energy was two orders of magnitude worse when using the HLLD flux instead of LHLLD. We also quantified the efficiency of IESS over a standard SSP-RK2 as a function of $\mathcal{M}_\mathrm{son}$. When the maximum sonic Mach number of the flow is below $(2-5)\times 10^{-2}$, IESS becomes considerably more efficient than explicit time-steppers.

In the third experiment we considered the growth of a Kelvin--Helmholtz instability under the effects of a weak  magnetic field parallel to the shear flow, which is known to generate more complex vortex structures than the case with a strong magnetic field \citep{frank1996}. The second-order convergence of the $y$-direction kinetic energy was checked using both the LHLLD and HLLD solvers for different shear velocities, such that the corresponding sonic Mach number ranged from $10^{-4}$ to $10^{-1}$. Again, we observed that the amount of dissipation was virtually independent of $\mathcal{M}_\mathrm{son}$ for LHLLD, while the numerical solution obtained with HLLD was progressively more degraded as the Mach number was further decreased. This test showed  that HLLD needed  twice or four times the resolution to be as accurate as LHLLD when $\mathcal{M}_\mathrm{son}\sim10^{-3}$ and $\mathcal{M}_\mathrm{son}\sim10^{-2}$, while it only needed $20\%$ more resolution at $\mathcal{M}_\mathrm{son}\sim10^{-1}$.

In our fourth test we simulated the rise of a hot bubble in an adiabatic stratification in MHSE. The initial magnetic field was horizontal and uniform. Different entropy perturbations were considered to test the capabilities of the MHD scheme in modeling slow flows in steep stratifications (in this case, the vertical domain spanned 4.6 pressure scale heights). Overall, the entropy content of the bubble was very well preserved and all results converged upon grid refinement. Thanks to the deviation method, we were able to successfully simulate the rise of the bubble for entropy perturbations as low as $10^{-7}$, leading to typical sonic Mach numbers of $3\times10^{-4}$. A relation between the rise velocity of the bubble and the entropy perturbation has been obtained on the basis of physical considerations. We show that the results obtained with this MHD scheme could satisfy that relation even on coarse grids, which suggests that discretization errors arising from the background stratification did not play any significant role in the evolution of the bubble. For comparison, we also ran the same setup at an intermediate entropy perturbation without well-balancing. The unbalanced states led to the generation of large pressure jumps at the cell interfaces, which launched strong waves that degraded the numerical solution at low resolution. Even when the magnitude of these errors was progressively reduced at higher resolution, they were still significant on the finest grid.

Lastly, we ran a fully convective box on a 3D Cartesian grid (with $2 N \times N \times 2 N$ grid cells) to simulate an SSD. The initial stratification was in MHSE and it resembled the thermodynamic conditions found in oxygen shell burning in a massive star \citep{andrassy2022}. Convection was driven by a heat source placed at the bottom of the box, and a weak seed magnetic field was planted in the system to initiate the dynamo. By changing the rate of energy injection, we were able to study different velocity regimes. In particular, we simulated three different cases, with $\mathcal{M}_\mathrm{son}\sim10^{-3}$, $10^{-2}$, and $10^{-1}$. We only considered moderate grid resolutions ($N=32,64,96$). In the kinematic phase, the magnetic field energy was exponentially amplified on the smallest scales of the turbulent flow, with a higher growth rate in finer grids, which is consistent with SSD amplification. When the Lorentz force started to affect the evolution of the fluid, the saturated nonlinear phase began. Because of the use of a low-Mach Riemann solver, the amount of magnetic energy amplified (compared to the kinetic energy content of the flow) did not depend on the sonic Mach number of convection and achieved sub-equipartition values for the resolutions considered in this study ($E_\mathrm{M}/E_\mathrm{K}\simeq0.1-0.2$).

Overall, the results obtained in these tests demonstrate that the numerical methods implemented in SLH can accurately and efficiently tackle a variety of MHD processes that act in stellar interiors, in regimes that are inaccessible to conventional methods.

\begin{acknowledgements}
The work of GL and FKR is supported by the German Research
Foundation (DFG) through the grant RO 3676/3-1.
The work of CB and CK is supported by DFG
through the grant KL 566/22-1.
PVFE was supported by the U.S. Department of Energy
through the Los Alamos National Laboratory (LANL). LANL is operated by
Triad National Security, LLC, for the National Nuclear Security Administration
of the U.S. Department of Energy (Contract No. 89233218CNA000001).
GL, RA, JH, GW, and FKR acknowledge support by the Klaus Tschira
Foundation. This work has been assigned a document release number LA-UR-22-27864.
\end{acknowledgements}

\bibliographystyle{aa}
\bibliography{lowmachmhd}

\begin{thebibliography}{113}
\expandafter\ifx\csname natexlab\endcsname\relax\def\natexlab#1{#1}\fi

\bibitem[{{Andrassy} {et~al.}(2020){Andrassy}, {Herwig}, {Woodward}, \&
  {Ritter}}]{andrassy2020}
{Andrassy}, R., {Herwig}, F., {Woodward}, P., \& {Ritter}, C. 2020, \mnras,
  491, 972

\bibitem[{{Andrassy} {et~al.}(2022){Andrassy}, {Higl}, {Mao}, {Moc{\'a}k},
  {Vlaykov}, {Arnett}, {Baraffe}, {Campbell}, {Constantino}, {Edelmann},
  {Goffrey}, {Guillet}, {Herwig}, {Hirschi}, {Horst}, {Leidi}, {Meakin},
  {Pratt}, {Rizzuti}, {R{\"o}pke}, \& {Woodward}}]{andrassy2022}
{Andrassy}, R., {Higl}, J., {Mao}, H., {et~al.} 2022, \aap, 659, A193

\bibitem[{{Augustson} {et~al.}(2016){Augustson}, {Brun}, \&
  {Toomre}}]{augustson2016}
{Augustson}, K.~C., {Brun}, A.~S., \& {Toomre}, J. 2016, \apj, 829, 92

\bibitem[{{Aydemir} \& {Barnes}(1985)}]{aydemir1985}
{Aydemir}, A.~Y. \& {Barnes}, D.~C. 1985, Journal of Computational Physics, 59,
  108

\bibitem[{{Balsara}(2004)}]{balsara2004}
{Balsara}, D.~S. 2004, \apjs, 151, 149

\bibitem[{{Balsara} \& {Spicer}(1999)}]{balsara1999}
{Balsara}, D.~S. \& {Spicer}, D.~S. 1999, Journal of Computational Physics,
  149, 270

\bibitem[{Berberich {et~al.}(2021)Berberich, Chandrashekar, \&
  Klingenberg}]{berberich2019}
Berberich, J.~P., Chandrashekar, P., \& Klingenberg, C. 2021, Computers \&
  Fluids, 219, 104858

\bibitem[{{Brackbill} \& {Barnes}(1980)}]{brackbill1980}
{Brackbill}, J.~U. \& {Barnes}, D.~C. 1980, Journal of Computational Physics,
  35, 426

\bibitem[{Brandenburg(2011)}]{brandenburg2011}
Brandenburg, A. 2011, The Astrophysical Journal, 741, 92

\bibitem[{{Brandenburg}(2014)}]{brandenburg2014}
{Brandenburg}, A. 2014, \apj, 791, 12

\bibitem[{{Brandenburg} \& {Subramanian}(2005)}]{brandenburg2005}
{Brandenburg}, A. \& {Subramanian}, K. 2005, \physrep, 417, 1

\bibitem[{{Brown} {et~al.}(2010){Brown}, {Browning}, {Brun}, {Miesch}, \&
  {Toomre}}]{brown2010}
{Brown}, B.~P., {Browning}, M.~K., {Brun}, A.~S., {Miesch}, M.~S., \& {Toomre},
  J. 2010, \apj, 711, 424

\bibitem[{{Browning}(2008)}]{browning2008}
{Browning}, M.~K. 2008, \apj, 676, 1262

\bibitem[{{Browning} {et~al.}(2006){Browning}, {Miesch}, {Brun}, \&
  {Toomre}}]{browning2006}
{Browning}, M.~K., {Miesch}, M.~S., {Brun}, A.~S., \& {Toomre}, J. 2006, \apjl,
  648, L157

\bibitem[{Brun \& Browning(2017)}]{brun2017}
Brun, A. \& Browning, M. 2017, Living Reviews in Solar Physics, 14, 4

\bibitem[{{Brun} {et~al.}(2005){Brun}, {Browning}, \& {Toomre}}]{brun2005}
{Brun}, A.~S., {Browning}, M.~K., \& {Toomre}, J. 2005, \apj, 629, 461

\bibitem[{{Brun} {et~al.}(2004){Brun}, {Miesch}, \& {Toomre}}]{brun2004}
{Brun}, A.~S., {Miesch}, M.~S., \& {Toomre}, J. 2004, \apj, 614, 1073

\bibitem[{{Canivete Cuissa} \& {Teyssier}(2022)}]{cuissa2022}
{Canivete Cuissa}, J.~R. \& {Teyssier}, R. 2022, arXiv e-prints,
  arXiv:2206.06824

\bibitem[{{Cargo} \& {Gallice}(1997)}]{cargo1997}
{Cargo}, P. \& {Gallice}, G. 1997, Journal of Computational Physics, 136, 446

\bibitem[{{Chac{\'o}n}(2008)}]{chacon2008}
{Chac{\'o}n}, L. 2008, Physics of Plasmas, 15, 056103

\bibitem[{{Chandrasekhar}(1961)}]{chandrasekhar1961}
{Chandrasekhar}, S. 1961, {Hydrodynamic and hydromagnetic stability}

\bibitem[{{Charbonneau}(2013)}]{charbonneau2013}
{Charbonneau}, P. 2013, {Solar and Stellar Dynamos}, Solar and Stellar Dynamos:
  Saas-Fee Advanced Course 39 Swiss Society for Astrophysics and Astronomy,
  Saas-Fee Advanced Courses, Volume 39. ISBN 978-3-642-32092-7. Springer-Verlag
  Berlin Heidelberg, 2013

\bibitem[{{Charlton} {et~al.}(1990){Charlton}, {Holmes}, {Lynch}, {Carreras},
  \& {Hender}}]{charlton1990}
{Charlton}, L.~A., {Holmes}, J.~A., {Lynch}, V.~E., {Carreras}, B.~A., \&
  {Hender}, T.~C. 1990, Journal of Computational Physics, 86, 270

\bibitem[{{Colella} \& {Woodward}(1984)}]{colella1984}
{Colella}, P. \& {Woodward}, P.~R. 1984, Journal of Computational Physics, 54,
  174

\bibitem[{{Courant} {et~al.}(1928){Courant}, {Friedrichs}, \&
  {Lewy}}]{courant1928}
{Courant}, R., {Friedrichs}, K., \& {Lewy}, H. 1928, Mathematische Annalen,
  100, 32

\bibitem[{{Dai} \& {Woodward}(1998)}]{dai1998}
{Dai}, W. \& {Woodward}, P.~R. 1998, \apj, 494, 317

\bibitem[{{Dedner} {et~al.}(2002){Dedner}, {Kemm}, {Kr{\"o}ner}, {Munz},
  {Schnitzer}, \& {Wesenberg}}]{dedner2002}
{Dedner}, A., {Kemm}, F., {Kr{\"o}ner}, D., {et~al.} 2002, Journal of
  Computational Physics, 175, 645

\bibitem[{{Dumbser} {et~al.}(2019){Dumbser}, {Balsara}, {Tavelli}, \&
  {Fambri}}]{dumbser2019}
{Dumbser}, M., {Balsara}, D.~S., {Tavelli}, M., \& {Fambri}, F. 2019,
  International Journal for Numerical Methods in Fluids, 89, 16

\bibitem[{{Edelmann}(2014)}]{edelmann2014a}
{Edelmann}, P.~V.~F. 2014, Dissertation, Technische Universit\"at M\"unchen

\bibitem[{{Edelmann} {et~al.}(2021){Edelmann}, {Horst}, {Berberich},
  {Andrassy}, {Higl}, {Leidi}, {Klingenberg}, \& {R{\"o}pke}}]{edelmann2021a}
{Edelmann}, P.~V.~F., {Horst}, L., {Berberich}, J.~P., {et~al.} 2021, \aap,
  652, A53

\bibitem[{Edelmann {et~al.}(2019)Edelmann, Ratnasingam, Pedersen, Bowman, Prat,
  \& Rogers}]{edelmann2019}
Edelmann, P. V.~F., Ratnasingam, R.~P., Pedersen, M.~G., {et~al.} 2019, The
  Astrophysical Journal, 876, 4

\bibitem[{{Edelmann} \& {R\"{o}pke}(2016)}]{edelmann2016b}
{Edelmann}, P.~V.~F. \& {R\"{o}pke}, F.~K. 2016, in {JUQUEEN} {E}xtreme
  {S}caling {W}orkshop 2016, ed. D.~Br\"ommel, W.~Frings, \& B.~J.~N. Wylie,
  JSC Internal Report No. FZJ-JSC-IB-2016-01, 63--67

\bibitem[{{Edelmann} {et~al.}(2017){Edelmann}, {R{\"o}pke}, {Hirschi},
  {Georgy}, \& {Jones}}]{edelmann2017a}
{Edelmann}, P.~V.~F., {R{\"o}pke}, F.~K., {Hirschi}, R., {Georgy}, C., \&
  {Jones}, S. 2017, \aap, 604, A25

\bibitem[{{Einfeldt} {et~al.}(1991){Einfeldt}, {Roe}, {Munz}, \&
  {Sjogreen}}]{einfeldt1991}
{Einfeldt}, B., {Roe}, P.~L., {Munz}, C.~D., \& {Sjogreen}, B. 1991, Journal of
  Computational Physics, 92, 273

\bibitem[{{Evans} \& {Hawley}(1988)}]{evans1988}
{Evans}, C.~R. \& {Hawley}, J.~F. 1988, \apj, 332, 659

\bibitem[{{Fambri}(2021)}]{fambri2021}
{Fambri}, F. 2021, International Journal for Numerical Methods in Fluids, 93,
  3447

\bibitem[{Featherstone \& Hindman(2016)}]{featherstone2016}
Featherstone, N. \& Hindman, B. 2016, The Astrophysical Journal, 818, 32

\bibitem[{{Featherstone} {et~al.}(2009){Featherstone}, {Browning}, {Brun}, \&
  {Toomre}}]{featherstone2009}
{Featherstone}, N.~A., {Browning}, M.~K., {Brun}, A.~S., \& {Toomre}, J. 2009,
  \apj, 705, 1000

\bibitem[{{Felipe} {et~al.}(2010){Felipe}, {Khomenko}, \&
  {Collados}}]{felipe2010}
{Felipe}, T., {Khomenko}, E., \& {Collados}, M. 2010, \apj, 719, 357

\bibitem[{{Frank} {et~al.}(1996){Frank}, {Jones}, {Ryu}, \&
  {Gaalaas}}]{frank1996}
{Frank}, A., {Jones}, T.~W., {Ryu}, D., \& {Gaalaas}, J.~B. 1996, \apj, 460,
  777

\bibitem[{{Fuchs} {et~al.}(2009){Fuchs}, {Mishra}, \& {Risebro}}]{fuchs2009}
{Fuchs}, F.~G., {Mishra}, S., \& {Risebro}, N.~H. 2009, Journal of
  Computational Physics, 228, 641

\bibitem[{{Gardiner} \& {Stone}(2005)}]{gardiner2005}
{Gardiner}, T.~A. \& {Stone}, J.~M. 2005, Journal of Computational Physics,
  205, 509

\bibitem[{{Gardiner} \& {Stone}(2008)}]{gardiner2008}
{Gardiner}, T.~A. \& {Stone}, J.~M. 2008, Journal of Computational Physics,
  227, 4123

\bibitem[{{Gastine} \& {Wicht}(2012{\natexlab{a}})}]{gastine2012a}
{Gastine}, T. \& {Wicht}, J. 2012{\natexlab{a}}, \icarus, 219, 428

\bibitem[{{Gastine} \& {Wicht}(2012{\natexlab{b}})}]{gastine2012b}
{Gastine}, T. \& {Wicht}, J. 2012{\natexlab{b}}, \icarus, 219, 428

\bibitem[{{Ghizaru} {et~al.}(2010){Ghizaru}, {Charbonneau}, \&
  {Smolarkiewicz}}]{ghizaru2010}
{Ghizaru}, M., {Charbonneau}, P., \& {Smolarkiewicz}, P.~K. 2010, \apjl, 715,
  L133

\bibitem[{Glasser {et~al.}(1999)Glasser, Sovinec, Nebel, Gianakon, Plimpton,
  Chu, Schnack, \& the NIMROD~Team}]{glasser1999}
Glasser, A.~H., Sovinec, C.~R., Nebel, R.~A., {et~al.} 1999, Plasma Physics and
  Controlled Fusion, 41, A747

\bibitem[{{Glatzmaier}(1984)}]{glatzmaier1984}
{Glatzmaier}, G.~A. 1984, Journal of Computational Physics, 55, 461

\bibitem[{{Glatzmaier}(1985)}]{glatzmaier1985}
{Glatzmaier}, G.~A. 1985, \apj, 291, 300

\bibitem[{Godunov \& Bohachevsky(1959)}]{godunov1959}
Godunov, S.~K. \& Bohachevsky, I. 1959, {Matemati{\v c}eskij sbornik}, 47(89),
  271

\bibitem[{{Harned} \& {Kerner}(1985)}]{harned1985}
{Harned}, D.~S. \& {Kerner}, W. 1985, Journal of Computational Physics, 60, 62

\bibitem[{{Higl} {et~al.}(2021){Higl}, {M{\"u}ller}, \& {Weiss}}]{higl2021}
{Higl}, J., {M{\"u}ller}, E., \& {Weiss}, A. 2021, \aap, 646, A133

\bibitem[{{Horst} {et~al.}(2020){Horst}, {Edelmann}, {Andr{\'a}ssy},
  {R{\"o}pke}, {Bowman}, {Aerts}, \& {Ratnasingam}}]{horst2020a}
{Horst}, L., {Edelmann}, P.~V.~F., {Andr{\'a}ssy}, R., {et~al.} 2020, \aap,
  641, A18

\bibitem[{{Horst} {et~al.}(2021){Horst}, {Hirschi}, {Edelmann}, {Andrassy}, \&
  {Roepke}}]{horst2021a}
{Horst}, L., {Hirschi}, R., {Edelmann}, P.~V.~F., {Andrassy}, R., \& {Roepke},
  F.~K. 2021, \aap, 653, A55

\bibitem[{{Hotta}(2017)}]{hotta2017}
{Hotta}, H. 2017, \apj, 843, 52

\bibitem[{{Hotta} {et~al.}(2015){Hotta}, {Rempel}, \& {Yokoyama}}]{hotta2015}
{Hotta}, H., {Rempel}, M., \& {Yokoyama}, T. 2015, \apj, 803, 42

\bibitem[{{Iskakov} {et~al.}(2007){Iskakov}, {Schekochihin}, {Cowley},
  {McWilliams}, \& {Proctor}}]{iskakov2007}
{Iskakov}, A.~B., {Schekochihin}, A.~A., {Cowley}, S.~C., {McWilliams}, J.~C.,
  \& {Proctor}, M.~R.~E. 2007, \prl, 98, 208501

\bibitem[{{Jardin}(2012)}]{jardin2012}
{Jardin}, S.~C. 2012, Journal of Computational Physics, 231, 822

\bibitem[{{Jones} {et~al.}(2009){Jones}, {Kuzanyan}, \& {Mitchell}}]{jones2009}
{Jones}, C.~A., {Kuzanyan}, K.~M., \& {Mitchell}, R.~H. 2009, Journal of Fluid
  Mechanics, 634, 291

\bibitem[{{K{\"a}pyl{\"a}}(2011)}]{kapyla2010}
{K{\"a}pyl{\"a}}, P.~J. 2011, Astronomische Nachrichten, 332, 43

\bibitem[{{K{\"a}pyl{\"a}}(2019)}]{kapyla2019}
{K{\"a}pyl{\"a}}, P.~J. 2019, \aap, 631, A122

\bibitem[{{K{\"a}pyl{\"a}}(2021)}]{kapyla2021}
{K{\"a}pyl{\"a}}, P.~J. 2021, \aap, 651, A66

\bibitem[{{K{\"a}pyl{\"a}} {et~al.}(2020){K{\"a}pyl{\"a}}, {Gent}, {Olspert},
  {K{\"a}pyl{\"a}}, \& {Brandenburg}}]{kapyla2020}
{K{\"a}pyl{\"a}}, P.~J., {Gent}, F.~A., {Olspert}, N., {K{\"a}pyl{\"a}}, M.~J.,
  \& {Brandenburg}, A. 2020, Geophysical and Astrophysical Fluid Dynamics, 114,
  8

\bibitem[{{K{\"a}pyl{\"a}} {et~al.}(2008){K{\"a}pyl{\"a}}, {Korpi}, \&
  {Brandenburg}}]{kapyla2008}
{K{\"a}pyl{\"a}}, P.~J., {Korpi}, M.~J., \& {Brandenburg}, A. 2008, \aap, 491,
  353

\bibitem[{{K{\"a}pyl{\"a}} {et~al.}(2011){K{\"a}pyl{\"a}}, {Mantere}, \&
  {Brandenburg}}]{kapyla2011}
{K{\"a}pyl{\"a}}, P.~J., {Mantere}, M.~J., \& {Brandenburg}, A. 2011,
  Astronomische Nachrichten, 332, 883

\bibitem[{{K{\"a}pyl{\"a}} {et~al.}(2012){K{\"a}pyl{\"a}}, {Mantere}, \&
  {Brandenburg}}]{kapyla2012}
{K{\"a}pyl{\"a}}, P.~J., {Mantere}, M.~J., \& {Brandenburg}, A. 2012, \apjl,
  755, L22

\bibitem[{{K{\"a}pyl{\"a}} {et~al.}(2013){K{\"a}pyl{\"a}}, {Mantere}, {Cole},
  {Warnecke}, \& {Brandenburg}}]{kapyla2013}
{K{\"a}pyl{\"a}}, P.~J., {Mantere}, M.~J., {Cole}, E., {Warnecke}, J., \&
  {Brandenburg}, A. 2013, \apj, 778, 41

\bibitem[{{Karak, B. B.} {et~al.}(2015){Karak, B. B.}, {K\"apyl\"a, P. J.},
  {K\"apyl\"a, M. J.}, {Brandenburg, A.}, {Olspert, N.}, \& {Pelt,
  J.}}]{karak2015}
{Karak, B. B.}, {K\"apyl\"a, P. J.}, {K\"apyl\"a, M. J.}, {et~al.} 2015, A\&A,
  576, A26

\bibitem[{{Kazantsev}(1968)}]{kazantsev1968}
{Kazantsev}, A.~P. 1968, Soviet Journal of Experimental and Theoretical
  Physics, 26, 1031

\bibitem[{{Khomenko} \& {Collados}(2006)}]{khomenko2006}
{Khomenko}, E. \& {Collados}, M. 2006, \apj, 653, 739

\bibitem[{{Kippenhahn} {et~al.}(2013){Kippenhahn}, {Weigert}, \&
  {Weiss}}]{kippenhahn2013}
{Kippenhahn}, R., {Weigert}, A., \& {Weiss}, A. 2013, {Stellar Structure and
  Evolution}

\bibitem[{{Kolmogorov}(1941)}]{kolmogorov1941}
{Kolmogorov}, A. 1941, Akademiia Nauk SSSR Doklady, 30, 301

\bibitem[{{Kupka} \& {Muthsam}(2017)}]{kupka2017}
{Kupka}, F. \& {Muthsam}, H.~J. 2017, Living Reviews in Computational
  Astrophysics, 3, 1

\bibitem[{{Lerbinger} \& {Luciani}(1991)}]{lerbinger1991}
{Lerbinger}, K. \& {Luciani}, J.~F. 1991, Journal of Computational Physics, 97,
  444

\bibitem[{LeVeque(2002)}]{leveque2002}
LeVeque, R.~J. 2002, Finite Volume Methods for Hyperbolic Problems, Cambridge
  Texts in Applied Mathematics (Cambridge University Press)

\bibitem[{{Londrillo} \& {del Zanna}(2004)}]{londrillo2004}
{Londrillo}, P. \& {del Zanna}, L. 2004, Journal of Computational Physics, 195,
  17

\bibitem[{{L{\"u}tjens} \& {Luciani}(2010)}]{lutjens2010}
{L{\"u}tjens}, H. \& {Luciani}, J.-F. 2010, Journal of Computational Physics,
  229, 8130

\bibitem[{{Masada} {et~al.}(2013){Masada}, {Yamada}, \&
  {Kageyama}}]{masada2013}
{Masada}, Y., {Yamada}, K., \& {Kageyama}, A. 2013, \apj, 778, 11

\bibitem[{{Matthaeus} \& {Brown}(1988)}]{matthaeus1988}
{Matthaeus}, W.~H. \& {Brown}, M.~R. 1988, Physics of Fluids, 31, 3634

\bibitem[{{Meneguzzi} {et~al.}(1981){Meneguzzi}, {Frisch}, \&
  {Pouquet}}]{meneguzzi1981}
{Meneguzzi}, M., {Frisch}, U., \& {Pouquet}, A. 1981, \prl, 47, 1060

\bibitem[{{Mestel}(1999)}]{mestel1999}
{Mestel}, L. 1999, {Stellar magnetism}

\bibitem[{Miczek(2013)}]{miczek2013a}
Miczek, F. 2013, Dissertation, Technische Universit\"at M\"unchen

\bibitem[{{Miczek} {et~al.}(2015){Miczek}, {R{\"o}pke}, \&
  {Edelmann}}]{miczek2015a}
{Miczek}, F., {R{\"o}pke}, F.~K., \& {Edelmann}, P.~V.~F. 2015, \aap, 576, A50

\bibitem[{{Mignone} \& {Del Zanna}(2021)}]{mignone2021}
{Mignone}, A. \& {Del Zanna}, L. 2021, Journal of Computational Physics, 424,
  109748

\bibitem[{{Minoshima} {et~al.}(2020){Minoshima}, {Kitamura}, \&
  {Miyoshi}}]{minoshima2020}
{Minoshima}, T., {Kitamura}, K., \& {Miyoshi}, T. 2020, \apjs, 248, 12

\bibitem[{Minoshima \& Miyoshi(2021)}]{minoshima2021}
Minoshima, T. \& Miyoshi, T. 2021, Journal of Computational Physics, 446,
  110639

\bibitem[{{Miyoshi} \& {Kusano}(2005)}]{miyoshi2005}
{Miyoshi}, T. \& {Kusano}, K. 2005, Journal of Computational Physics, 208, 315

\bibitem[{Müller(2020)}]{muller2020}
Müller, B. 2020, Living Reviews in Computational Astrophysics, 6, 3

\bibitem[{{Pietarila Graham} {et~al.}(2010){Pietarila Graham}, {Cameron}, \&
  {Sch{\"u}ssler}}]{graham2010}
{Pietarila Graham}, J., {Cameron}, R., \& {Sch{\"u}ssler}, M. 2010, \apj, 714,
  1606

\bibitem[{{Powell}(1994)}]{powell1994}
{Powell}, K.~G. 1994, {Approximate Riemann solver for magnetohydrodynamics
  (that works in more than one dimension)}

\bibitem[{{Powell} {et~al.}(1999){Powell}, {Roe}, {Linde}, {Gombosi}, \& {De
  Zeeuw}}]{powell1999}
{Powell}, K.~G., {Roe}, P.~L., {Linde}, T.~J., {Gombosi}, T.~I., \& {De Zeeuw},
  D.~L. 1999, Journal of Computational Physics, 154, 284

\bibitem[{Rempel(2005)}]{rempel2005}
Rempel, M. 2005, The Astrophysical Journal, 622, 1320

\bibitem[{{Rempel}(2014)}]{rempel2014}
{Rempel}, M. 2014, \apj, 789, 132

\bibitem[{Rempel(2018)}]{rempel2018}
Rempel, M. 2018, The Astrophysical Journal, 859, 161

\bibitem[{{Riva} \& {Steiner}(2022)}]{riva2022}
{Riva}, F. \& {Steiner}, O. 2022, \aap, 660, A115

\bibitem[{{Rogers} {et~al.}(2013){Rogers}, {Lin}, {McElwaine}, \&
  {Lau}}]{rogers2013}
{Rogers}, T.~M., {Lin}, D.~N.~C., {McElwaine}, J.~N., \& {Lau}, H.~H.~B. 2013,
  \apj, 772, 21

\bibitem[{{Schekochihin} {et~al.}(2004){Schekochihin}, {Cowley}, {Taylor},
  {Maron}, \& {McWilliams}}]{schekochihin2004}
{Schekochihin}, A.~A., {Cowley}, S.~C., {Taylor}, S.~F., {Maron}, J.~L., \&
  {McWilliams}, J.~C. 2004, \apj, 612, 276

\bibitem[{Schekochihin {et~al.}(2007)Schekochihin, Iskakov, Cowley, McWilliams,
  Proctor, \& Yousef}]{schekochihin2007}
Schekochihin, A.~A., Iskakov, A.~B., Cowley, S.~C., {et~al.} 2007, New Journal
  of Physics, 9, 300

\bibitem[{Schnack {et~al.}(1987)Schnack, Barnes, Mikic, Harned, \&
  Caramana}]{schnack1987}
Schnack, D., Barnes, D., Mikic, Z., Harned, D.~S., \& Caramana, E. 1987,
  Journal of Computational Physics, 70, 330

\bibitem[{{Seta} \& {Federrath}(2020)}]{seta2020}
{Seta}, A. \& {Federrath}, C. 2020, \mnras, 499, 2076

\bibitem[{Shu \& Osher(1988)}]{shu1988}
Shu, C.-W. \& Osher, S. 1988, Journal of Computational Physics, 77, 439

\bibitem[{{Smolarkiewicz} \& {Charbonneau}(2013)}]{smolarkiewicz2013}
{Smolarkiewicz}, P.~K. \& {Charbonneau}, P. 2013, Journal of Computational
  Physics, 236, 608

\bibitem[{{Stone} {et~al.}(2008){Stone}, {Gardiner}, {Teuben}, {Hawley}, \&
  {Simon}}]{stone2008}
{Stone}, J.~M., {Gardiner}, T.~A., {Teuben}, P., {Hawley}, J.~F., \& {Simon},
  J.~B. 2008, \apjs, 178, 137

\bibitem[{Strang(1968)}]{strang1968}
Strang, G. 1968, SIAM Journal on Numerical Analysis, 5, 506

\bibitem[{Sutherland(2010)}]{sutherland2010}
Sutherland, B.~R. 2010, Internal Gravity Waves (Cambridge University Press)

\bibitem[{{Timmes} \& {Swesty}(2000)}]{timmes2000}
{Timmes}, F.~X. \& {Swesty}, F.~D. 2000, \apjs, 126, 501

\bibitem[{Toro(2009)}]{toro2009a}
Toro, E.~F. 2009, Riemann Solvers and Numerical Methods for Fluid Dynamics: A
  Practical Introduction (Berlin Heidelberg: Springer)

\bibitem[{{T{\'o}th}(2000)}]{toth2000}
{T{\'o}th}, G. 2000, Journal of Computational Physics, 161, 605

\bibitem[{{Viallet} {et~al.}(2011){Viallet}, {Baraffe}, \&
  {Walder}}]{viallet2011}
{Viallet}, M., {Baraffe}, I., \& {Walder}, R. 2011, \aap, 531, A86

\bibitem[{{Viviani} {et~al.}(2019){Viviani}, {K{\"a}pyl{\"a}}, {Warnecke},
  {K{\"a}pyl{\"a}}, \& {Rheinhardt}}]{viviani2019}
{Viviani}, M., {K{\"a}pyl{\"a}}, M.~J., {Warnecke}, J., {K{\"a}pyl{\"a}},
  P.~J., \& {Rheinhardt}, M. 2019, \apj, 886, 21

\bibitem[{{V{\"o}gler} {et~al.}(2005){V{\"o}gler}, {Shelyag}, {Sch{\"u}ssler},
  {Cattaneo}, {Emonet}, \& {Linde}}]{vogler2005}
{V{\"o}gler}, A., {Shelyag}, S., {Sch{\"u}ssler}, M., {et~al.} 2005, \aap, 429,
  335

\bibitem[{{Warnecke} {et~al.}(2016){Warnecke}, {K{\"a}pyl{\"a}},
  {K{\"a}pyl{\"a}}, \& {Brandenburg}}]{warnecke2016}
{Warnecke}, J., {K{\"a}pyl{\"a}}, P.~J., {K{\"a}pyl{\"a}}, M.~J., \&
  {Brandenburg}, A. 2016, \aap, 596, A115

\bibitem[{{Yadav} {et~al.}(2016){Yadav}, {Christensen}, {Wolk}, \&
  {Poppenhaeger}}]{yadav2016}
{Yadav}, R.~K., {Christensen}, U.~R., {Wolk}, S.~J., \& {Poppenhaeger}, K.
  2016, \apjl, 833, L28

\end{thebibliography}

\begin{appendix}

\section{Magnetized Kelvin--Helmholtz instability}
\label{appendix:khi}

This appendix explores the effects of the grid resolution and  strength of the initial magnetic field on the evolution of the Kelvin--Helmholtz instability shown in Sect.~\ref{sec:khi}.

\begin{figure*}[h!]
  \centering
  \includegraphics[width=\textwidth]{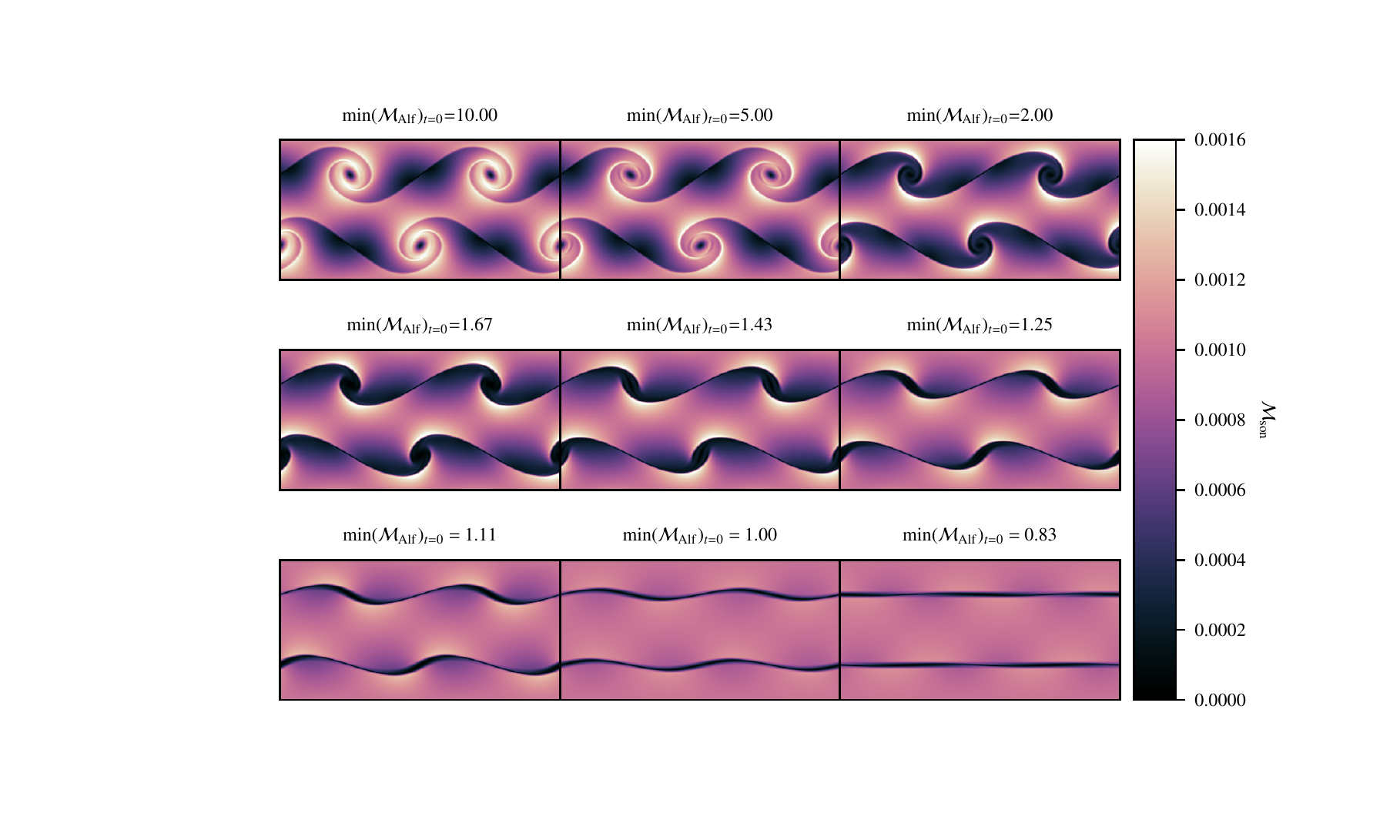}
  \caption{Distribution of the sonic Mach number in the simulations of the Kelvin--Helmholtz instability at $t/t_{\max}=1/6$ for different values of the initial magnetic field, $B_x$, computed as $B_x=\sqrt{\gamma}\alpha$, with $\alpha=(0.1,0.2,0.5,0.6,0.7,0.9,0.9,1.0,1.2)$. These simulations were performed on a $512\times256$ grid with $M_x=10^{-3}$. The title in each panel is the corresponding minimum Alfv\'en Mach number of the flow at $t=0$. For a strong enough initial magnetic field, the magnetic stresses prevent the growth of the instability.}
\label{fig:C-stability}
\end{figure*}

\begin{figure*}[h!]
  \centering
  \includegraphics[width=\textwidth]{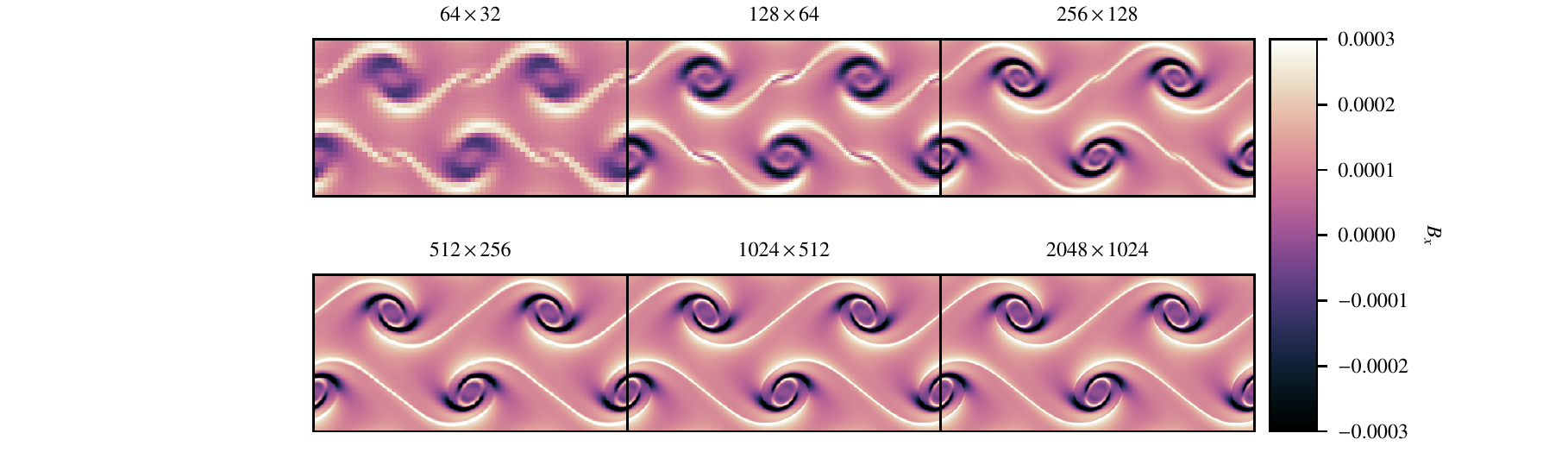}
  \caption{Distribution of $B_x$ in the simulations of the Kelvin--Helmholtz instability at $t/t_{\max}=1/6$ for different grid resolutions, starting from the initial conditions described in Sect.~\ref{sec:khi} with $\mathcal{M}_x=10^{-3}$. On grids with $N\leq128$, numerical discretization errors generate grid-scale vorticity, which leads to the growth of secondary instabilities in the regions between the primary rolls. This effect does not appear in better converged simulations.}
\label{fig:C-resolution}
\end{figure*}

\begin{figure*}[h!]
  \centering
  \includegraphics[width=\textwidth]{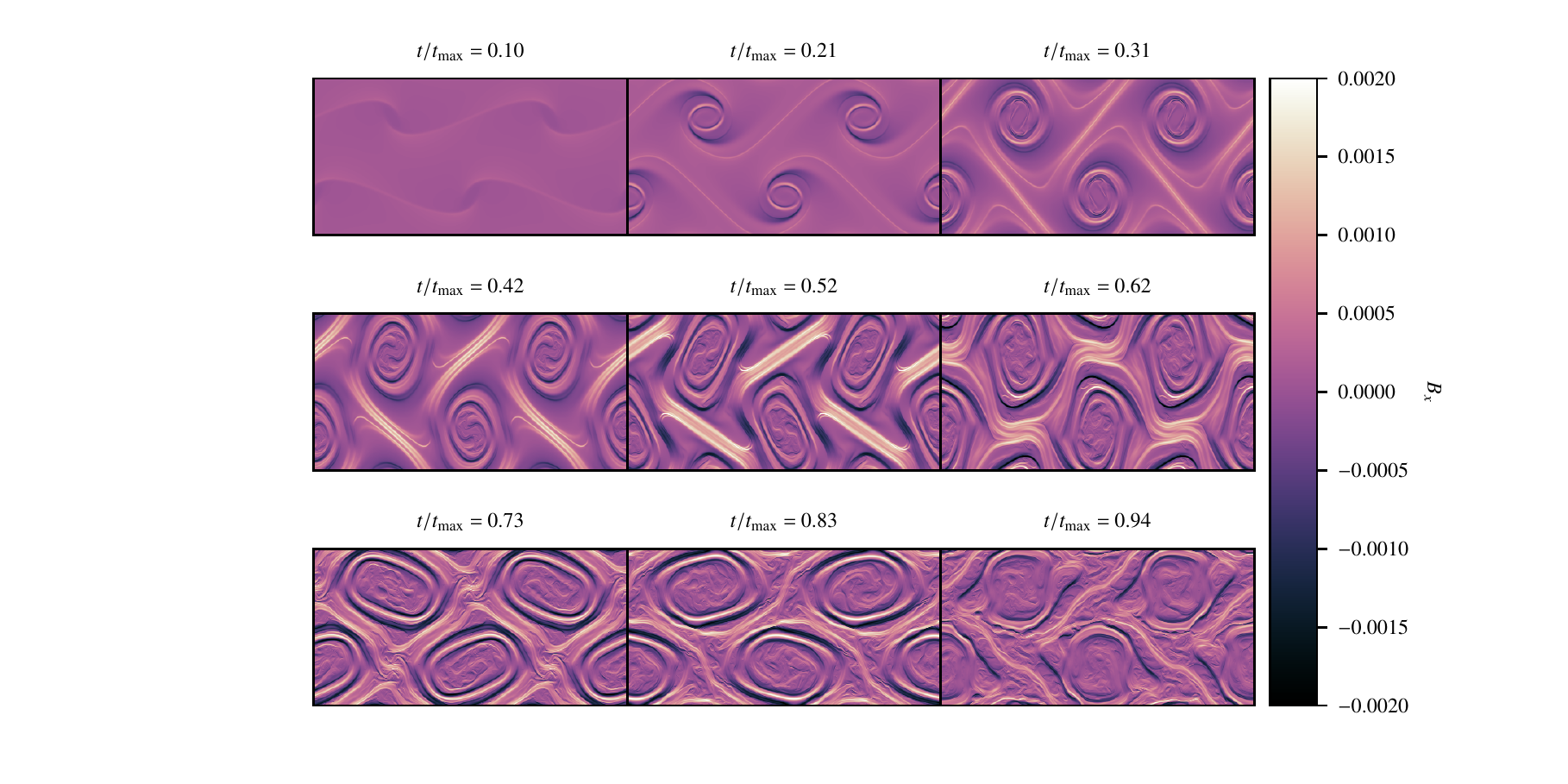}
  \caption{Time evolution of $B_x$ in the simulations of the Kelvin--Helmholtz instability starting from the initial conditions described in Sect.~\ref{sec:khi} with $\mathcal{M}_x=10^{-3}$. The grid is $2048\times1024$.}
\label{fig:C-reconnection}
\end{figure*}

\section{Hot bubble}

Here, we extend the study described in Sect.~\ref{sec:hot-bubble}. In particular, we show the dependence of the entropy fluctuations, $\mathcal{M}_\mathrm{son}$, and $p_B$ on the magnitude of the initial entropy perturbation $(\Delta A/ A)_{t=0}$.

\begin{figure*}[h!]
  \centering
  \includegraphics[width=\textwidth]{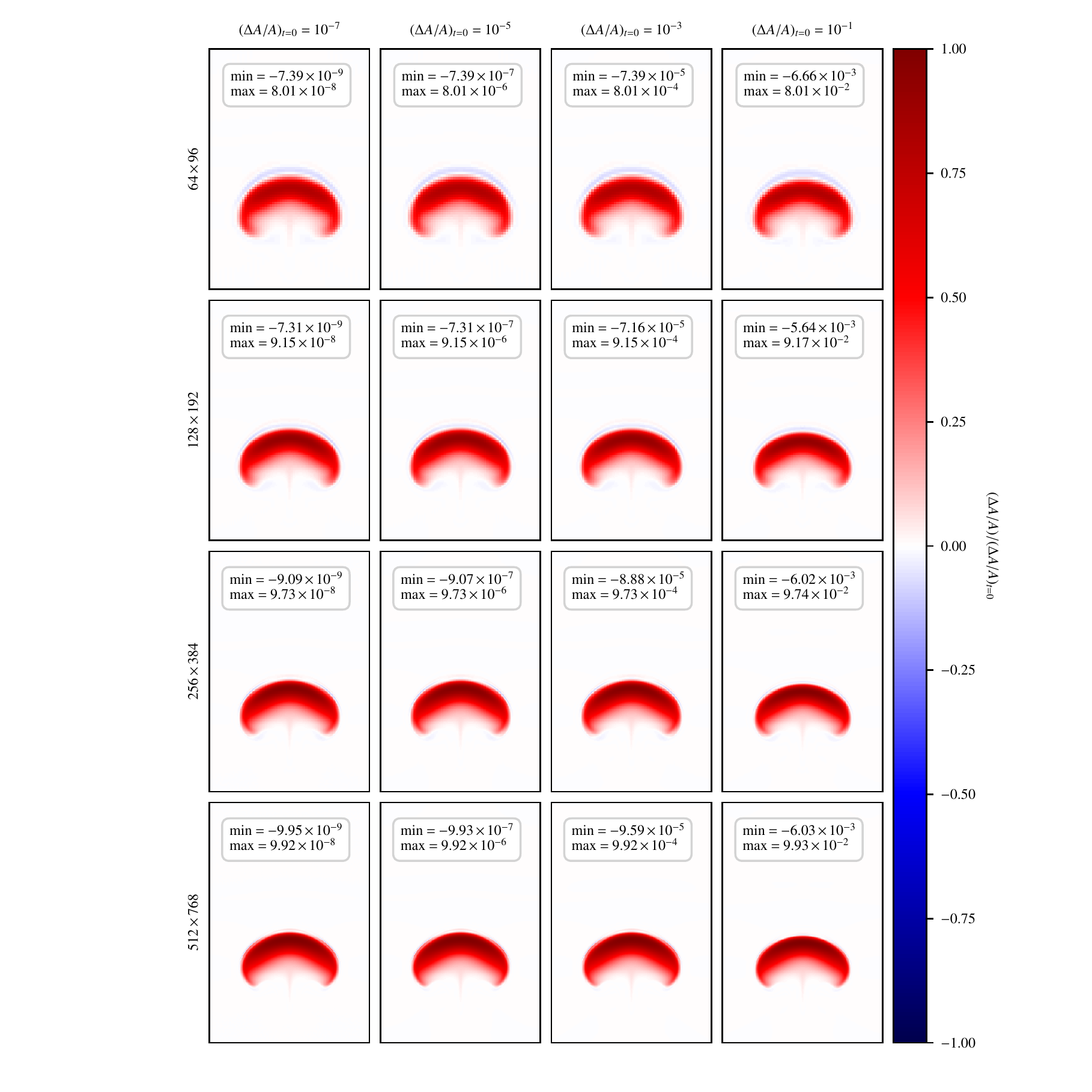}
  \caption{Final distribution of the entropy fluctuations, $\Delta A/A$, of the hot bubble for different values of $(\Delta A/A)_{t=0}$ and grid resolutions. Each panel is rescaled by the corresponding value of $(\Delta A/A)_{t=0}$. The insets provide the minimum and maximum values of the entropy fluctuations in each plot. }
\label{fig:entropy}
\end{figure*}

\begin{figure*}[h!]
  \centering
  \includegraphics[width=\textwidth]{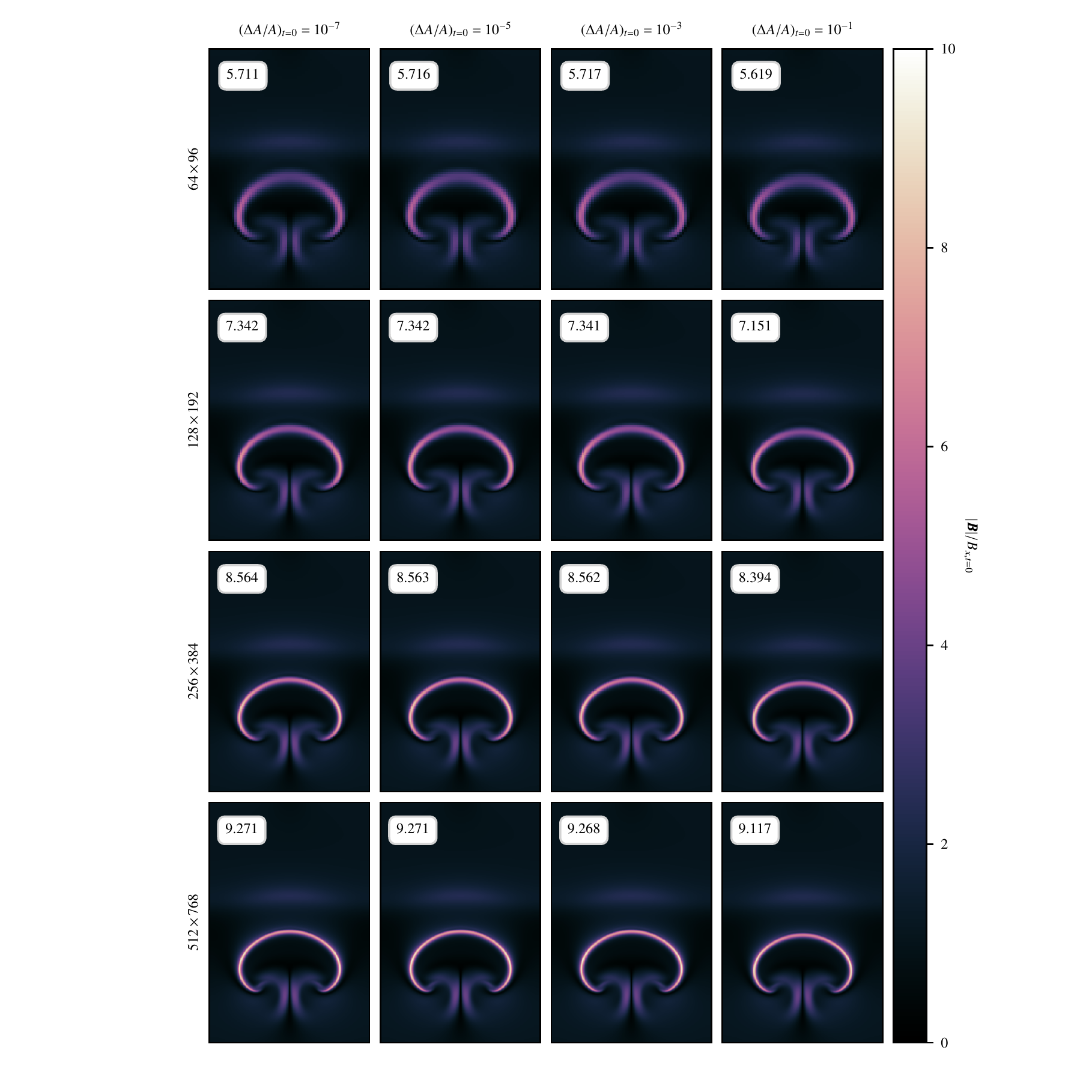}
  \caption{Final distribution of $|\bm{B}|/B_{x,t=0}$ for different values of $(\Delta A/A)_{t=0}$ and grid resolutions in the simulations of the hot bubble. The insets show the maximum ratio in each panel. The amount of numerical resistivity decreases upon grid refinement, which leads to the generation of narrower stripes with stronger magnetic fields.}
\label{fig:magnetic-pressure}
\end{figure*}

\begin{figure*}[h!]
  \centering
  \includegraphics[width=\textwidth]{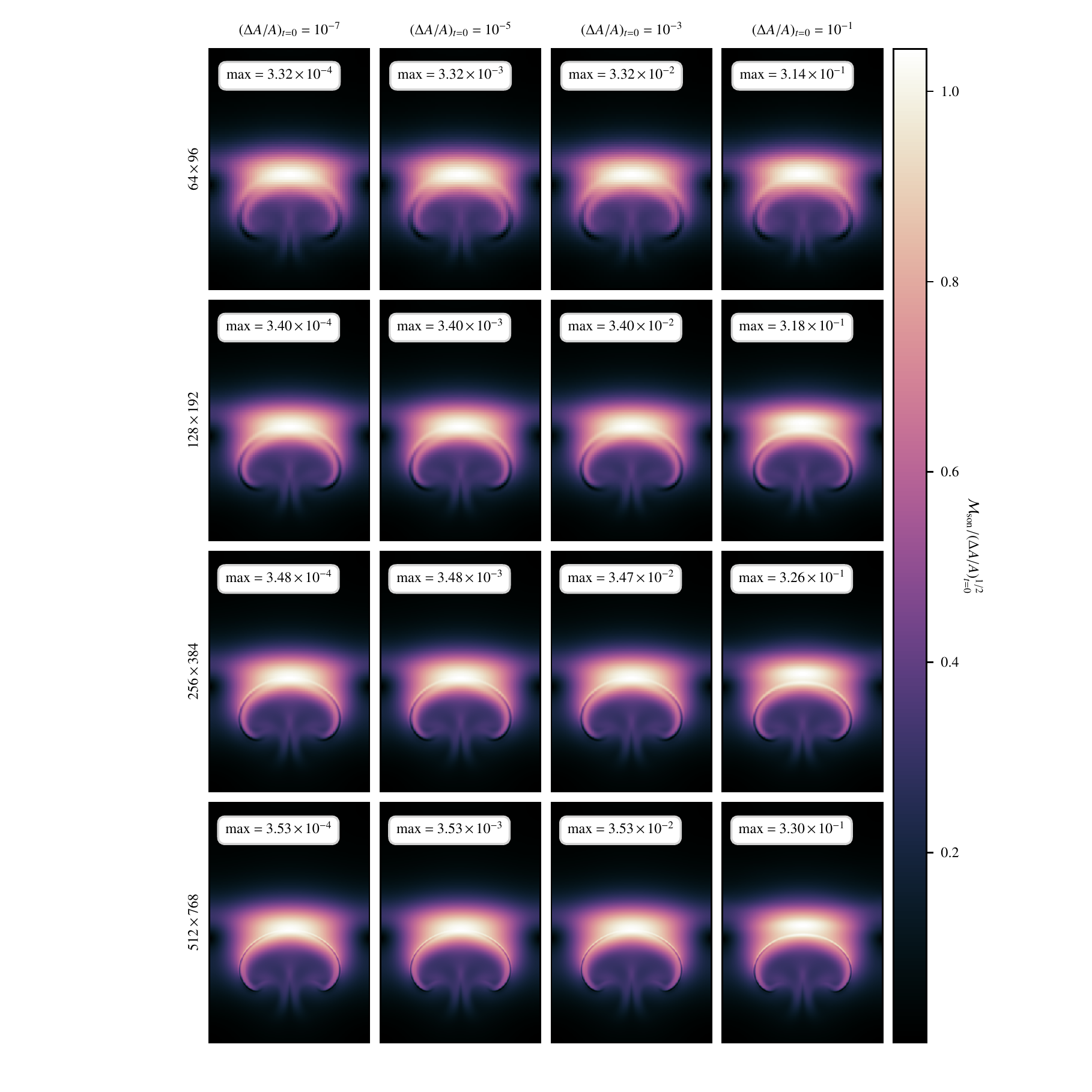}
  \caption{Final distribution of the sonic Mach number of the hot bubble for different values of $(\Delta A/A)_{t=0}$ and different grid resolutions. Each panel is rescaled by the corresponding value of $(\Delta A/A)^{1/2}_{t=0}$. The insets show the maximum sonic Mach number. An entropy perturbation of $(\Delta A/A)_{t=0}=0.1$ drives flows that are far from the low-Mach-number regime. In this case,  effects of compressibility caused by the high ram pressure of the bubble are large enough to cause a $6-7\%$ deviation from the theoretical scaling discussed in Sect.~\ref{sec:hot-bubble}.   }
\label{fig:mach}
\end{figure*}

\section{Small-scale dynamo}

In this section we extend the analysis of the SSD test described in Sect.~\ref{sec:ssd}. In particular, we show 1D vertical averages of the velocity and magnetic field distributions, the time evolution of the numerical divergence of the magnetic field, vertical cuts of the sonic Mach number distribution and the time evolution of the total magnetic energy in the kinematic phase.

\begin{figure*}
  \centering
  \includegraphics[width=\textwidth]{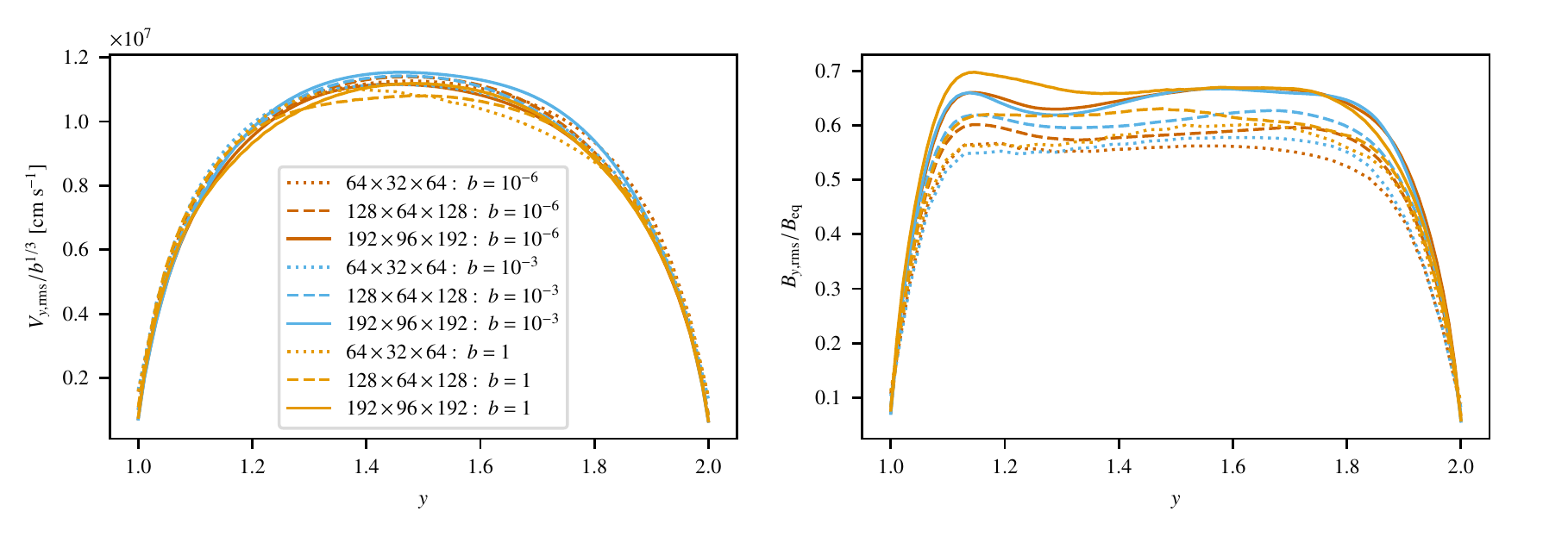}
  \caption{Vertical profiles of $V_y$ (left) and $B_y$ (right) in the simulations of the SSD, averaged over $20<t/\tau_\mathrm{conv}<40$. Different colors represent different grid resolutions, while different line styles are used for representing different values of $b$. The vertical velocity is rescaled by $b^{2/3}$ to remove the dependence of the energy injection rate, while $B_y$ is rescaled by the corresponding equipartition value, $B_{\mathrm{eq}}=\sqrt{\rho}|\mathbf{V}|_\mathrm{rms}$.}
\label{fig:profiles}
\end{figure*}

\begin{figure*}
  \centering
  \includegraphics[width=\textwidth]{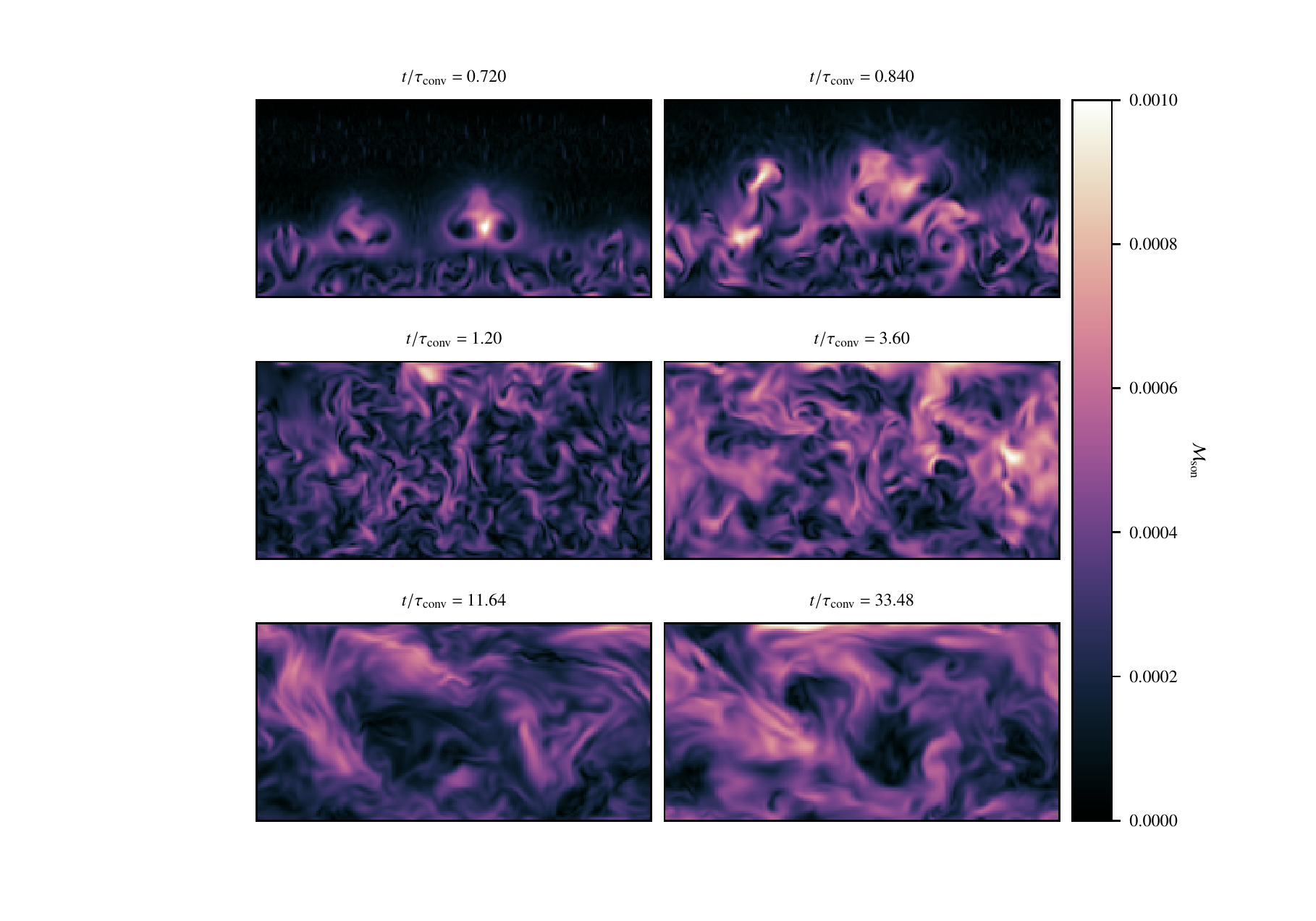}
  \caption{Vertical cuts of the sonic Mach number taken at $z=0$ in the simulations of the SSD run with $b=10^{-6}$ on the $192\times96\times192$ grid. Each panel is taken at a different moment in time, which is given in the title. As the SSD approaches the saturated regime, the small-scale structures in the velocity field are damped by the Lorentz force, and vertical motions mainly happen in the form of large-scale flows.}
\label{fig:vcuts}
\end{figure*}

\begin{figure*}
  \centering
  \includegraphics[width=\textwidth]{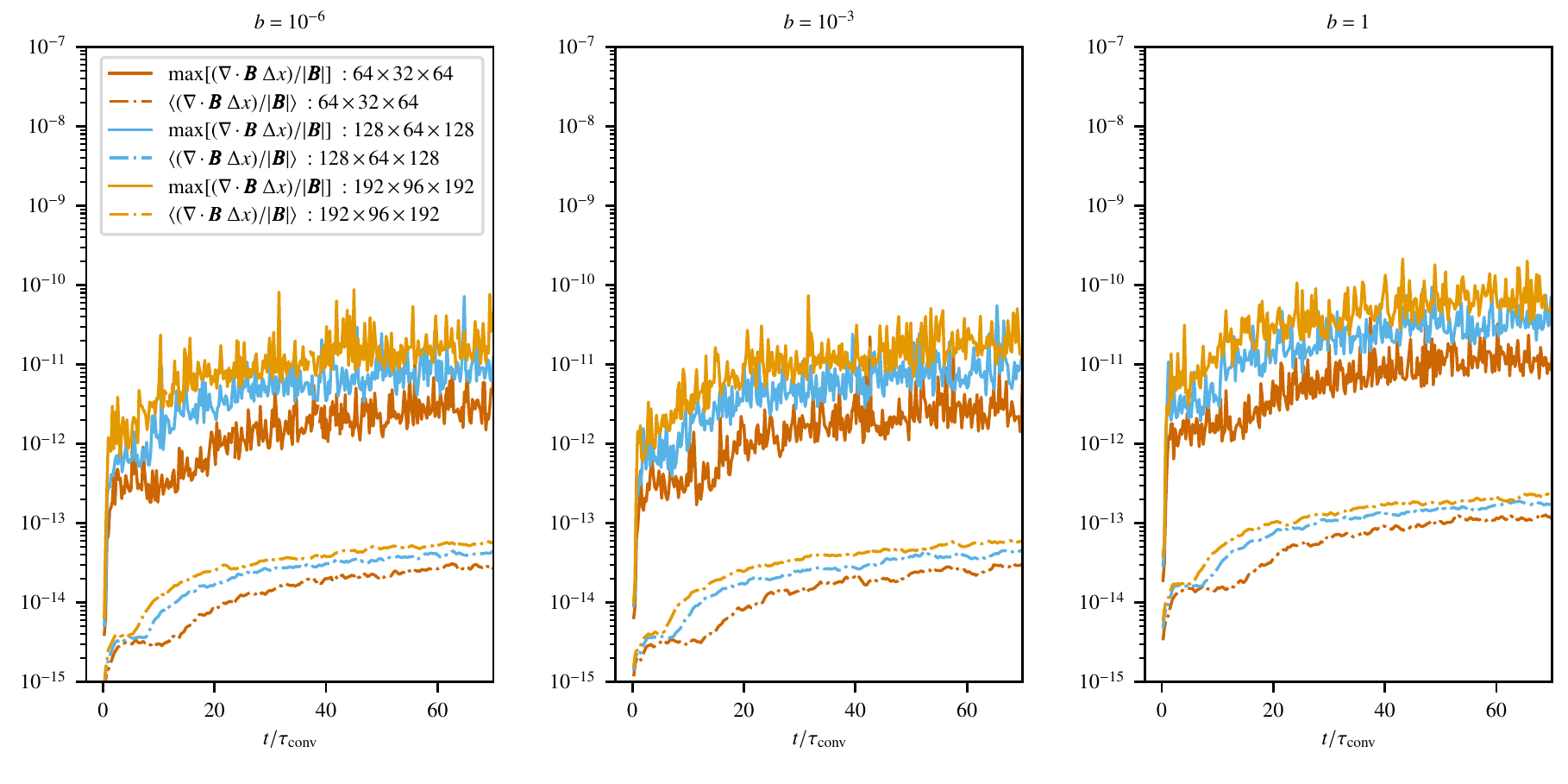}
  \caption{Time evolution of the maximum (solid line) and mean (dot-dashed) relative divergence of the magnetic field in the simulations of the SSD. Since the induction equation is solved using a staggered constrained transport method, the update on the magnetic field keeps the divergence in Eq. \ref{divb} within round-off error. Although these errors accumulate in time, by the end of the simulation magnetic monopoles are still irrelevant to the dynamics of the SSD.}
\label{fig:divb}
\end{figure*}

\begin{figure}
  \centering
  \includegraphics[width=0.5\textwidth]{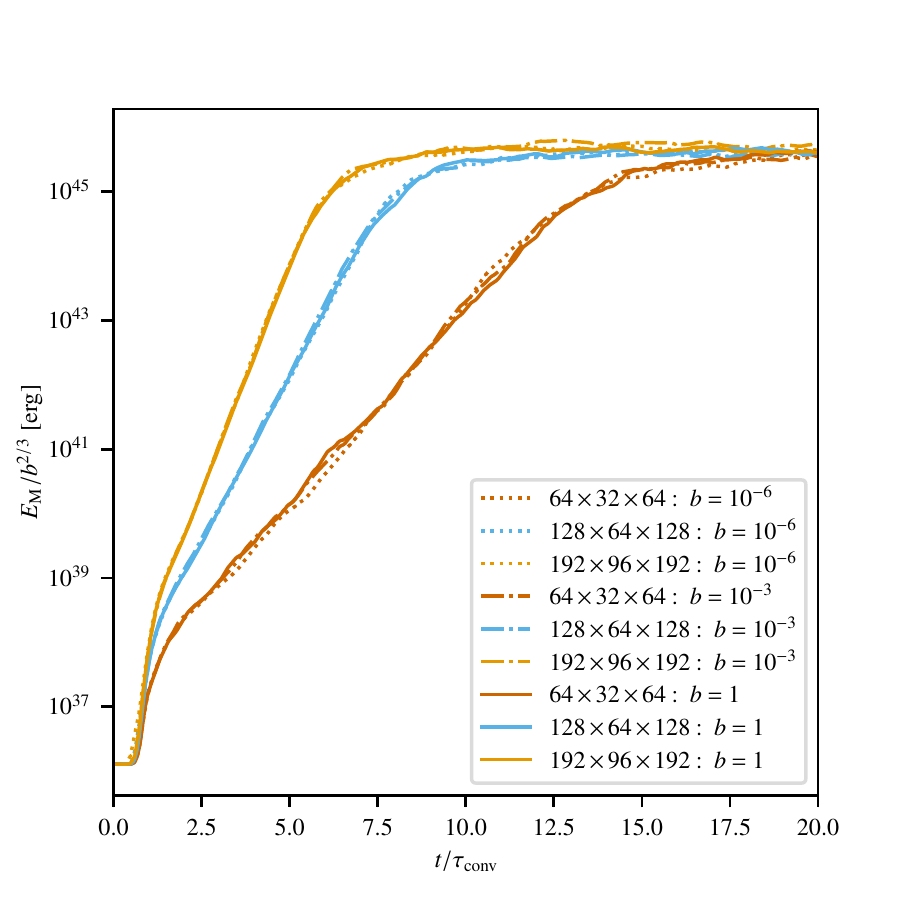}
  \caption{Time evolution of the magnetic energy in the simulations of the SSD up to $t/\tau_\mathrm{conv}=20$. Each line style represents a specific value of the boost factor, $b$, while different colors are used for different numbers of grid cells.}
\label{fig:time-evol-lin}
\end{figure}

\end{appendix}
\end{document}